\documentclass[useAMS,usenatbib,usegraphicx]{mn2e}
\usepackage{ulem} %for remove text by bars
\usepackage[dvips]{graphicx,color}% Include figure files
\usepackage{dcolumn}% Align table columns on decimal point
\usepackage{bm}% bold math
\newcommand{\bvector}[1]{\mbox{\boldmath $#1$}}

\def\aap{A\&A}

\def\apj{ApJ}

\def\mnras{MNRAS}

\def\pasp{PASP}

\usepackage{color}

\title[Rotational equilibria by Lagrangian variational principle]
{Rotational equilibria by Lagrangian variational principle: 
toward multi-dimensional stellar evolutions}

\author[N.Yasutake, K. Fujisawa,  and S. Yamada]
       {Nobutoshi Yasutake$^1$,
       Kotaro Fujisawa$^2$,
       Shoichi Yamada$^2$\\
       $^1$Physics Department, Chiba Institute of Technology, 
       Shibazono 2-1-1, Narashino, Chiba, 275-0023, Japan\\
       $^2$Advanced Research Institute for Science and Engineering, Waseda University, 
       Okubo 3-4-1, Shinjuku, Tokyo, 169-8555, Japan
}

\begin{document}

\maketitle

\begin{abstract}
We have developed a new formulation to obtain self-gravitating, axisymmetric configurations in permanent rotation. The formulation is based on the Lagrangian variational principle with a triangulated mesh. It treats not only barotropic but also baroclinic equations of state. We compare the various stellar equilibria obtained by our new scheme with those by Hachisu's self-consistent field scheme for the barotropic case, and those by Fujisawa's self-consistent field scheme for the baroclinic case. Included in these rotational configurations are those with shellular-type rotations, which are commonly assumed in the evolution calculation of rotating stars. Although radiation processes, convections and meridional flows have not been taken into account in this study, we have in mind the application of this method to the two-dimensional evolution calculations of  rotating stars, for which the Lagrangian formulation is best suited.
\end{abstract}


\section{INTRODUCTION}
Stellar evolution theory is well established especially for spherically symmetric stars, in which there is no rotation and no magnetic field. After the works of pioneers of the field~\citep{b2fh57,cameron57,hayashi61}, Henyey proposed in his seminal papers a method, which later became the defacto standard for stellar evolution calculations~\citep{henyey64}. The method is stable and capable of self-consistently incorporating various physical processes that occur in the stellar interior. It has been modified and extended a lot approximately in the last half century to accommodate rotation\citep{maeder00, woosley02}. 

In stellar evolution calculations, we need to obtain stellar structures consistent with, nuclear reactions (and/or molecule formations), radiative transport of energy, convective and circular motions as well as a realistic equation of state. Since the time-scale of stellar evolution is normally much longer than the dynamical time-scale, it is well approximated by series of time-independent configurations in hydrostatic equilibrium. In the presence of rotation, this is not simple. One of the difficulties is to obtain rotational equilibria for a given angular momentum distribution. What is more, however, we do not know a priori what the distribution of angular momentum looks like in the stellar interior although the problem has been studied theoretically since the 19th century by Carl Jacobi, Richard Dedekind, Peter Lejeune Dirichlet, and Bernhard Riemann to mention a few \citep{chandrasekhar}. It was in the last century that some methods to obtain rotating hydrostatic  equilibria were proposed, and it was mathematically shown that cylindrical distributions of specific angular momentum are realized for EOSs, in which matter pressure is a function of density alone. This type of EOS is referred to ``{\it barotrope}"~\citep{ostriker, hachisu86}. 

The EOS is {\it not barotropic} in the realistic stellar interior, but {\it baroclinic} in general, i.e., pressure depends on density, entropy and chemical compositions~\footnote{For compact stars such as white dwarfs and neutron stars, the EOS may be approximately {\it barotropic}, since the temperature is negligibly low.}. Then the numerical construction of rotational equilibria becomes much more difficult compared with the case for barotropes, since the first integral of the hydrostatic equations that is available in the barotropic case no longer exists in the baroclinic case~\citep{uryu94, uryu95, roxburgh06, espinosa07, espinosa13, rieutord16}. Even if some rotational configurations are somehow obtained, it is a highly non-trivial issue how to make a sequence out of them that represents stellar evolutions appropriately. Note that in the presence of rotation fluid elements composing rotating stars could change their positions non-radially in complicated ways as the stars evolve in time even without convection. It would be highly difficult to describe such displacements of fluid elements with fluxes on a fixed numerical mesh from a Eulerian point of view, since the motions are extremely slow. The Lagrangian formulation will be hence more appropriate just as in the spherically symmetric case. And this is exactly the idea in this paper. 
%以下、根拠を示してないので削除
%Note that our formulation can save the numerical costs compared with the other particle-based methods to obtain the Eulerian values, e.g. pressure, density, and temperature, since 
We introduced in this paper a triangulated mesh with each node representing a fluid element approximately. Starting from an arbitrary reference configuration, we seek for the Lagrangian displacement for each node that gives a rotational equilibrium as a result. In so doing, the variational principle is employed. The method has its own difficulty, though. 
% in searching for solutions of hydrostatic equilibria; it is necessary to eliminate the gauge freedom in the displacement, otherwise, we can not identifies the solution. In this study, we introduce the Mote-Calro technique, which enables us to neglect this gauge freedom.

In this paper, we construct a couple of configurations in rotational equilibrium with different angular momentum distributions in order to demonstrate the capability of our new scheme. Included in them is the so-called shellular-type rotations. It has been argued that in the stellar interior turbulence is anisotropic, operating more strongly within each isobaric surface than in its normal direction, and that a uniform rotation should be realized in the isobar as a consequence. Such a rotation law is referred to as ``shellular rotation" rather than cylindrical rotation \citep{zahn92, meynet97}. 
It should be stressed, however, that no rotational equilibrium with the shellular rotation was constructed numerically, not to mention analytically \footnote{The obvious exception is uniformly rotating stars.}. Very recently Fujisawa, one of the authors of this paper, has succeeded in producing some configurations with shellular-type rotation\citep{fujisawa15}. \footnote{By "shellular-type" we mean that the configuration is not perfectly but almost shellular with iso-angular-velocity surfaces being nearly aligned with isobars.} It is a nice demonstration of performance of our new scheme to reconstruct these configurations.

The organization of the paper is as follows. 
In Section 2, we describe the formulation in detail: the Lagrangian variational principle, on which our scheme (referred to as the YFY scheme hereafter) is based, will  be reviewed first; then the finite discretization on the triangulated grid will be explained and the handling of self-gravity is also mentioned; finally the minimization of the energy function with the Monte Carlo technique, which is analogous to those utilized, e.g., in nuclear physics, will be described. In Section 3, we demonstrate the nice performance of YFY scheme, constructing some rotational configurations and comparing them with the those obtained by other numerical schemes, including Fujisawa's self-consistent field scheme mentioned above. In the last section, we summarize this study and discuss possible extensions of our scheme as future works.

\section{Formulation}
Rotational equilibria should satisfy the following equations:
(i) the force balance equation:   
\begin{eqnarray}
\nabla P + \rho \nabla \phi - \frac{\rho j^2}{\varpi^3} {\bf e_{\varpi}}= 0, \label{eq:equilibrium}
\end{eqnarray}
where $P$, $\rho$, $\phi$, $j$, $\varpi$, ${\bf e_{\varpi}}$ are pressure, density, gravitational potential, specific angular momentum,  the distance from the rotation axis, and the unit vector perpendicular to the rotation axis, respectively; \\
(ii) the Poisson equation:
\begin{eqnarray}
\Delta \phi = 4 \pi G \rho,\label{eq:poisson}
\end{eqnarray}
where $G$ is the gravitational constant; these equations are supplemented with the equations of state~(EOS).\\
%\begin{eqnarray}
%P=K(\bvector{r}) \rho^{1+1/N}, \label{eq:eos}
%\label{eq:eos}
%\end{eqnarray}
%where $K$ is a coefficient depended on the \bvector{r}, and $N$ is a polytropic-type index. 

Although equations (1), (2) look simple, especially for a very realistic EOS, numerically solving them may not be so easy. Most of previous works assumed a barotropic EOS, in which pressure is a function of density alone, and for which it is mathematically shown that rotation is uniform on concentric cylinders and, as a consequence, there is a first integral of equation (1).
The realistic equation of state, however, is not barotropic but baroclinic, in which case pressure depends on density, entropy and chemical compositions. 
Then there is no first integral available any longer and the rotation law becomes unknown a priori. In this paper, we do not use the force balance equation, equation (1). Instead the variational principle is utilized as described in detail below.

\subsection{Lagrangian variational principle} 
\label{subsec:lagrangian}
We first describe the Lagrangian variational principle, which is obeyed by rotational equilibria. We obtain the energy functional, which is minimized by a rotational equilibrium and will be discretized on a triangulated mesh later for numerical calculations. The variational principle itself is not our own invention but is actually given in a textbook \citep{tassoul78}, although it is a core of the entire formulation proposed in this paper.
 
Suppose an arbitrary reference configuration with density, specific entropy and  specific angular momentum distributions $({\rho}(\bvector{\bar{r}}), {s}(\bvector{\bar{r}}), {j}(\bvector{\bar{r}}))$, where \bvector{\bar{r}} is a spatial coordinate and will also serve as a Lagrange coordinate as will become clear shortly. We, then, introduce a diffeomorphic mapping: 
$\bvector{{r}}=\bvector{f}(\bvector{\bar{r}})$, which is identified with the Lagrangian displacement that deforms the reference configuration to another configuration. Then the following relation should hold if the mass were to be conserved in the deformation: 
\begin{equation}
{\rho}(\bvector{\bar{r}})~dV(\bvector{\bar{r}})= \rho(\bvector{r})~dV(\bvector{{r}}),
\label{eq:mass_cnsv}
\end{equation} 
where $\rho(\bvector{\bar{r}})$ is a density at $\bvector{\bar{r}}$ and $dV(\bvector{\bar{r}})$ is an infinitesimal volume around $\bvector{\bar{r}}$ in the reference configuration whereas $\rho(\bvector{{r}})$ and $dV(\bvector{{r}})$ are the density and volume at the corresponding position in the mapped configuration.The volume element is transformed with the Jacobian $J$ of the mapping as 
\begin{equation}
dV(\bvector{\bar{r}})= {J}(\bvector{f}^{-1})~dV(\bvector{{r}}), 
\end{equation} 
the density after the displacement is expressed as 
\begin{equation}
\rho(\bvector{r}) = {\rho}(\bvector{\bar{r}}) {J}(\bvector{f}^{-1}) .
\end{equation} 

Introducing another diffeomorphism $\bvector{g}$, we consider the consecutive mapping $\bvector{g \cdot f}$, which we also refer to as the product of maps $\bvector{g}$ and $\bvector{f}$ and interpret as a Lagrangian displacement:
\begin{equation}
\bvector{\bar{r}} \stackrel{\bvector{f}}{\to} \bvector{{r}} \stackrel{\bvector{g}}{\to}\bvector{{r}'},
\label{eq:mapping}
\end{equation} 
where $\bvector{{r}'}$ is the coordinate after the mapping $\bvector{g}$.

Using the formula for the product of Jacobians, 
\begin{equation}
{J}((\bvector{g\cdot f})^{-1}) = {J}(\bvector{f}^{-1}) \cdot  {J}(\bvector{g}^{-1}),
\end{equation} 
we obtain the density after the mapping shown in equation (\ref{eq:mapping}) as
\begin{equation}
\rho(\bvector{r'}) = {\rho}(\bvector{\bar{r}}) {J}((\bvector{g \cdot f})^{-1} )
= {\rho}(\bvector{\bar{r}}) {J}(\bvector{f}^{-1} ){J}(\bvector{g}^{-1} )
= \rho(\bvector{r}) {J}(\bvector{g}^{-1} ).
\end{equation} 
This equation implies that the intermediate configuration obtained after the mapping $\bvector{f}$ served as the reference configuration for the configuration after the mapping $\bvector{g}$ and one can forget about the original reference configuration. Employing this fact, we consider infinitesimal displacements alone in the following:
\begin{equation}
\delta \bvector{f}: \bvector{r} \rightarrow \bvector{r'} = \bvector{r} + \bvector{\xi}
\end{equation} 
where $\bvector{\xi}$ is the so-called Lagrangian displacement vector, an infinitesimal vector that connects the original point with its image by the map. The Jacobian for the (inverse) infinitesimal mapping is given as 
\begin{equation}
J (\delta \bvector{f}^{-1})= J ^{-1} (\delta \bvector{f}) = 1- \mbox{div}(\bvector{\xi}).
\end{equation} 
Then the Lagrangian variation of density $\mathit{\Delta} \rho$ is obtained as 
\begin{equation}
\mathit{\Delta} \rho \equiv \rho(\bvector{r'}) - \rho(\bvector{r}) 
= (J (\delta \bvector{f}^{-1})-1)\rho(\bvector{r})
%= - \textgt{div} (\delta \bvector{r}) \rho(\bvector{r}) .
= - \mbox{div} (\bvector{\xi}) \rho(\bvector{r}) .
\label{eq:del_rho}
\end{equation}

In this study, we consider the variational principle, in which rotational equilibria are obtained. Supposing that there is no Lagrangian variation in specific entropy and angular momentum for the above infinitesimal displacement,
\begin{eqnarray}
\Delta s =0, \label{eq:del_s} \\
\Delta j =0, \label{eq:del_j} 
\end{eqnarray}
we consider the variation of the following functional, to which we refer as the energy functional \footnote{This is regarded as a functional of the map $\bvector{f}$. Its variation is a functional of $\delta \bvector{f}$ and hence of $\bvector{\xi}$ as shown shortly in the following.} hereafter:
\begin{equation}
E(\bvector{\xi}) = \int\!\! \varepsilon \rho \, dV + \frac{1}{2} \! \int \!\! \rho \phi \, dV + \int \!\! \frac{1}{2} \, \rho \! \left ( \frac{j}{\varpi} \right )^2 dV. 
\label{eq:e}
\end{equation} 
In this expression the integration domain is the stellar interior, and $\varepsilon$ and $\phi$ denote  the specific internal energy and gravitational potential, respectively; $\varpi$ is the distance from the rotational axis. We consider the variation of each term on the right hand side of equation (\ref{eq:e}) in turn. The variation of the first term is evaluated as follows:
\begin{eqnarray}
 \delta \left( \int\!\! \varepsilon \rho \, dV \right)
&=&  \int\!\! \mathit{\Delta} \varepsilon \, \rho \, dV
= \int\!\! \left( \frac{\partial \varepsilon}{\partial  \rho} \right)_s \mathit{\Delta} \rho \, \rho \, dV 
\nonumber \\
&=& \int\!\! \left( \frac{P}{\rho^2} \right) \mathit{\Delta} \rho \, \rho \, dV 
= -\int\!\! P  \, \mbox{div}(\bvector{\xi}) \,  dV  
\nonumber \\
&=& \int\!\!  \nabla P  \cdot \bvector{\xi} \,  dV.
\label{eq:E_1st}
\end{eqnarray}
Here $\mathit{\Delta}\varepsilon$ is the Lagrangian variation of internal energy, which is given by the Lagrangian variation of density upon using the assumption that there is no entropy variation, equation (\ref{eq:del_s}). The final expression is obtained with the use of equation (\ref{eq:del_rho}).

The second term is evaluated as follows:
\begin{eqnarray}
\delta \left (\frac{1}{2} \! \int \!\! \rho \phi \, dV \right)
&=& \frac{1}{2} \! \int \!\! \left( \delta \rho \cdot \phi + \rho \cdot \delta \phi  \right)\, dV 
\nonumber \\
&=& \frac{1}{2} \! \int \!\! \left( -\nabla (\rho \, \bvector{\xi}) \cdot \phi 
         + \rho \cdot \delta \phi  \right)\, dV \nonumber \\
&=& \frac{1}{2} \! \int \!\! \left( \rho \, \bvector{\xi} \cdot \nabla \phi + \rho \cdot \delta \phi  \right)\, dV, 
\label{eq:E_2nd0}
\end{eqnarray}  
where the second equality is obtained with an employment of the following relation:
\begin{eqnarray}
\delta \rho = \Delta \rho - \xi \nabla \rho.
\end{eqnarray}
$\delta \phi$ in the last term of the final expression is the Eulerian variation of the gravitational potential, which is evaluated from the Poisson equation,
$ \Delta \phi = 4\pi G \rho,$ and its solution with the Green's function,
\begin{eqnarray}
\phi = - \int dV' \rho(\bvector{r'}) / {|\bvector{r-r'}|} .
\end{eqnarray}
 Then $\int \!\!  \rho \cdot \delta \phi \, dV$ is given as 
\begin{eqnarray}
\int \!\!  \rho \cdot \delta \phi \, dV
&=& -G \int \!\!  \frac{\delta \rho(\bvector{r'}) \rho(\bvector{r})} {|\bvector{r-r'}|} \, dV' \, dV
\nonumber \\
&=& \int \!\!  \delta \rho(\bvector{r'}) \phi(\bvector{r'})\, dV' 
 = \! \int \!\! \rho \, \bvector{\xi} \cdot \nabla \phi  \, dV. 
\end{eqnarray}  
Equation (\ref{eq:E_2nd0}) is hence expressed as 
\begin{eqnarray}
\delta \left (\frac{1}{2} \! \int \!\! \rho \phi \, dV \right)
= \! \int \!\!  \rho \, \bvector{\xi} \cdot \nabla \phi \, dV. 
\label{eq:E_2nd}
\end{eqnarray}

The variation in the last term of equation (\ref{eq:e}) is calculated as  
\begin{eqnarray}
\delta \left( \int \!\! \frac{1}{2} \, \rho \! \left ( \frac{j}{\varpi} \right )^2 dV \right) 
&=& \int \!\! \frac{1}{2} \, \rho \mathit{\Delta} \! \left ( \frac{j^2}{\varpi^2} \right ) dV
\nonumber \\
&=& \int \!\! \,   \frac{1}{2} \, j^2 \, \bvector{\xi} \cdot \frac{\partial}{\partial \bvector{r}} \left( \frac{1}{\varpi^2}  \right) \, \rho \, dV \nonumber \\
&=& - \int \!\! \,  \frac{j^2}{\varpi^3} \, \bvector{\xi} \cdot {\bf e_{\varpi}} \, \rho \, dV.
\label{eq:E_3rd}
\end{eqnarray}
In the second equality we employed the assumption that there is no Lagrangian variation in the specific angular momentum, equation~(\ref{eq:del_j}).

We are now ready to derive the Euler-Lagrange equation for the energy functional in equation  (\ref{eq:e}). All the variations given by equations (\ref{eq:E_1st}), (\ref{eq:E_2nd}) and (\ref{eq:E_3rd}) are expressed with the Lagrangian displacement vector $\xi$ and the Euler-Lagrange equation corresponds to its coefficient in $\delta E$. The resultant equation is obviously equation (1). This implies that the rotational equilibrium is a stationary point of the functional and that it may be obtained not by solving equation (1) but by minimizing the energy functional somehow\footnote{The rotational equilibrium may not be a minimum point of the energy functional, which will be indeed the case if the configuration is convectively unstable. Possible treatments of such cases will be discussed elsewhere.}.

In the above formulation of the variational principle, we treated the gravitational potential $\phi$ as a functional of density, formally solving the Poisson equation with the Green's function. It is possible, however, to treat the gravitational potential also as an independent variational variable. The energy functional should be modified then to 
\begin{eqnarray}
E(\bvector{\xi}) & = & \int\!\! \varepsilon \rho \, dV + \int \!\! \rho \phi \, dV  +  \int \!\! \frac{1}{2} \, \rho \! \left ( \frac{j}{\varpi} \right )^2 \! dV \nonumber \\
&& + \frac{1}{4 \pi G} \int \frac{1}{2} (\nabla \phi)^2 dV.
\label{eq:e2}
\end{eqnarray} 
In this case the Poisson equation for the gravitational potential itself is obtained as a Euler-Lagrange equation with respect to $\phi$. Note that the factor of 1/2 disappears from the second term, reflecting the fact that $\phi$ is an independent variable. The problem with this formula from a practical point of view is that the fourth term has a non-compact support, i.e., the integral region extends to infinity. This problem may be alleviated, however, by imposing an appropriate boundary condition at the stellar surface~\footnote{This is nothing but the evaluation somehow of the integral outside the star. See section 2.4.}. In the following, we employ the first formulation, in which the gravitational potential is treated as a functional of density through the solution of the Poisson equation. The second form of the energy functional, equation (\ref{eq:e2}), is employed, however, to derive a discretized Poisson equation in section 2.3.

\subsection{Discretization of the Energy Functional on the Triangulated Mesh}
\label{subsec:triangle}

The Lagrangian variational principle presented above is not our own invention but has been known for many years \citep{tassoul78}. The following implementation, however, is our original idea. We adopt a finite element method: the meridian section of an axisymmetric star is covered by a triangulated mesh, and approximate the energy functional on this mesh. Various quantities are assigned to each grid point or node. They are moved to artificially deform the star. We explain the method of these procedures in this subsection. 

%----- paragragh on FIG.1------
The triangulated mesh consists of nodes and edges, with adjacent nodes being connected by an edge. The so-called adjacency matrix, which is commonly used in graph theory and gives the relation between neighboring nodes with its ($i$, $j$) component $A_{ij}$ being unity (zero) if the $i$-th and $j$-th nodes are (not) connected with an edge, is convenient and will be used in evaluating various quantities numerically. Note that this matrix is symmetric. 
%メッシュの作り方は重要でないので上のように変更。
% Let us start the explanation from a creation of a triangulated mesh. First, suppose to put nodes on space, numbering them randomly as shown in Fig.~\ref{fig:tri01}-(a). Second, we connect the nodes with edges making triangular cells as shown in Fig.~\ref{fig:tri01}-(b). Note that the way of triangulation is not unique. 
To each node we assign coordinates, mass, specific angular momentum, entropy and volume fractions of various elements \footnote{In this paper we ignore the fractions of elements for simplicity.}. The latter four are conserved quantities as long as we do not consider nuclear reactions and transport of energy and angular momentum and are fixed in the node shifts, 

\begin{eqnarray}
\rho_i = m_i / V_i = m_i / (\frac{1}{3}\sum_{cell \in i} V_{\Delta}) .
\label{eq:element}
\end{eqnarray} 
In this expression, $m_i$ and $\rho_i$ are the mass and density assigned to the $i$-th node, respectively, and the summation runs over the cells surrounding the $i$-th node whose volumes are denoted by $V_\Delta$.

%---- energy functional -----
The energy functional, equation (\ref{eq:e}) is approximated on this mesh as: 
\begin{equation}
E_{\rm FEM}(\bvector{r}_i) = \sum_{i}\varepsilon _i m_i + \frac{1}{2}\sum_{i}\phi _i m_i + \sum_{i}\frac{1}{2} \!\left( \frac{j_i}{\varpi _i} \right )^2 \! m_i,
\label{eq:eq2}
\end{equation}
which is now actually a function of the coordinates of all nodes, $\bvector{r}_i$, with the subscript specifying the individual node. This corresponds to an approximate evaluation of the Jacobian $J(\bvector{r}_i)$ at each node. The minimization of this energy gives the coordinates of nodes in the rotational equilibrium for given distributions of mass, specific entropy and angular momentum on the triangulated mesh. We then get the resultant density at each node through equation  (\ref{eq:element}). It should be obvious that the specific internal energy $\varepsilon _i$ is also regarded as a function of the coordinates of nodes, since it is a function of density and specific entropy $s_i$, the former of which is obtained once the node positions are determined as just mentioned and the latter is fixed. As for the gravitational potential $\phi _i$, equation (\ref{eq:poisson}) is solved numerically on the same triangulated mesh for the density obtained this way. It can be hence regarded also as a function of the coordinates of nodes.

\subsection{Poisson equation for gravitational potential}
As mentioned just now, in our formulation, the gravitational potential is regarded as a functional, or a function in the discretized version, of density by solving the Poisson equation. As also noted earlier in section~\ref{subsec:lagrangian}, we employ the energy functional given by equation (\ref{eq:e2}) to derive the discretized version of Poisson equation. 

We first approximate the second and fourth terms on the right hand side of the equation on the triangulated mesh as
\begin{eqnarray}
%\sum_k \int_{\rm{cell}}~~ \hspace{-3mm} \left( (\frac{1}{4\pi G}) \frac{1}{2} \nabla \cdot \nabla \phi -  \phi \rho \right) dV, 
\sum_k \int_{\rm{cell}}~~ \hspace{-3mm} \left( (\frac{1}{4\pi G}) \frac{1}{2} \nabla \phi \cdot \nabla \phi +  \phi \rho \right) dV, 
\label{eq:poisson2}
\end{eqnarray} 
where the volume integral is done for each triangular cell and summed over all cells.
The integrand is approximated linearly in each cell as
\begin{eqnarray} 
\phi =\alpha_1^\phi z + \alpha_2^\phi \varpi + \alpha_3^\phi ,
\end{eqnarray}
where the coefficients, $\alpha$'s, are given in terms of the values of the gravitational potential at three nodes of the cell as 
\begin{eqnarray}
{ \alpha}_a^\phi={ T}_{ab}^{-1}{ \phi}_b.  
\nonumber
\end{eqnarray}
In this expression, the subscripts $a$ and $b$ specify the node of the cell and the matrix $T_{ab}$ is determined by the geometry of the mesh alone and given in Appendix A. Then the derivative of $\phi$ in the integrand of equation~(\ref{eq:poisson2}) is approximated as
\begin{eqnarray}
\nabla \phi =(\alpha_1^\phi,\alpha_2^\phi,0).
\end{eqnarray}
Using this approximation, we evaluate the first term of equation (\ref{eq:poisson2}) as
\begin{eqnarray}
&&  \sum_k \int_{\rm{cell}}  \nabla \phi \cdot \nabla \phi ~dV \nonumber \\
&=&  \sum_k \left( \alpha_1^\phi \alpha_1^\phi+  \alpha_2^\phi \alpha_2^\phi \right)\int_{\rm{cell}}~dV 
\nonumber \\
&=& \sum_k \sum_{a,b \in  k} \left(T_{1a}^{-1}T_{1b}^{-1}+T_{2a}^{-1}T_{2b}^{-1} \right) \phi_a \phi_b  V_\Delta. \label{1st_int}
\end{eqnarray} 
The second term of equation (\ref{eq:poisson2}) is, on the other hand, expressed as 
\begin{eqnarray}
\sum_k \int_{\rm{cell}} \phi \rho ~dV = \sum_i \phi_i m_i,
\label{2nd_int}
\end{eqnarray} 
in which the sum on the right hand side is taken over all nodes.
Collecting the terms, we obtain the expression for equation (\ref{eq:poisson2}) as 
\begin{eqnarray}
\sum_{k}  \sum_{a,b \in  k} \frac{1}{2} \left(T_{1a}^{-1}T_{1b}^{-1}+T_{2a}^{-1}T_{2b}^{-1} \right) V_\Delta  \cdot \phi_a \phi_b 
&=& \sum_i \phi_i m_i .
\nonumber \\
\end{eqnarray} 
The derivative of this expression with respective to $\phi_i$ gives the discretized Poisson equation as
\begin{eqnarray}
\sum_{j} \mathcal{M}_{ij} \phi_{j} = 4 \pi G m_i .
\nonumber 
\end{eqnarray} 
Here we introduce a symmetric matrix $\mathcal{M}_{ij}$ defined as
\begin{eqnarray}
\sum_{i,j} \mathcal{M}_{ij} \phi_{i} \phi_{j} = \sum_{k} \sum_{a,b \in  k} \frac{1}{2} \left(T_{1a}^{-1}T_{1b}^{-1}+T_{2a}^{-1}T_{2b}^{-1} \right) V_\Delta \phi_{a} \phi_{b},  
\end{eqnarray} 
in which the double sum on the left hand side runs over all nodes. Note that the matrix $\mathcal{M}_{ij}$ is determined by the mesh geometry alone. This approximate Poisson equation is solved algebraically as
\begin{eqnarray}
\phi_{i} &=& 4 \pi G  \mathcal{M}_{ij}^{-1} m_j.
\label{poisson3}
\end{eqnarray} 
In so doing the boundary condition is imposed Dirichlet-type at the stellar surface with the values of the gravitational potential being calculated with the Green's function (see below).

In order to validate the formulation, we compare it with another method. In fact, the Green's function method, which is employed to calculate the boundary values of the potential, is actually  applicable everywhere. Adopting the multi-pole expansion of the Green's function, we write the potential as
\begin{eqnarray}
\phi_i = -2\pi G\sum^\infty_{l=0} P_l(\cos \theta_i) \int^\pi_0 \sin{\theta '} d \theta' P_l (\cos \theta ') \nonumber\\
\times \left( \int^{r_i}_0 \frac{(r')^{l+2}}{r^{l+1}}dr'  + \int^\infty_{r_i} \frac{r'^l}{(r')^{l-1}}dr' \right) \rho_i(r', \theta') ,
\label{poisson4}
\end{eqnarray} 
where $P_l$'s are the Legendre polynomials. The integrals are evaluated numerically.

\begin{figure*}
\begin{center}
\includegraphics[width=20pc]{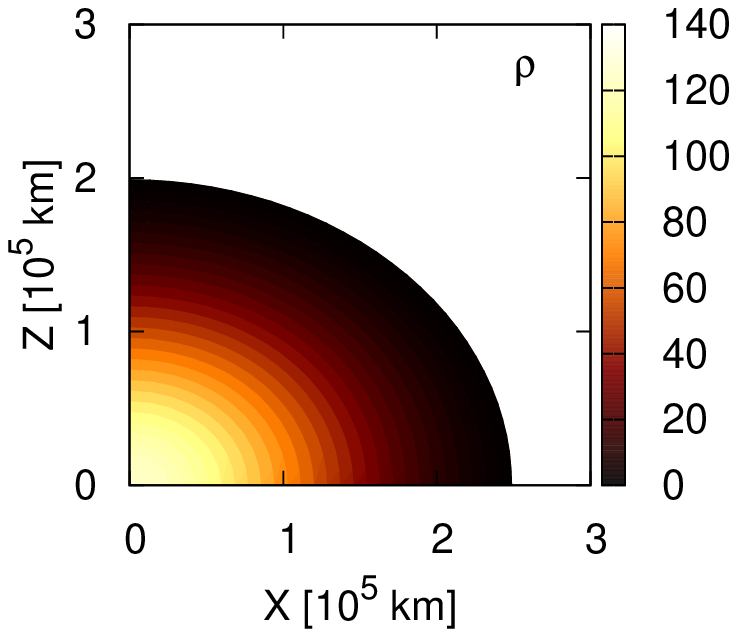}
\includegraphics[width=17.5pc]{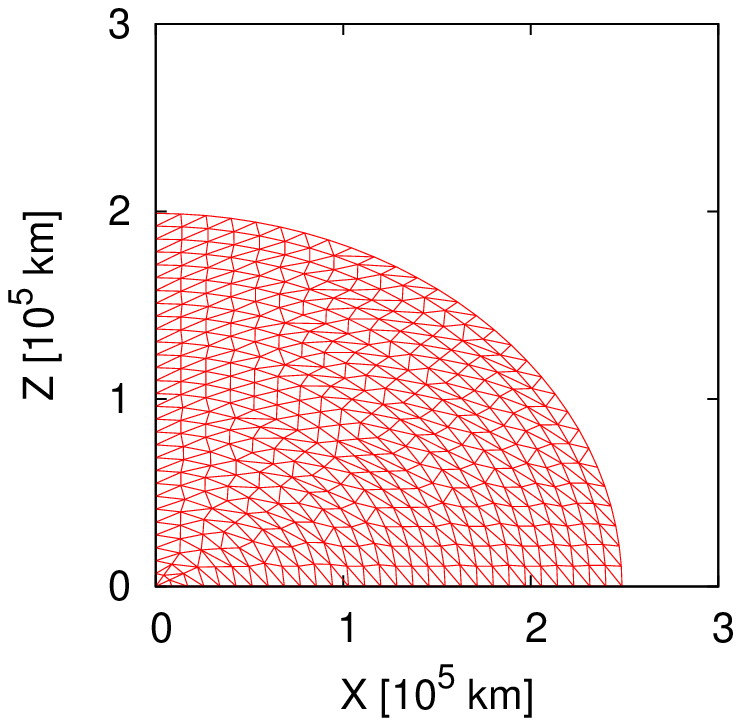}
\caption{\label{fig:density} (colour on-line). An example of density distribution~(left panel), and the edges and nodes~(right panel). The colour bar shows the density in cgs unit.
} 
\end{center}
\end{figure*}

\begin{figure*}
\begin{center}
\includegraphics[width=11.5pc]{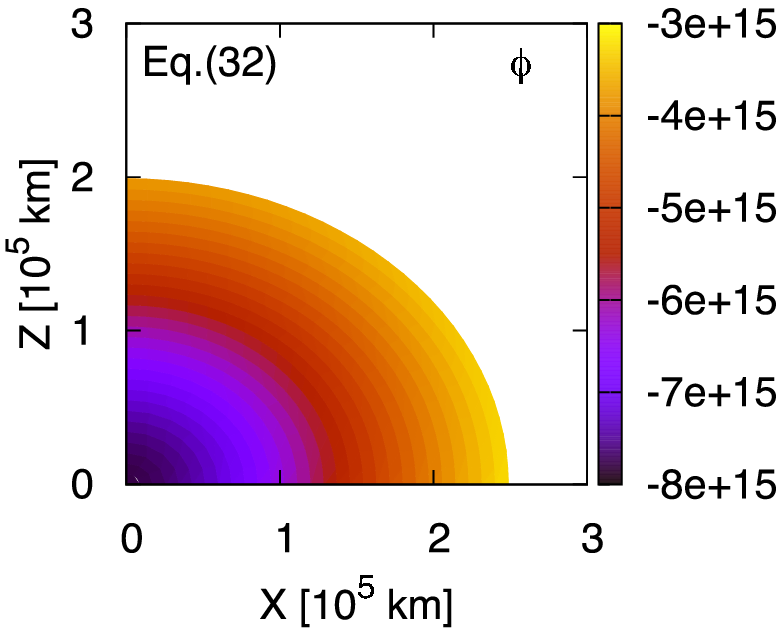}
\hspace{2mm}
\includegraphics[width=11.5pc]{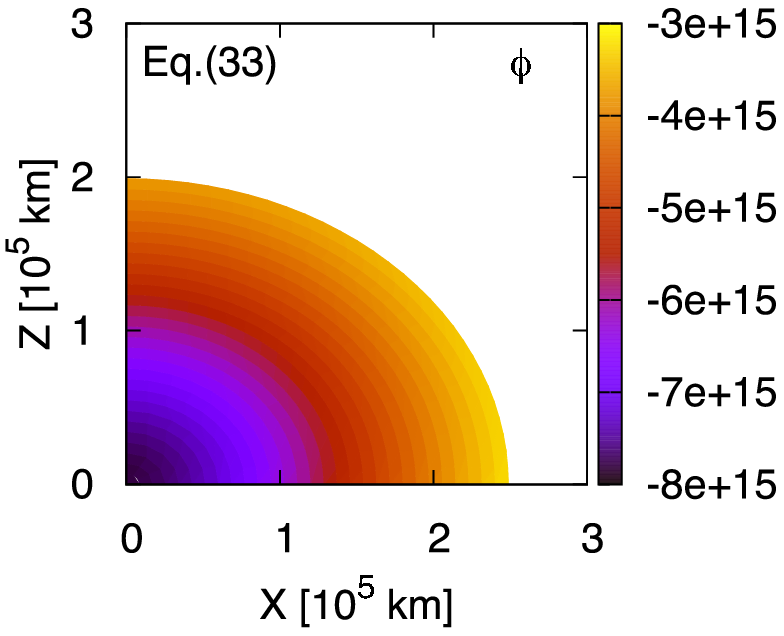}
\hspace{-10mm}
\includegraphics[width=16pc]{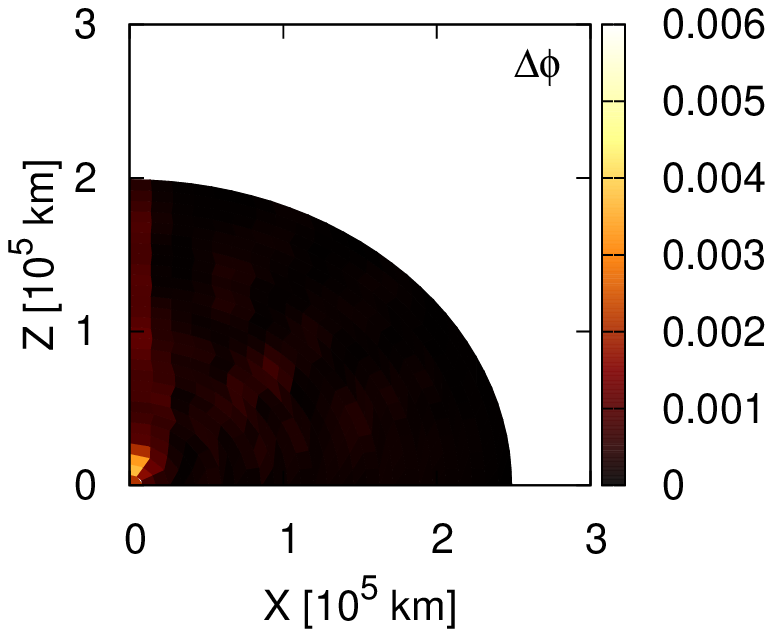}
\caption{\label{fig:potential} (colour on-line). Left, middle panels show the gravitational potential by the  expressions in equation (\ref{poisson3}) and equation (\ref{poisson4}) for the density profile given in Fig. \ref{fig:density}. The unit is in cgs. Right panel shows the difference ratio between the results of left and middle panels. The ratio is given by the difference divided by the result of middle one. 
} 
\end{center}
\end{figure*}

%----- paragragh on FIG.2,3------
The comparison is done for a non-spherical density distribution shown in the left panel of Fig.~\ref{fig:density}. The triangulated mesh we deploy is displayed with the nodes and edges in the right panel. The mass $m_i$ associated with node $i$ is obtained from the density $\rho_i$ at the node through equation (\ref{eq:element}). We evaluate the gravitational potential either via equation (\ref{poisson3}) or equation (\ref{poisson4}). The left and middle panels of Fig.~\ref{fig:potential} correspond to the numerical results for equations (\ref{poisson3}) and (\ref{poisson4}), respectively, which look almost identical to each other. 
As a matter of fact, the right panel of Fig.~\ref{fig:potential} demonstrates that the both results are in agreement within less than 1\% errors.
Here the error is defined as the absolute value of the ratio of the difference between the two results to the one obtained with equation (\ref{poisson4}).

 %----- FIG.1-----

%----- FIG.2-----

\subsection{Minimization of energy function}
We employ a Monte Carlo technique to search for minimum energy function. The reason is the following. It is well known from the perturbation theory based on the Lagrangian displacement vector that a class of {\it trivial} Lagrangian displacements make the distributions of physical quantities unchanged~\citep{friedman}. An example is the isentropic case, i.e. the specific entropy is uniform in the stellar interior. This is most easily understood from the fact that the Lagrangian displacement vector $\bvector{\xi} = 1/\rho \nabla \times \bvector{\eta}$ for an arbitrary axisymmetric vector field $\bvector{\eta}$, which would be obtained as a Euler-Lagrange equations for the discretized energy function, equation (\ref{eq:eq2}), does not change the density as a function of spatial position. This residual {\it gauge degree of freedom} renders the system of equations for $\bvector{r}_i$ indefinite and would make singular the matrix obtained by linearization if one were to employ the Newton-Raphson method.  This means that, without an appropriate gauge fixing, the formulation could not be applied to isentropic configurations, which are normally supposed to be the simplest case. Even if the specific entropy is not homogeneous, the energy function is rather insensitive to the displacements of nodes close to the rotation axis and/or to the surface in general, which implies that there are always small (i.e. almost singular) eigenvalues in the matrix of the linearized equations, which could hamper the employment of the Newton-Raphson method. 

The Monte Carlo method described below could naturally alleviate these difficulties. 
In this technique we displace randomly the grid points and search literally for the configuration that makes the energy function minimum. The problem of the gauge degree of freedom does not arise in the Monte Carlo method, since one solution is automatically picked up out of infinitely many possibilities by a dice. And that is sufficient indeed for our purpose because physical quantities are identical for any choice. Such an approach is not special and is indeed employed in other fields such as nuclear physics, in which deformed nuclei are constructed in a way that is analogous to the one we adopt in this paper.  

%``iteration process"
The actual minimization procedure is the following. Given an initial configuration, we move one of the nodes in the radial direction slightly and see if this shift lowers the value of the energy function or not. If not, we cancel the shift. The amount of dislocation is a random variable but is limited to $\sim 1 \%$ of the lengths of the edges attached to the node. We do the same things in the lateral direction and sweep the entire mesh. We repeat this process until the value of the energy function is no longer lowered compared with the one of 100 sweeps before. This condition is, however, not enough to get hydro-static equilibria, and we also need to check virial residuals, which will be described later. In the current version of the formulation, the connections of nodes by edges represented by the adjacency matrix $A_{ij}$ introduced in subsection \ref{subsec:triangle} are fixed through the minimization process. It is hence important to perform the initial triangulation properly so that the iteration process would lead to convergence. This is particularly the case when rotation is very rapid and the resultant rotational equilibrium is different and some triangular cells may become highly deformed. It may be necessary then to implement the  re-gridding somehow. 

The nodes on the outer boundary, which are referred to as the anchor nodes and have a vanishing mass, are not moved during this minimization process, since an expansion in the radial direction would continue for ever otherwise. In some cases, however, the outer boundary comes too close to or goes too far from the nearby active nodes. We then shift the anchor nodes appropriately and fix them again thereafter. We find that large deformations of triangular cells degrade accuracy of the energy function and in some cases generate a fictitious local minimum. To avoid such artifacts the re-gridding may be a solution as mentioned above but in the current version we apply a smoothing to the deformed portion of the grid, which has experienced more than 30 \% of change in area by a single Monte Carlo sweep. We know empirically that such a large variation in area is commonly accompanied by the appearance of a very acute angle in the cell.

We introduce the residual in the virial relation defined as 
\begin{equation}
V_C =  \left |  \frac{2\,T+W+3\! \int \! P dV}{W} \right |,
\label{eq:vc}
\end{equation}
in which $T$ and $W$ are the rotational and gravitational energies, respectively, 
and the third term is the volume integral of pressure. $V_C$ is equal to zero in the exact theory and is a measure of numerical accuracy commonly used in the literature (e.g. \citealt{eriguchi85}), and we also adopt it together with energy convergence. In this paper, we have typically achieved $V_C <  10^{-3}$ with $\sim$500 grid points, which is comparable previous studies where $V_C \sim O(10^{-4})$ was achieved (\citealt{eriguchi85}). It is possible to improve accuracy by deploying a larger number of nodes. 

%--------------------------
\section{Validation}  
%--------------------------

In this section we apply our new method to some test cases: barotropic configurations in subsection 3.1 and baroclinic ones in subsection 3.2. We compare the resultant configurations with ones obtained by other Eulerian schemes. In subsection \ref{subsec:rearrange}, we mention some technical issues such as the smoothing procedure. 

%%%%%%%%%%%%%%%%%%%%%%%%%%%%%%%%%%%%%%%
\subsection{Comparison of barotropic equilibria}  
\label{subsec:HSCF}
%%%%%%%%%%%%%%%%%%%%%%%%%%%%%%%%%%%%%%%
We begin with the barotropic case, in which pressure depends on density alone, $P=P(\rho)$. 
We consider the isentropic configurations as a representative of barotropes. Here we compare our solutions with those given by Hachisu's self-consistent field~(HSCF) scheme~\citep{hachisu86}, a well-established numerical method  for rapidly rotating barotropic configuration. 
%この段落は不要か、図を見せた後にしないと、何を言っているか分からない。
%The values of virial residual are achieved $V_C \sim O(10^{-4})$ for all cases. In all figures for YFY scheme, we do not show the outer boundary since we do not impose same boundary condition with HSCF scheme. In YFY scheme, the outer boundary is fixed as the anchor, which is just put depended on the inner density distribution. In this meaning, as for the outer boundary, it is a variational principle in a roundabout way.

In this section, we consider two types of rotation laws. As repeatedly mentioned, the barotropic configuration rotates cylindrically, i.e., the iso-angular velocity surfaces are concentric cylinders.
\\
(i) rigid rotation:
\begin{equation}
\Omega^2({\bf r}) =\Omega_0^2,
\label{eq:w_rig}
\end{equation} 
(ii) differential rotation: 
\begin{equation}
%Ω_0 / (r*sin(th)**2 + 0.9**2)^(0.5)
\Omega^2({\bf r}) =\frac{\Omega_0^2}{(\varpi/R_e)^2 + d^2},
\label{eq:w_dif}
\end{equation} 
where $\Omega ({\bf r})$ is an angular velocity at the position of ${\bf r}$. In equation  (\ref{eq:w_dif}) $R_e$ is the equatorial radius, $\Omega_0$ is a parameter that specifies the overall magnitude of angular velocity, and $d$ is a dimensionless parameter that determines the a degree of differential rotation: as $d$ increases, the angular velocity becomes uniform whereas in the opposite limit the rotational velocity becomes constant~\citep{eriguchi85}. 
We set $d=0.90$ in the following. 

It should be reminded that in our YFY scheme the distribution of angular velocity is not known a priori, since we assign a conserved specific angular momentum to each node of the triangulated mesh in the initial reference configuration, which is arbitrary, and the node position is changed later. This is not the case for the HSCF scheme, which is a Eulerian method and derives the density distribution on the fixed mesh, assuming from the beginning that the rotation is cylindrical. Note that even the cylindrical rotation is not guaranteed in the YFY scheme. It is hence a good diagnosis for our method. In order to make comparison between the two schemes, we take the following strategy: we first employ the HSCF scheme to obtain the rotational equilibrium for a given rotation law and cover the resultant configuration with a triangulated mesh and obtain the mass and specific angular momentum that should be assigned to each node; we then expand by 20 \% in the radial direction and use it as the initial reference configuration for our YFY scheme. If it works properly, the mesh should return to the original shape after the minimization of the energy function. Incidentally we have confirmed that if the artificial expansion of the mesh is not administered, the mesh does not change by the minimization as it should.

\begin{figure*}
\includegraphics[width=15pc]{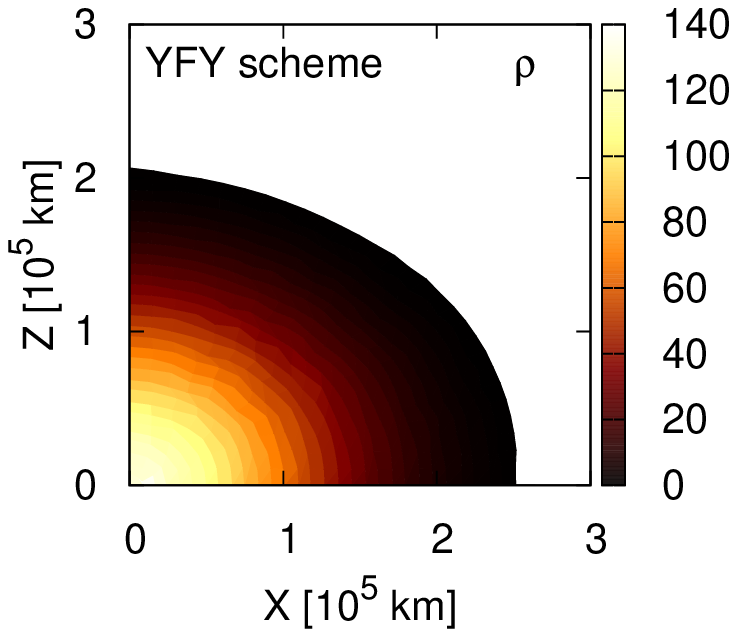}
\hspace{-13mm}
\includegraphics[width=15pc]{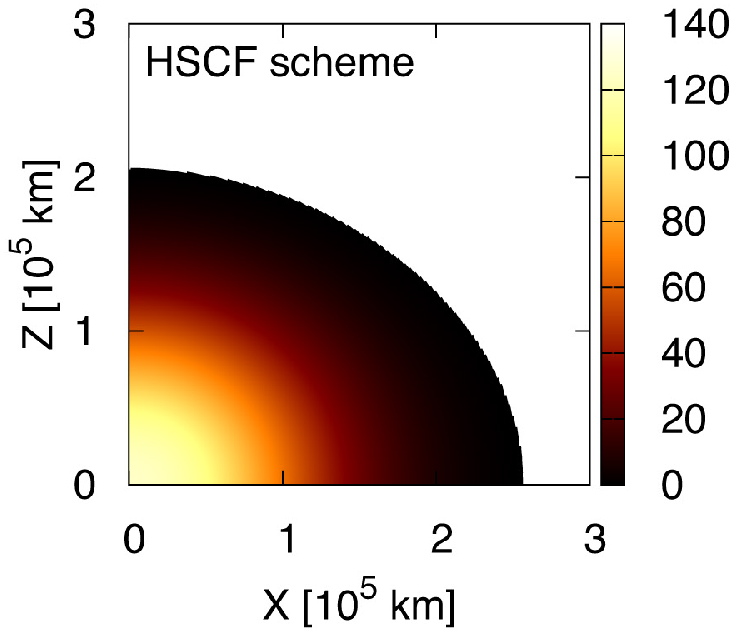}
\hspace{-13mm}
\includegraphics[width=15pc]{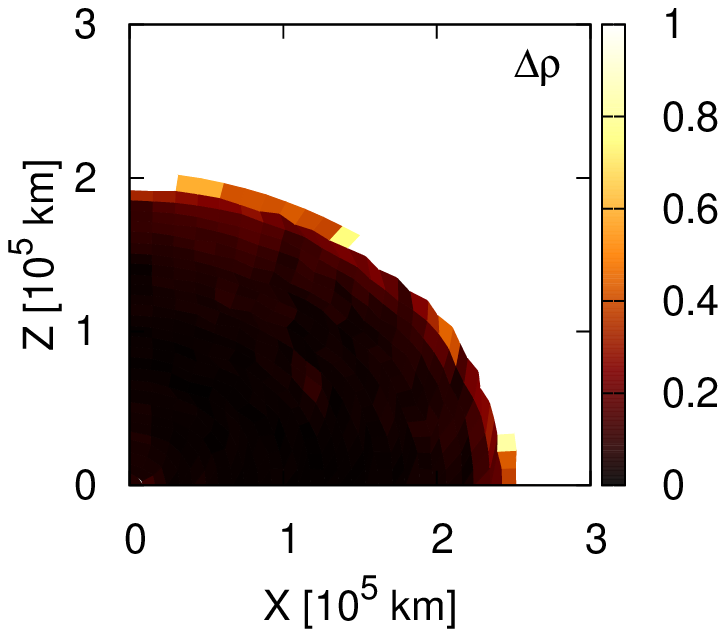}
\caption{\label{fig:d_rig} (colour on-line). 
The density distributions for a rigidly rotating barotropic configuration obtained with the YFY scheme(left panel), and the HSCF scheme(middle panel) as well as the relative differences thereof (right panel). The unit of density is given in cgs. 
The central density, equatorial radius, (uniform) specific entropy and ratio of rotational energy to gravitational one are $\rho_0 = 1.24 \times 10^2$ g cm$^{-3}$, $R_e = 2.57 \times 10^5$ km, $s=14.7 {\rm k_B}$ and  $T/|W| = 3.7 ~\% $. } 
\end{figure*}

%----- paragraph on FIG.4 -----
One of such comparisons is shown in Fig.~\ref{fig:d_rig}, where the density distributions obtained with the two schemes are presented for a rigidly rotation. Although HSCF scheme utilizes non-dimensional quantities are normalized with the maximum density $\rho_0$ and equatorial radius $R_e$, we give here rather arbitrarily specific values $\rho_0$ and $R_e$ as $\rho_0 = 1.24 \times 10^2$ g cm$^{-3}$ and $R_e = 2.57 \times 10^5$ km in order to facilitate comparison. 
The ratio of poler radius $R_p$ to the equatorial set to $R_p/R_e=0.80$ in this model, which corresponds to the ratio of rotational energy to gravitational one of $T/|W| = 3.7~\% $. 
As mentioned already, we assume a uniform entropy distribution in this model: the specific entropy is $s=14.7~{\rm k_B}$ with ${\rm k_B}$ being the Boltzmann constant if the matter is composed of hydrogens alone. 

The two configurations look quite similar to each other, which is also confirmed in the right panel in which relative differences are displayed as a contour color map: the density distributions agree with each other within an error of 5 \% in most places. A closer inspection reveals, however, that the deviation gets larger near the surface, where the density itself is quite low. The vicinity of the rotation axis is another region, where the accuracy is compromised. This is due to the fact that our scheme is based on the variational principle and it is intrinsically difficult to determine the configuration of the regions that do not contribute much to the energy functional. Some multi-layer treatment may be needed to handle this problem.
  
\begin{figure*}
\includegraphics[width=16pc]{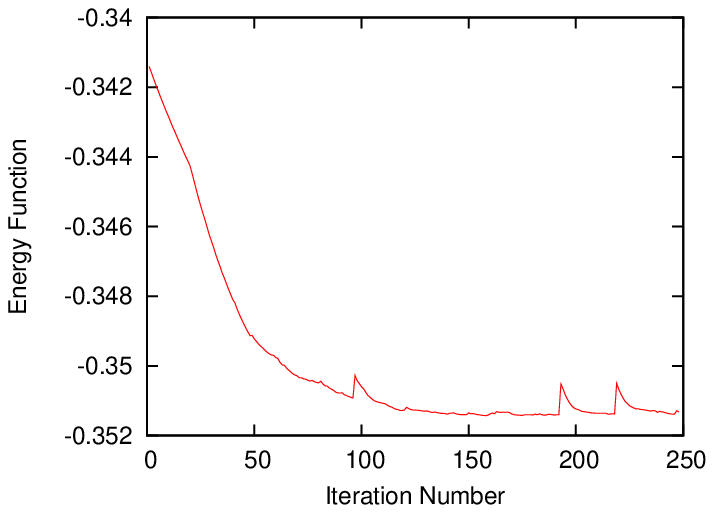}
\includegraphics[width=16pc]{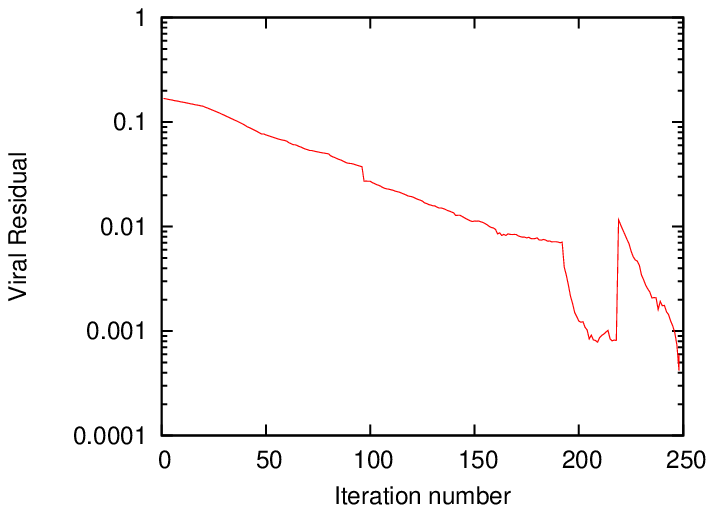}
\caption{\label{fig:his_rig} (colour on-line). 
The values of the energy function (left panel) and of the virial residual (right panel) for the rigid rotation given in Fig.~\ref{fig:d_rig}. The configuration settled to an rotational equilibrium
iteration process is converged after 248 sweeps, where the energy function is longer lowered and the virial residual reaches $<~10^{-3}$, and the iteration is terminated there.
The energy function is normalized by $G {\mbox M}_{\odot}^2/{\mbox R}_{\odot}$, in which ${\mbox R}_\odot$ is the solar radius. The spikes that are seen in the energy function from time to time are due to smoothings performed to avoid false local minima.} 
\end{figure*}

%----- paragraph on FIG.5 -----
In Fig.~\ref{fig:his_rig} we show how the energy function (left panel) and the virial residual (right panel) change as the Monte-Carlo sweep proceeds for the same model. The iteration process is terminated after 248 sweeps, at which point the value of the energy function is no longer lowered and, importantly, the virial residual is sufficiently small ($< \sim 10^{-3}$) simultaneously. One recognizes that there are occasional glitches (three major ones and many minor ones) in the left panel, at which the value of the energy function rises discontinuously. There are corresponding discontinuities in the right panel. These irregularities are produced by what we call smoothing which is a reconfiguration of the mesh we conduct when the mesh gets distorted too much. Without such a measure, the search is trapped in local energy minimum as we describe later. 

\begin{figure*}
\includegraphics[width=16pc]{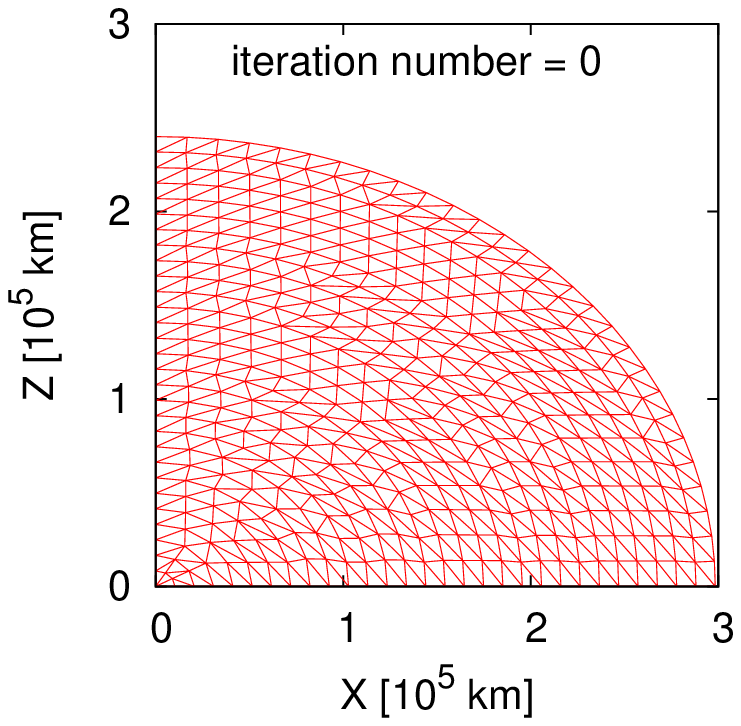}
\hspace{-15mm}
\includegraphics[width=16pc]{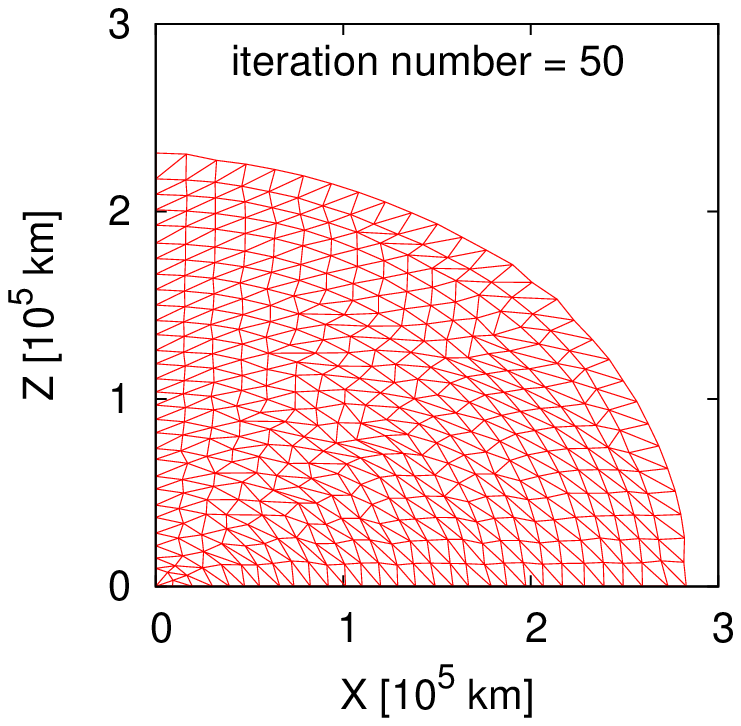}
\hspace{-15mm}
\includegraphics[width=16pc]{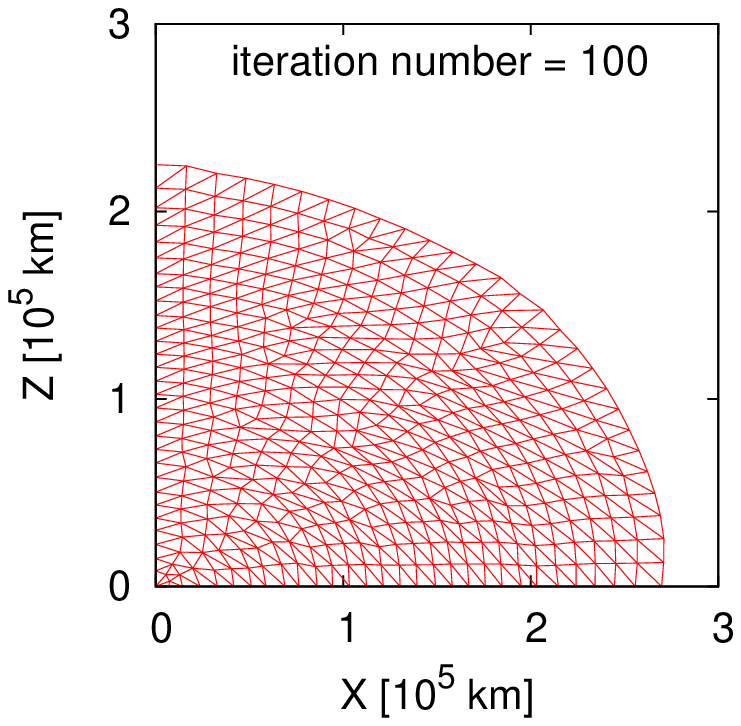}\\
\includegraphics[width=16pc]{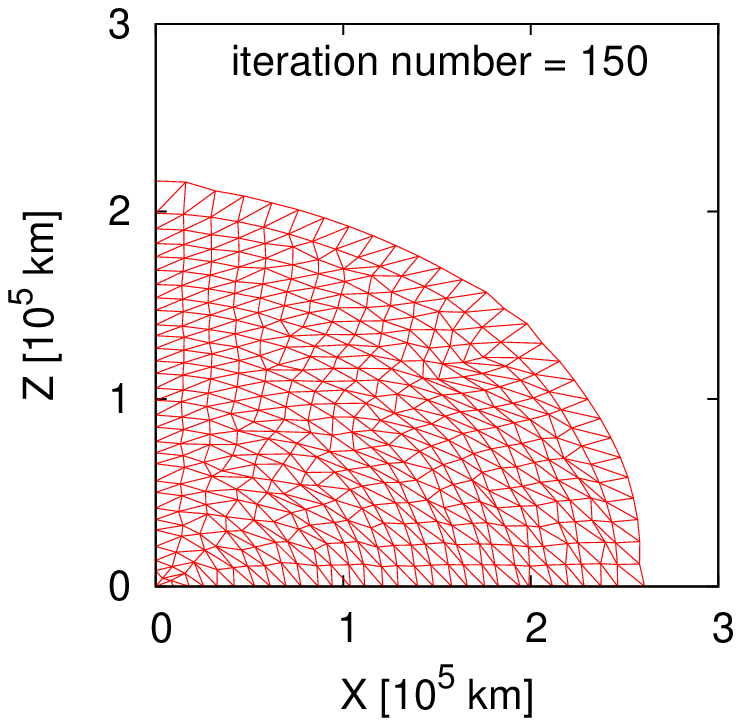}
\hspace{-15mm}
\includegraphics[width=16pc]{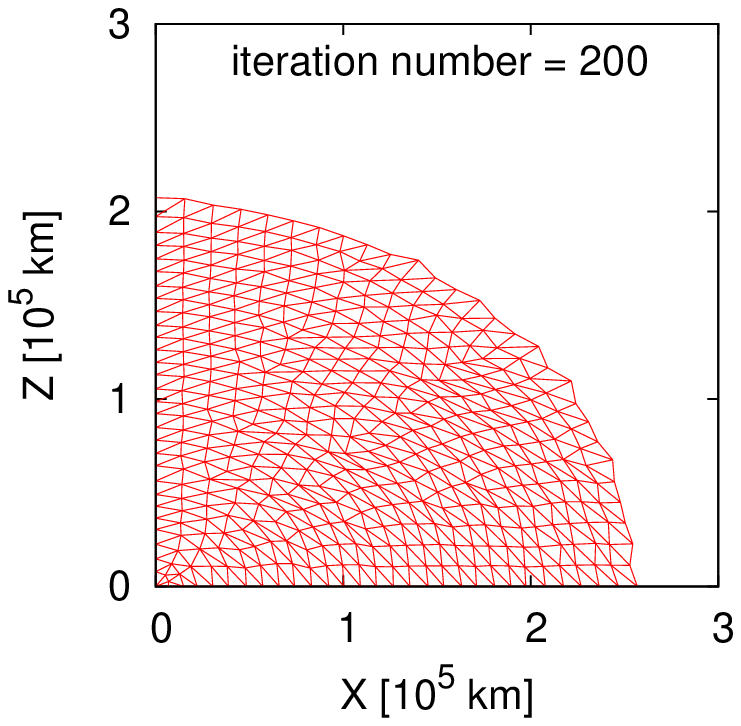}
\hspace{-15mm}
\includegraphics[width=16pc]{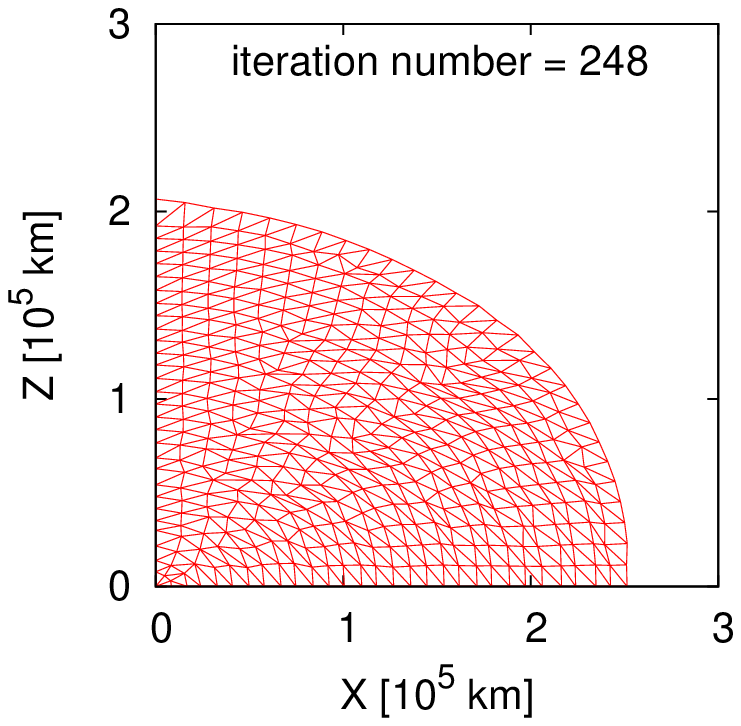}
\caption{\label{fig:edge_rig} (colour on-line). 
The change of configuration in the searching process for the solution given in Fig.\ref{fig:d_rig}. } 
\vspace{5mm}
\end{figure*}

%----- paragraph on FIG.6 -----
The corresponding change of the configuration is presented in Fig.~\ref{fig:edge_rig}. One can see a gradual shrink of the entire grid. Note that although the energy function looks settled to a minimum around 130 sweeps as shown in the left panel of Fig.~\ref{fig:his_rig}, the configuration is far from rotational equilibrium. This is an example of the false local minimum just mentioned. The fact that this is a fake is reflected in the value of the virial residual given in the right panel of Fig.~\ref{fig:d_rig}. This is the reason why we need to conduct the smoothing and the virial residual is an important diagnostic quantity to judge convergence.

 %----- FIG.3-----

%----- FIG.4-----

%----- FIG.5-----

\bigskip
\bigskip

 %%%%%%\subsubsection{Differential rotation} %%%%%
 
\begin{figure*}
\includegraphics[width=15pc]{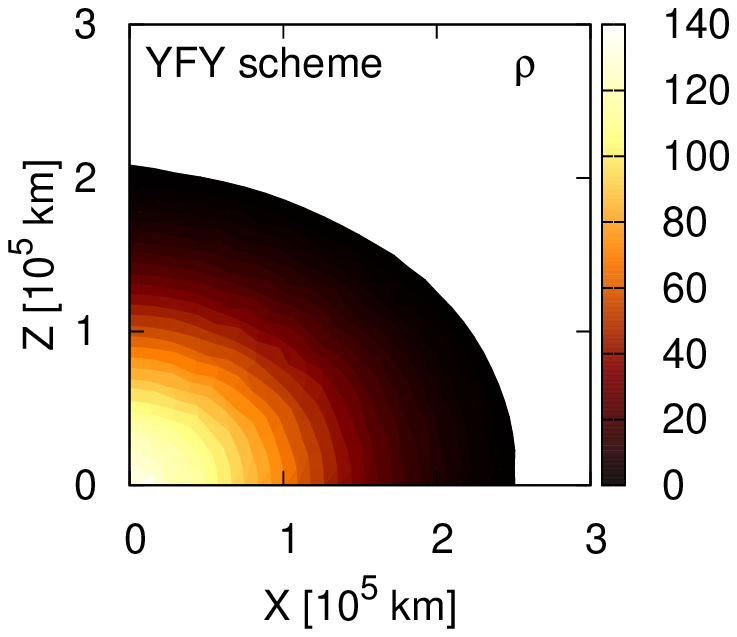}
\hspace{-13mm}
\includegraphics[width=15pc]{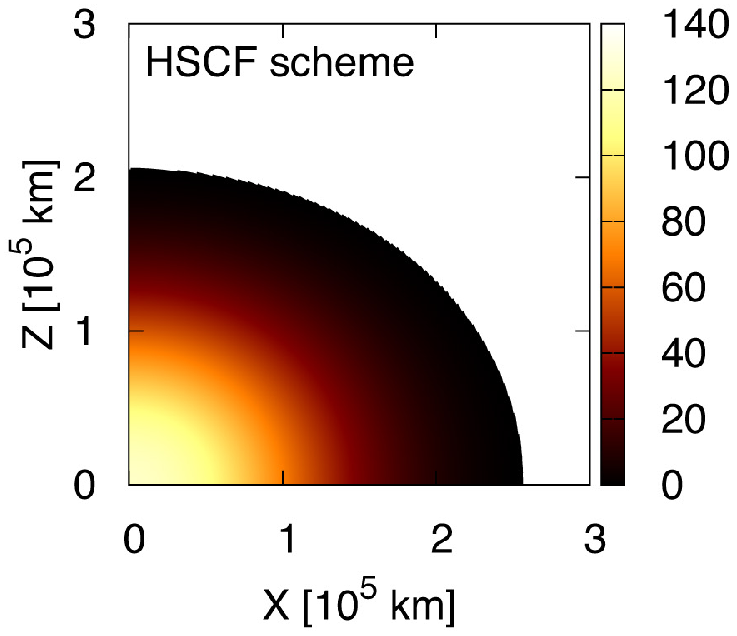}
\hspace{-13mm}
\includegraphics[width=15pc]{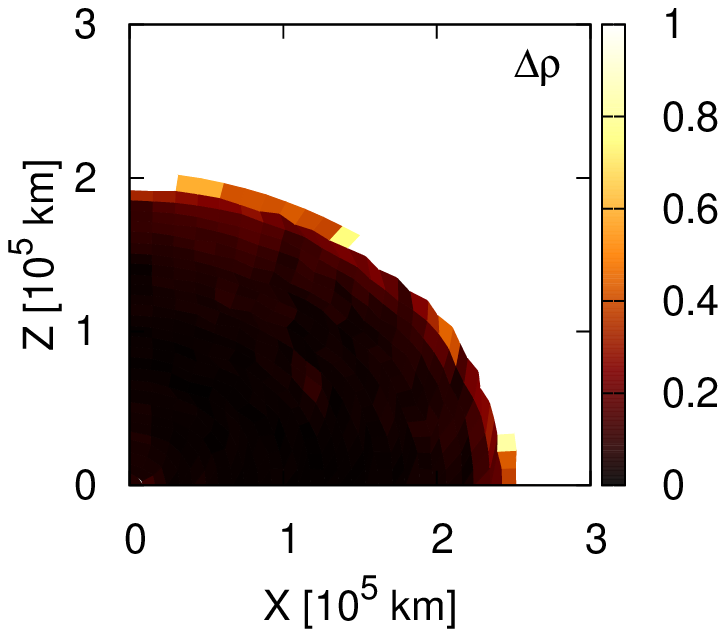}
\caption{\label{fig:d_dif} (colour on-line). 
Same as Fig.~\ref{fig:d_rig} but for the differential rotation obeying the law given in equation (\ref{eq:w_dif}). The values $\rho_0$, $R_e$, $s$, and $R_p/R_e$ are the same as those in Fig.~\ref{fig:d_rig}.} 
\end{figure*}

 %----- paragraph on FIG.7 -----
Next, we present in Fig.~\ref{fig:d_dif} an example of the differential rotation given by equation (\ref{eq:w_dif}). As in Fig.~\ref{fig:d_rig}, the left panel shows the result by the YFY scheme, whereas the middle panel is for the HSCF scheme. We set again the central density and equatorial radius to $\rho_0 = 1.24 \times 10^2$ g cm$^{-3}$ and $R_e = 2.57 \times 10^5$ km. 
%----- これは本当か？ -----> 14.7 kBと14.68の違いくらい。------
The specific entropy is constant in space and the same as for the rigid rotation in the previous case. Then the total mass is also unchanged. 
%--------------------------
The density distributions agree with each other within 5 \% in most regions, but not near the surface again. 

\begin{figure*}
\includegraphics[width=16pc]{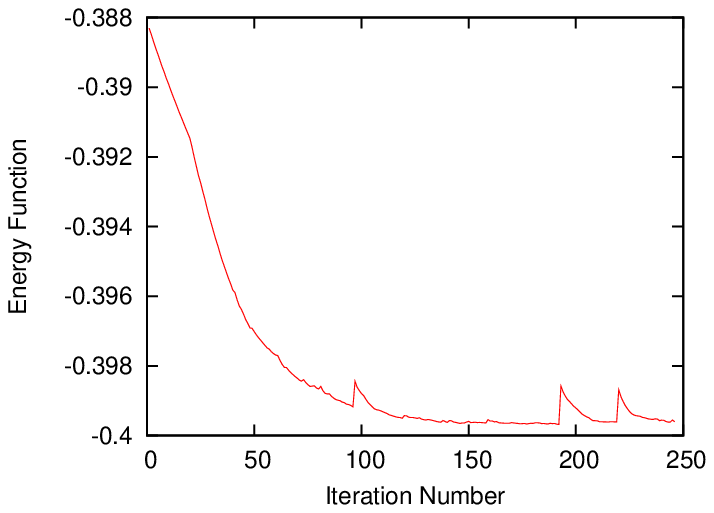}
\includegraphics[width=16pc]{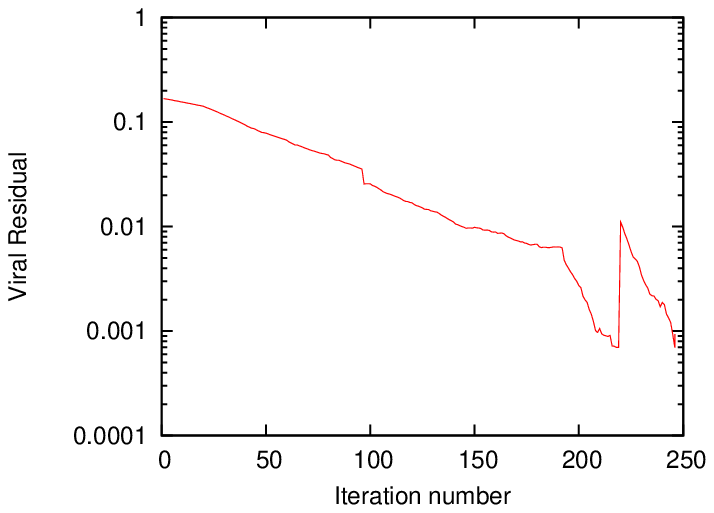}
\caption{\label{fig:his_dif} (colour on-line). 
Same as Fig.~\ref{fig:his_rig} but for the differential rotation in Fig.~\ref{fig:d_dif}. The iteration is 
terminated after 246 sweeps, at which point the energy functional is no longer lowered and the virial residual is sufficiently small $<~10^{-3}$.} 
\end{figure*}

%----- paragraph on FIG.8 -----
The behavior of the energy function (left panel) and virial residual (right panel) 
is qualitatively the same as what we saw in Fig.~\ref{fig:his_rig} for the rigid rotation.
The iteration process is terminated after 248 sweeps. These panels are very similar to Fig.~\ref{fig:his_dif}, though the rotation law is different. 

\begin{figure*}
\includegraphics[width=16pc]{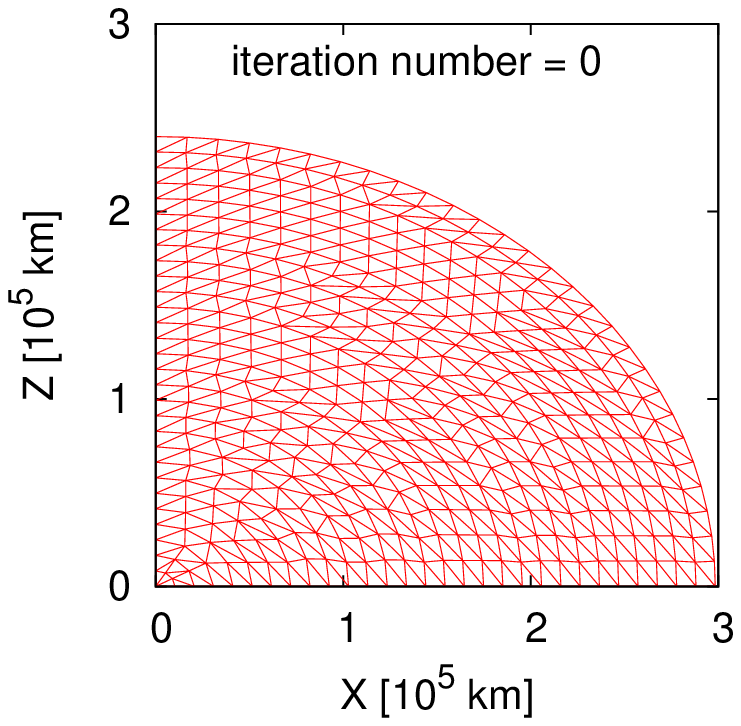}
\hspace{-15mm}
\includegraphics[width=16pc]{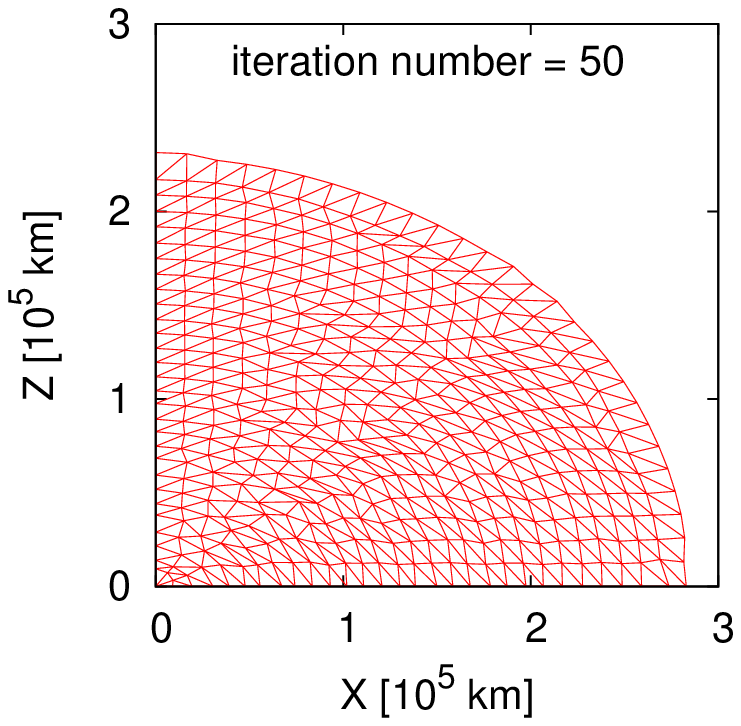}
\hspace{-15mm}
\includegraphics[width=16pc]{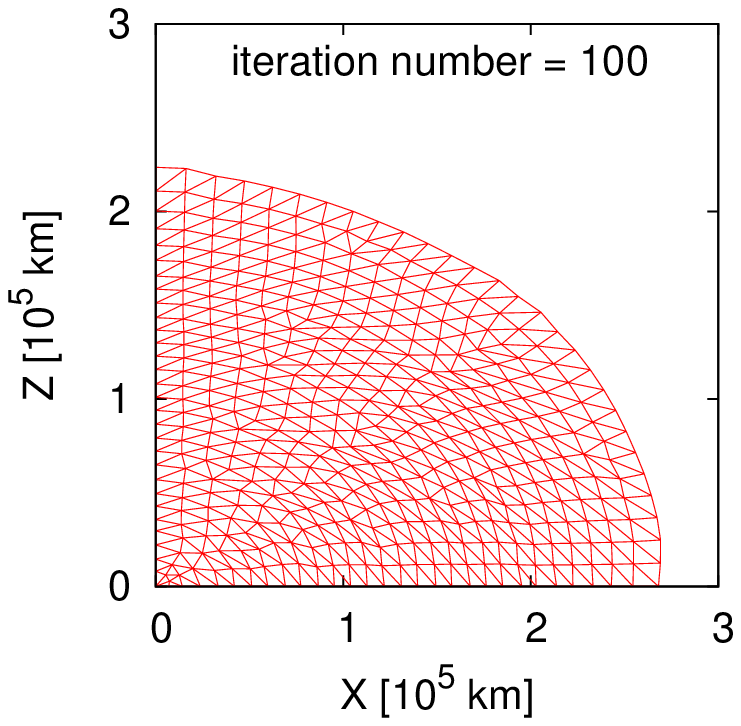}\\
\includegraphics[width=16pc]{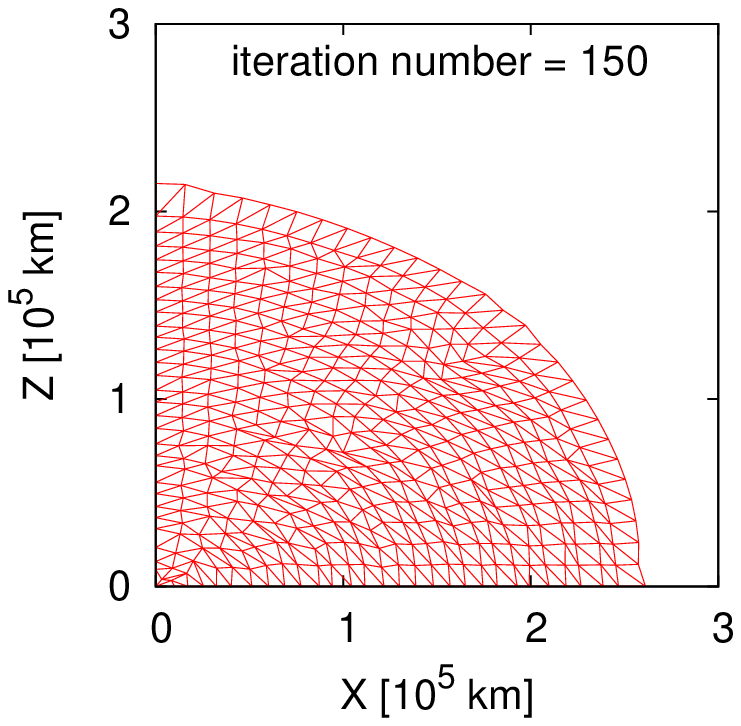}
\hspace{-15mm}
\includegraphics[width=16pc]{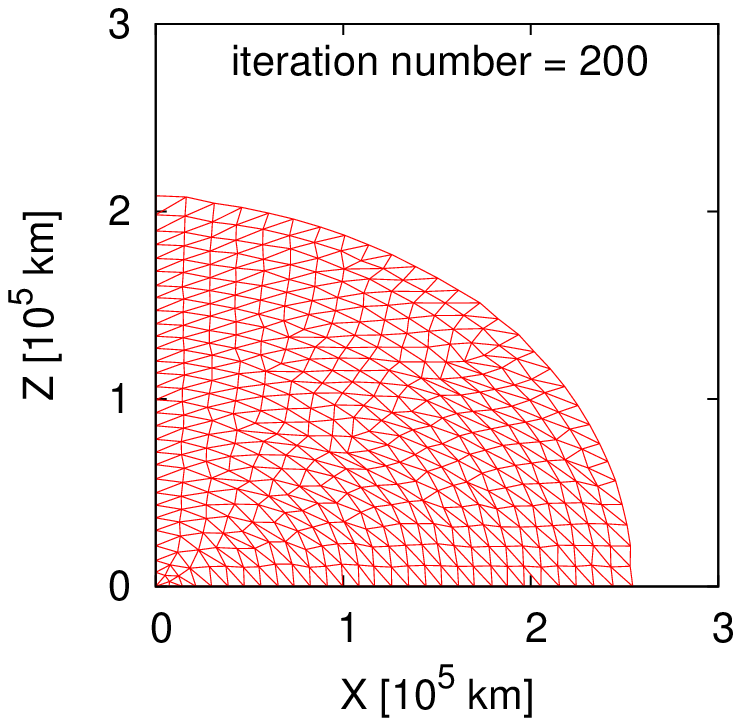}
\hspace{-15mm}
\includegraphics[width=16pc]{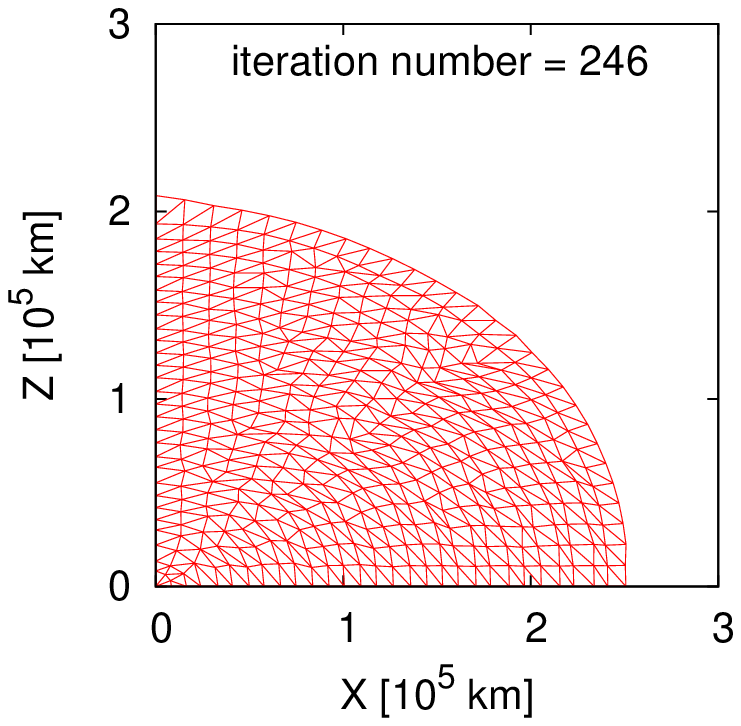}
\caption{\label{fig:edge_dif} (colour on-line). 
Same as Fig.\ref{fig:edge_rig} but for the same differential rotation as in Figs.~\ref{fig:d_dif}, \ref{fig:his_dif}.  
} 
\vspace{5mm}
\end{figure*}

%----- paragraph on FIG.9 -----
In Fig.~\ref{fig:edge_dif}, the logs of the stellar structures for a differential rotation given by equation (\ref{eq:w_dif}) are illustrated again. The structure changes from initial model---the results of HSCF scheme expanded by 20 \% to the radial direction--- to the hydro-static solution. The star continues to transform at 150 sweeps, although the energy function is converged in Fig.~\ref{fig:his_dif}.

%----- paragraph on FIG.10 -----
We also show the distributions of angular momentum in this model as $\Omega_0 = 1.6 \times 10^{-3}$  rad s$^{-1}$, and $T/|W|$ is 4.3 \%. As for the numerical accuracy, as we see some differences ($\sim 20 $\%) between both schemes especially around rotational axis because of the division of zero or small values. 

%-------------- subsec 3-1-3 --------------
%この部分は不要か、このセクションの最初に置くべき。実際、cylindrical rotationになることは既に述べて使っている。安定性は自明。
%  \subsubsection{The Solberg criterion, and the another condition} %Stability on barotropic stars  
%As closing this subsection, we focus on the stability conditions on barotropic stars.
%In stable homentropic ~(not baroclinic) stars, the angular momentum per unit mass must necessarily increase outward,
%\begin{eqnarray}
%\frac{d}{d \varpi} (\Omega^2 \varpi^4) > 0,
%\label{eq:solberg}
%\end{eqnarray}
%which is known as the Solberg criterion. The criterion is necessary (but adequate) condition for the stability on homentropic and axisymmetric stars. 
%
%We also note that, by simple derivation from the equations of motion~\citet{tassoul78}, the isobaric- and isopycnic-surfaces all coincide if and only if
%\begin{eqnarray}
%\frac{d \Omega}{d z} = 0. 
%\label{eq:dwdz=0}
%\end{eqnarray}
%Hence barotropic stars, in which the EOS is expressed as $P=P(\rho)$, must satisfy the above relation.
%Summarizing equations (\ref{eq:solberg}) and (\ref{eq:dwdz=0}), they are understood as, 
%``{\it rotating barotropic stars have cylindrical distributions on angular velocity, including uniform rotation}".  
%
% In rigid rotation models, the Solberg condition (\ref{eq:solberg}) is straightforwardly satisfied, and equation (\ref{eq:dwdz=0}) is also satisfied. As for the differential rotation, it is also clear that the cylindrical rotation is realized as it should be as shown in Fig.~\ref{fig:w_dif}: equation  (\ref{eq:solberg}) and (\ref{eq:dwdz=0}) are both satisfied.  
 It is emphasized that this is not a trivial thing for YFY  scheme. In fact, the rotation law is not known a priori in our formulation. The cylindrical rotation is {\it not assumed but obtained} as a result of computations. This is a strong vindication that our formulation works well indeed. 
 We assume the other rotational laws including a rigid rotation in the reference configuration, and find in all cases cylindrical rotations in the outcomes. 
 
%----- FIG.6-----

%----- FIG.7-----

%----- FIG.8-----

%----- FIG.9-----
\begin{figure*}
\includegraphics[width=15pc]{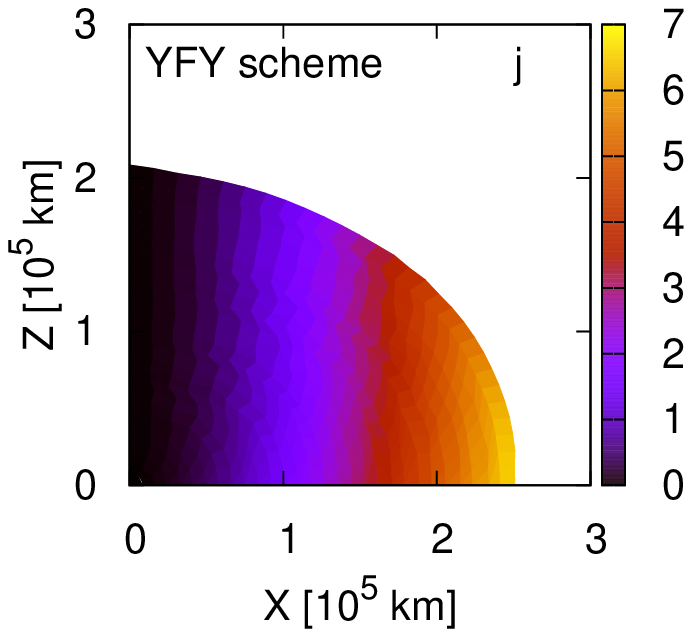}
\hspace{-13mm}
\includegraphics[width=15pc]{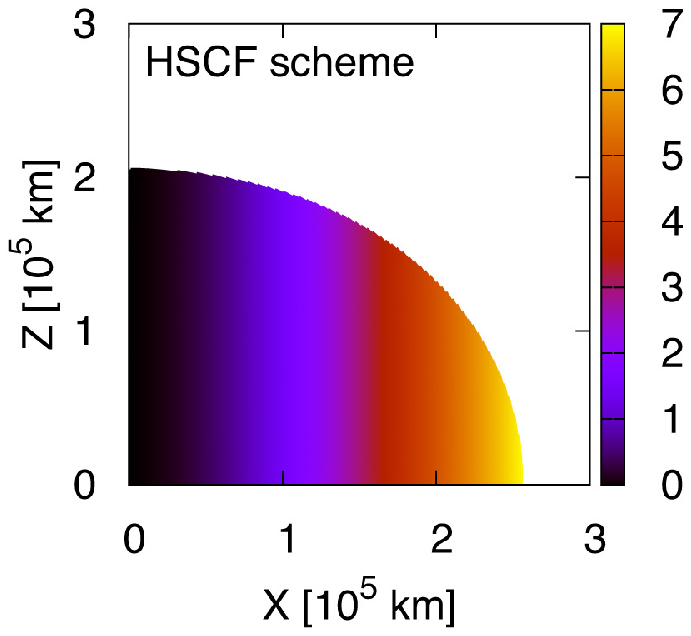}
\hspace{-13mm}
\includegraphics[width=15pc]{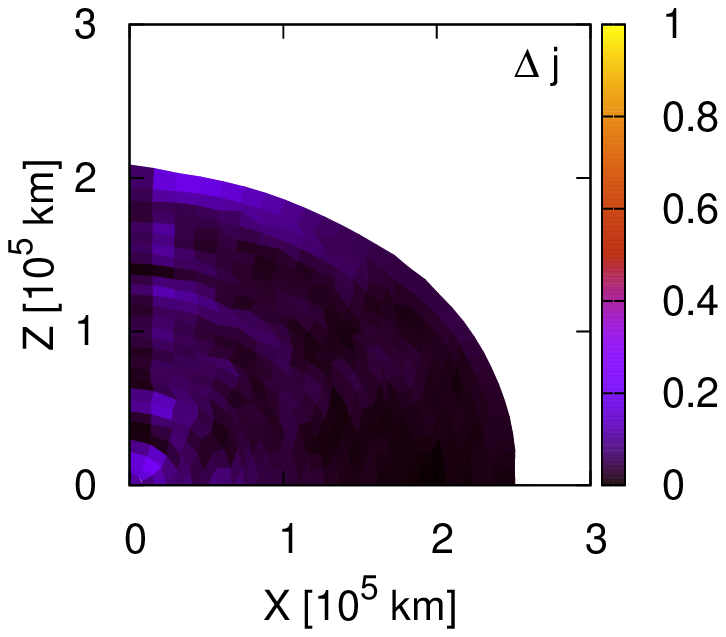}
\caption{\label{fig:j_dif} (colour on-line). Distribution of specific angular momentum 
for the same differential rotation as in Figs.~\ref{fig:d_dif}-\ref{fig:edge_dif}. The angular momentum is in unit of 10$^{17}$ cm$^2$ rad s$^{-1}$. For this model, $\Omega_0=1.6 \times 10^{-3}$ rad s$^{-1}$, and $T/|W| = 4.3 ~\% $. } 
\end{figure*}

%%%%%%%%%%%%%%%%%%%%%%%%%%%%%%%%%%%%%%%%%%%%%%%%
\subsection{Comparison of baroclinic equilibria}  
\label{subsec:FSCF}
%%%%%%%%%%%%%%%%%%%%%%%%%%%%%%%%%%%%%%%%%%%%%%%%

It is obvious that the EOS is generally baroclinic, i.e., pressure depends not only on density but also entropy and chemical abundance in stars. It is, then, necessary to handle such baroclinic equilibria for the study on the structures of rotating stars. Such investigations are not many in the literature, though~\citep{uryu94, espinosa07, espinosa13, fujisawa15, rieutord16}. In the stellar evolution calculation, the shellular rotation is commonly assumed as a rotation law. It is not obvious, however, whether there exist such hydrostatic equilibria or not, especially for rapid rotation. Recently, Fujisawa (2015), one of the authors of this paper, formulated a new self-consistent field scheme that is suitable for the construction of baroclinic equilibria. It is a Eulerian method and may be regarded as a (nontrivial) extension of the HSCF scheme we used in the previous subsection. It is probably superior to the formulation proposed in this paper in terms of accuracy. The advantage of the YFY scheme, however, is that it is a Lagrangian method and suitable to construct an evolutionary sequence out of an ensemble of rotational equilibria. Anyway, with two independent numerical schemes at hand to obtain baroclinic equilibria, it is a good opportunity for us to make comparisons and validate both of them simultaneously. 
%----- 以下はここで言う必要なし。-----
%In this context, the issue of our paper is to research possible hydrostatic equilibrium under arbitral mass, entropy and angular momentum distributions; i.e. arbitral density, thermal energy, and angular velocity distributions. We, therefore, remove out the effects of radiation process, convections and meridional flow for the purpose. It could be the constraints for the distributions of entropy indeed. In following, we compare our results~(YFY scheme) with the ones by Fujisawa's self-consistent field~(FSCF) scheme~\citep{fujisawa15}, solving the same equations in both methods, ignoring the radiation process, convection and meridional flow. 

%-------------- entropy distribution --------------
The baroclinicity is most easily introduced by an artificial modification of the entropy distribution. 
We employ the polytrope-like EOS, $p = K\rho^{\gamma}$ with $K$ being position-dependent,\\
(i) spherical iso-$K$ surfaces:
\begin{equation}
K({\bf r})=K_0  \biggl( 1 + e~\frac{r^2}{R_e^2} \biggr),
\label{eq:K_brcl}
\end{equation} 
(ii) oblate iso-$K$ surfaces:
\begin{equation}
K({\bf r})=K_0  \bigg\{ 1 + e_1  \biggl(1 + e_2  P_2(\cos{\theta}) \biggr) \frac{r^2}{R_e^2} \bigg\}, 
%K     = K0 * ( 1 + 0.45 * (1 + 0.8 * P2(\cos th)) * r**2)x
\label{eq:K_shell}
\end{equation}
where, the Legendre polynomials are denoted by $P_n$ with $\theta$ being colattitude and employed to give non-sphericity to $K$ distribution. Note that $K$ corresponds to a (constant) specific entropy in the original polytropic EOS and, hence, the modifications above may be regarded as an introduction of entropy-dependence to pressure. Two dimensionless parameters $e$ and $e_1$ specify radial non-uniformity whereas $e_2$ gives colatitudinal dependency. 
We set $e=0.35$, $e_1=0.45$ and $e_2=0.80$ here. These models are highly artificial admittedly, and more realistic models, in which the entropy distributions are determined self-consistently with energy generations and radiative transfer, will be presented in the forthcoming papers. 

In the FSCF scheme, the rotation law is given on the equatorial plane and the angular velocity distribution is solved self-consistently. We hence employ the rotation law given in equation (\ref{eq:w_dif}) only on the equatorial plane. The resultant configurations are expanded radially by 20 \% and used as reference configurations for the YFY scheme just as in the previous sub section. 

\begin{figure*}
\includegraphics[width=15pc]{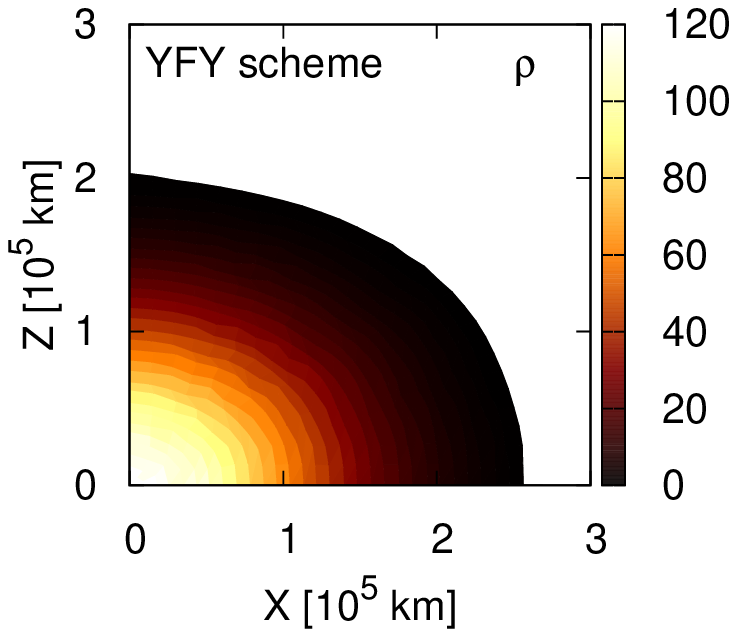}
\hspace{-13mm}
\includegraphics[width=15pc]{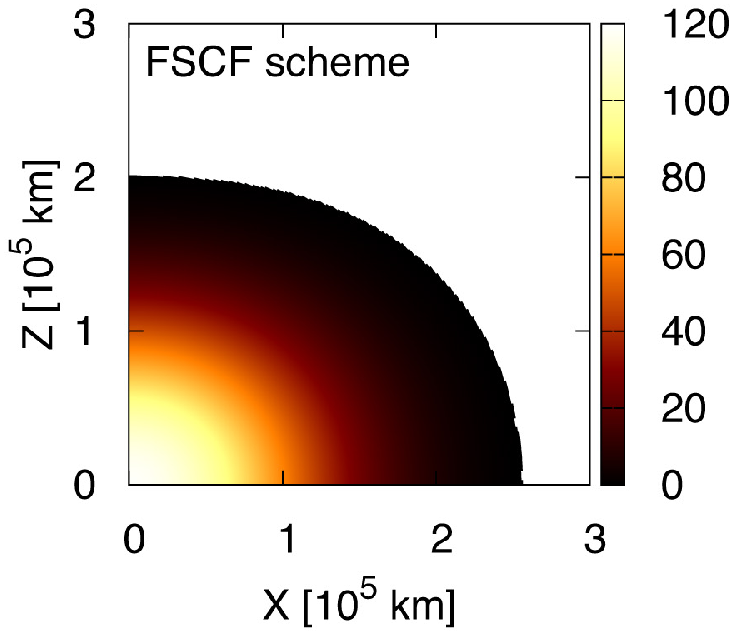}
\hspace{-13mm}
\includegraphics[width=15pc]{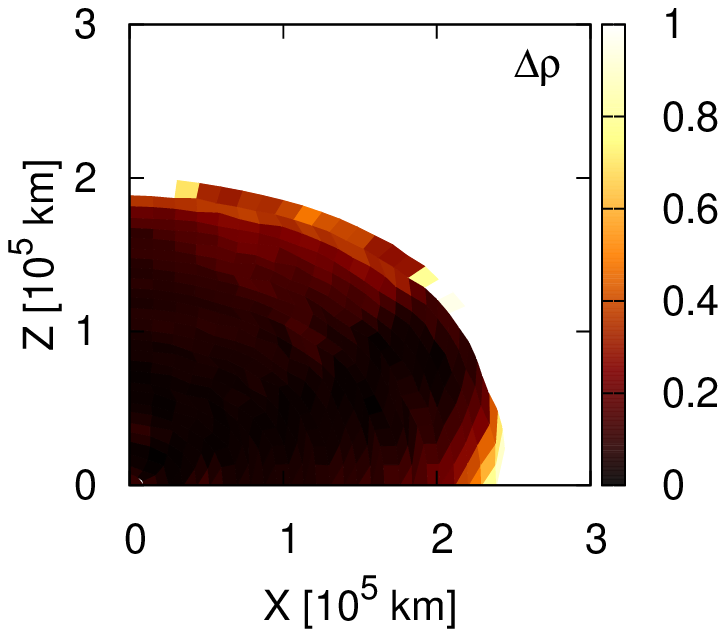}
\caption{\label{fig:d_brcl} (colour on-line). 
Same as Fig.~\ref{fig:d_rig} but for a baroclinic configuration with the spherical isentropic surfaces given in equation (\ref{eq:K_brcl}). The maximum density and equatorial radius normalization factors $\rho_0$ and $R_e$ on FSCF scheme are set to $\rho_0 = 1.19 \times 10^2$ g cm$^{-3}$ and $ R_e=2.57 \times 10^5$ km, respectively, and the constant $K_0$ in equation (\ref{eq:K_brcl}) is chosen to be $5.49 \times 10^{13}$ in cgs unit.
} 
\end{figure*}

%----- paragraph on FIG.11 -----
We first present the results for case (i) with the spherical isentropic surfaces given in equation (\ref{eq:K_brcl}). In Fig.~\ref{fig:d_brcl}, we show isopycnic color-contours, in which the left panel is for the YFY scheme and the middle panel for the FSCF scheme. Just as in the HSCF scheme, the FSCF scheme deals with non-dimensional quantities normalized with the maximum density $\rho_0$ and equatorial radius $R_e$. To facilitate the comparison with our scheme, we set $\rho_0 = 1.19 \times 10^2$ g cm$^{-3}$ and $ R_e=2.57 \times 10^5$ km. The values of $K_0$ and $\Omega_0$ chosen to be $5.49 \times 10^{13}$ in cgs unit and $1.9 \times 10^{-3}$ rad s$^{-1}$, respectively. This value of $\Omega_0$ is a few hundred  times the solar angular velocity. 
We find again a reasonable agreement in most regions as is evident in the right panel of the figure, in which the relative difference is shown. Note that the number of the grid points employed in the FSCF method, $N_r=513$ and $N_\theta=257$, is much larger than that deployed in the YFY method, $N=489$. It is also apparent from the same figure that the deviation gets larger in the vicinity of the surface for the reason mentioned earlier.
%---- これは要らない。むしろ上のように結果を述べる。----
%Note again that, in the panel of YFY scheme, we do not show the outer boundary since we do not impose the same outer boundary condition with FSCF scheme as described. We, therefore, show the one inner shell from the outer boundary with the dashed line in YFY scheme. We plot in the panel for YFY scheme only the active grid points and the outer boundary of the mesh is not displayed for the reason given previously in the comparison of the barotropic models.

\begin{figure*}
\includegraphics[width=16pc]{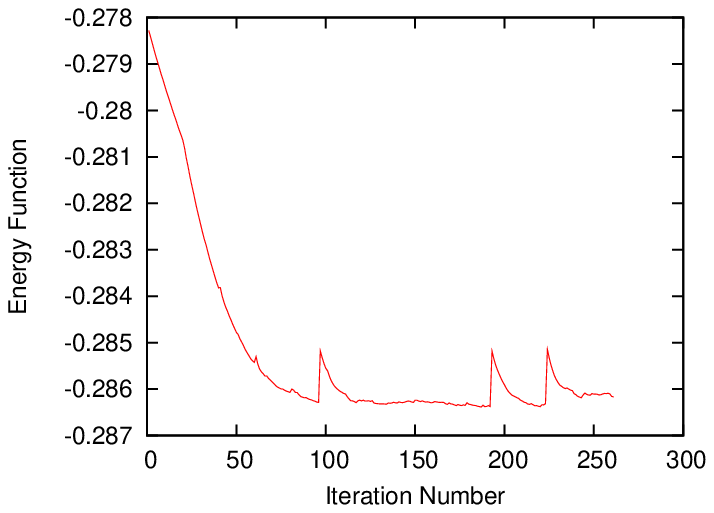}
\includegraphics[width=16pc]{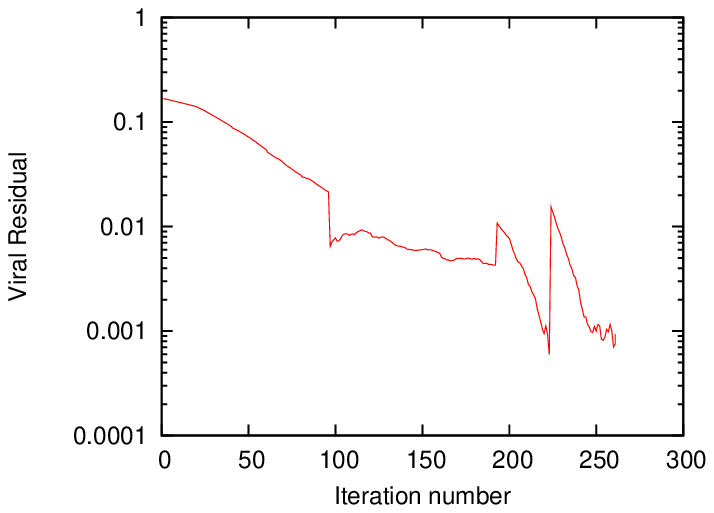}
\caption{\label{fig:his_brcl} (colour on-line). 
Same as Fig.~\ref{fig:his_rig} but for the baroclinic configuration presented in Fig.~\ref{fig:d_brcl}. The iteration is terminated at 261 sweeps. } 
\end{figure*}

\begin{figure*}
\includegraphics[width=16pc]{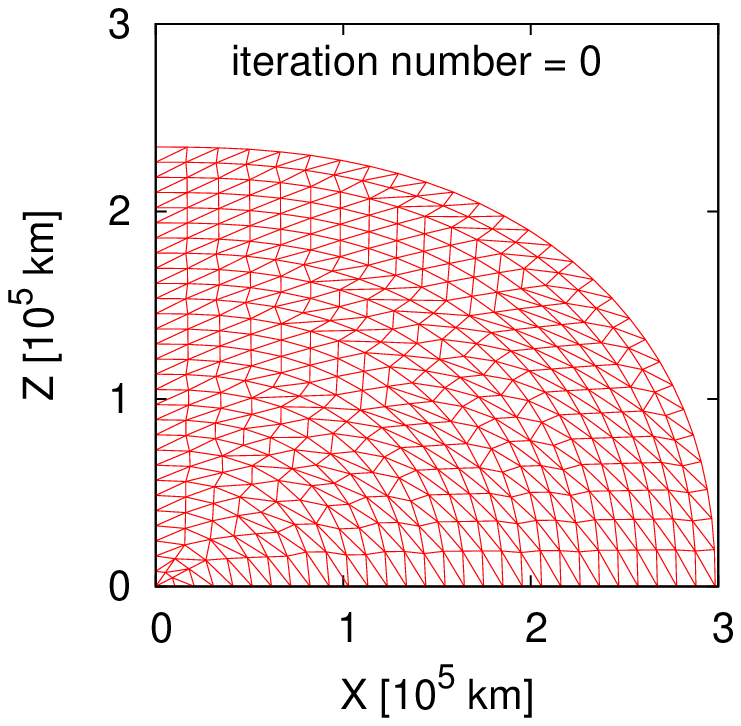}
\hspace{-15mm}
\includegraphics[width=16pc]{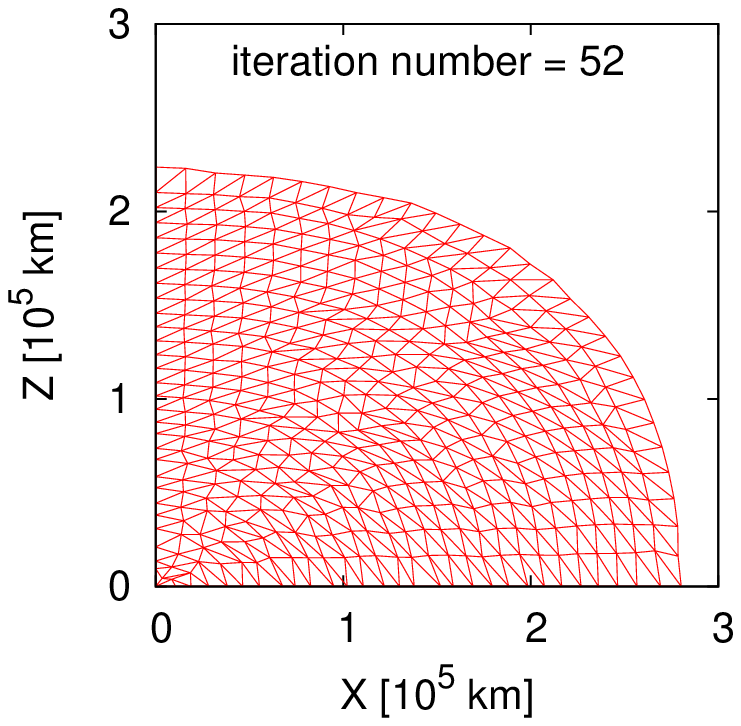}
\hspace{-15mm}
\includegraphics[width=16pc]{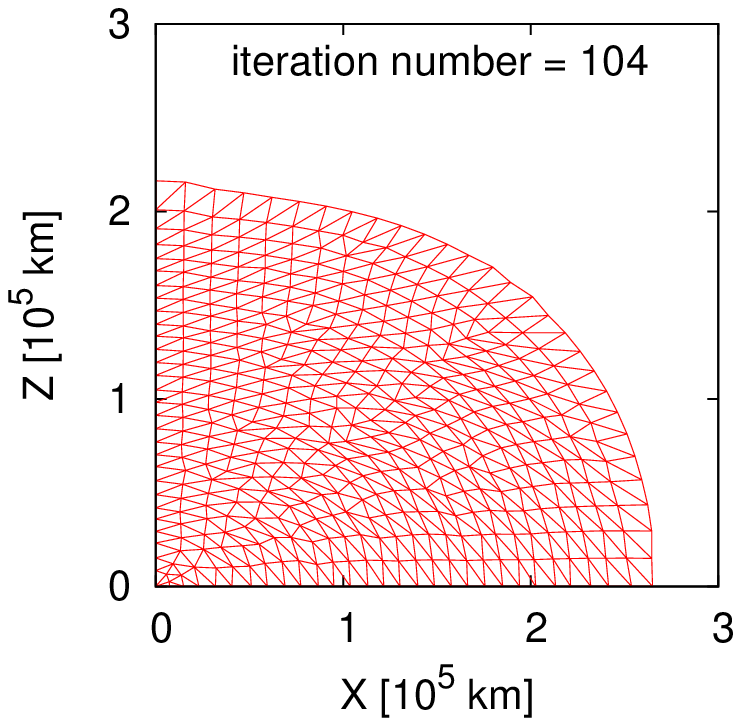}\\
\includegraphics[width=16pc]{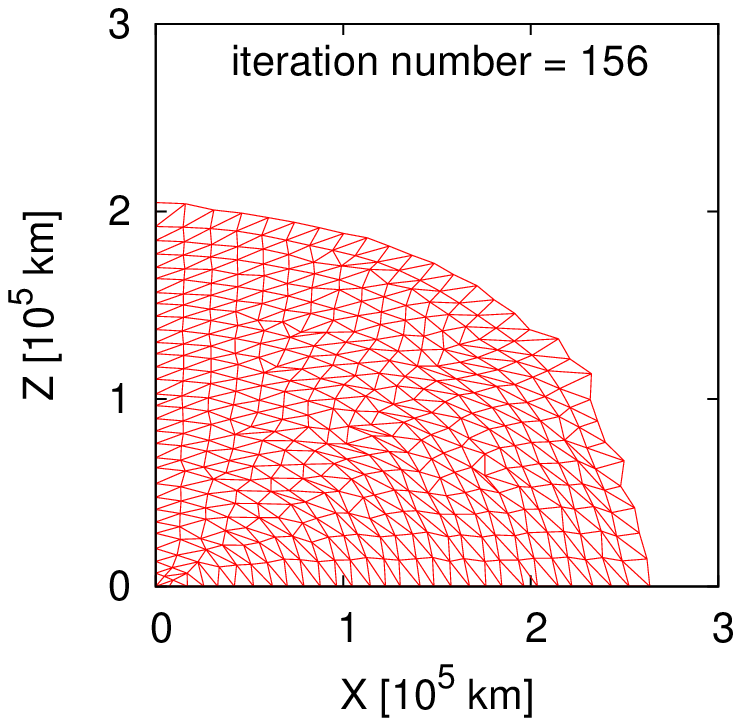}
\hspace{-15mm}
\includegraphics[width=16pc]{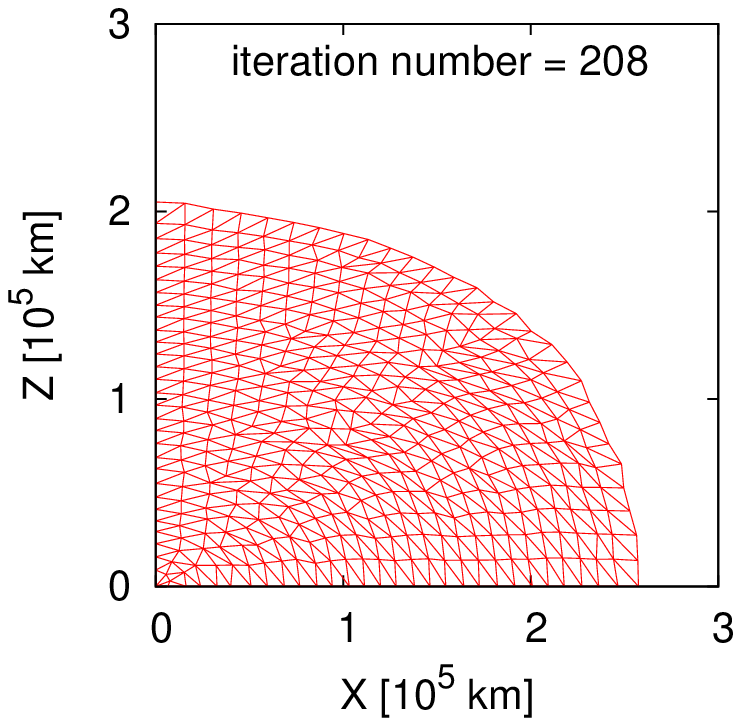}
\hspace{-15mm}
\includegraphics[width=16pc]{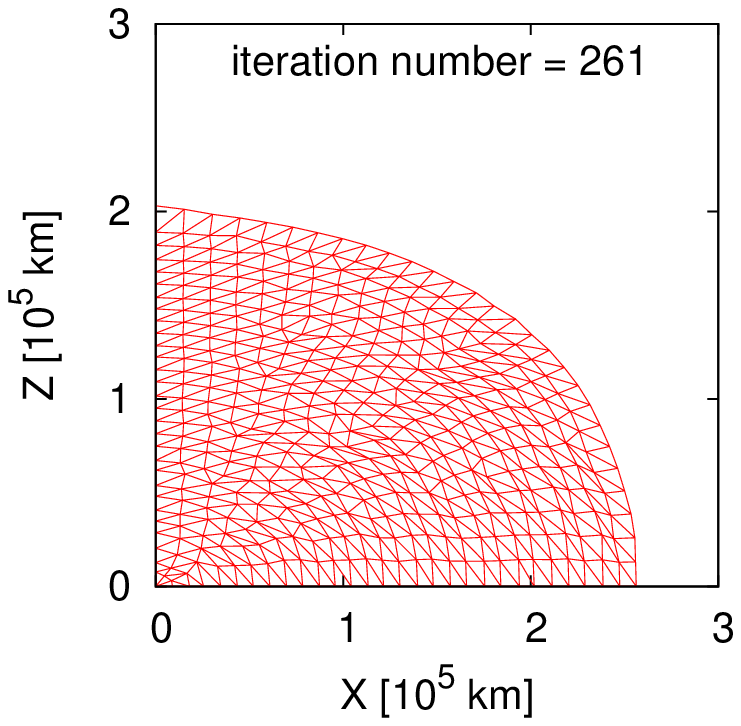}
\caption{\label{fig:edge_brcl} (colour on-line). 
Same as Fig.~\ref{fig:edge_rig} but for the baroclinic configuration presented in Figs.~\ref{fig:d_brcl} and \ref{fig:his_brcl}. 
} 
\vspace{5mm}
\end{figure*}

%----- paragraph on FIG.12,13 -----I
In Figs.~\ref{fig:his_brcl} and  \ref{fig:edge_brcl} we present the values of the energy function and virial residual as a function of the number of sweeps and the corresponding change of the configuration, respectively, just as in the barotropic case. It is observed that the energy function is settled to a certain value after 130 sweeps. It is apparent from the value of virial residual $\sim 10^{-2}$ that this is a false minimum. The true one is reached at 261 sweeps after a couple of shuffles by the smoothings, which are marked by both major and minor spikes in the energy function shown in the left panel of Fig. 12. This demonstrates again that the virial residual is an important diagnostic measure for true minimum and the smoothing is an indispensable process to escape the trap of false minima.

\bigskip

\begin{figure*}
\includegraphics[width=18pc]{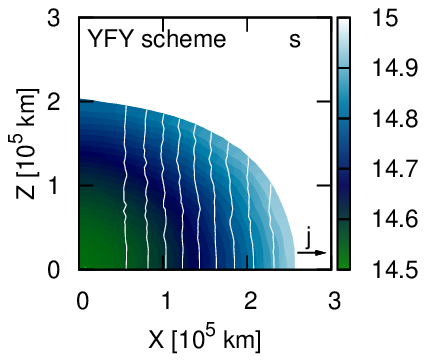}
\includegraphics[width=18pc]{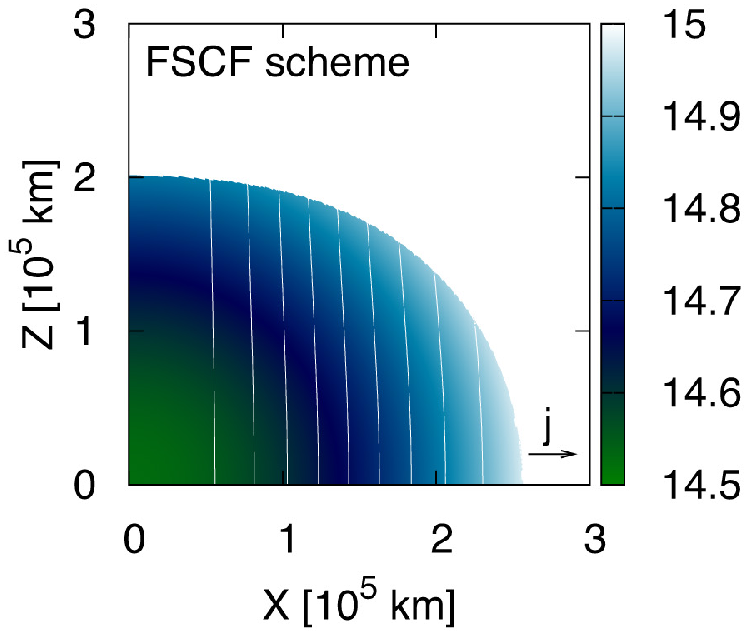} \\
\vspace{-10mm}
\includegraphics[width=18pc]{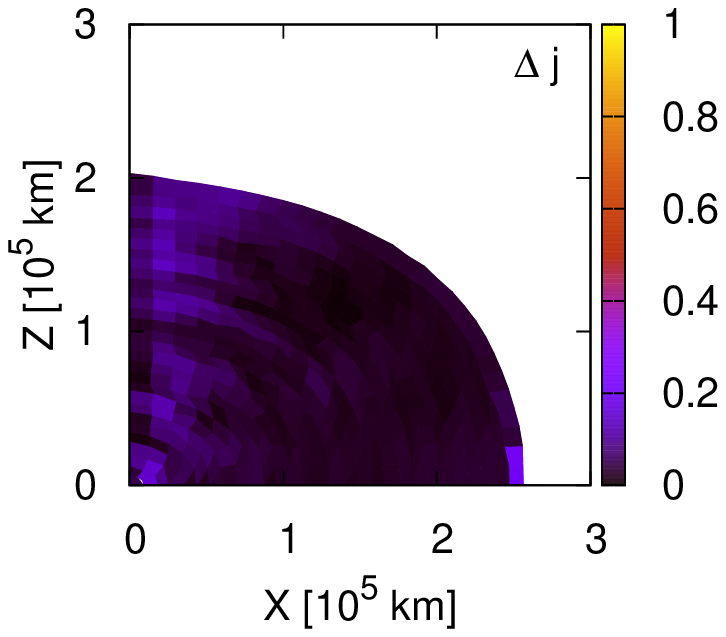}
\includegraphics[width=18pc]{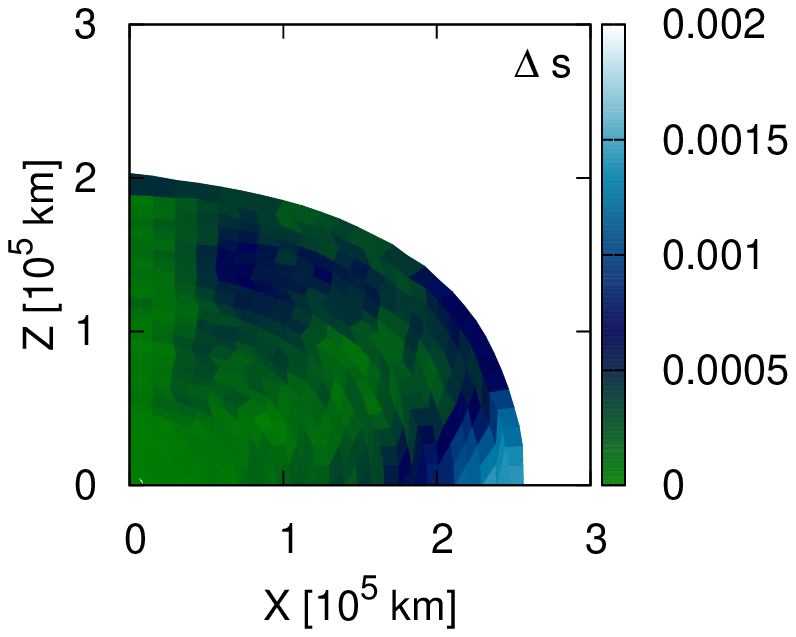}
\caption{\label{fig:s_brcl} (colour on-line). 
(Upper panels) iso-angular momentum contours on top of the entropy color maps for the baroclinic 
configuration shown in Figs.~\ref{fig:d_brcl}-\ref{fig:his_brcl}. The results for the YFY scheme and the FSCF scheme are displayed in the left and right panels, respectively.  The arrows indicate the direction, in which the angular momentum increases. (Lower panels) the relative differences of specific angular momentum (left panel) and entropy (right panel) between the results obtained with the schemes.} 
\end{figure*}

%----- paragraph on FIG.14 -H{\o}iland criterion ----
It is well known that the stability of rotational equilibria against axisymmetric perturbations can be judged by the Solberg and H{\o}iland criterion which is summarized as follows: stable configurations satisfy (i) the specific entropy $s$, and (ii) the specific angular momentum $j$ increases from the poles to the equator on  the isentropic surface. The rotational configuration we obtained above satisfies these  conditions as can be understood from Fig. \ref{fig:s_brcl}, in which we display the iso-angular momentum surfaces on top of the density color-contours; the arrows in the figure indicate the direction, in which the angular momentum increases.
% It is clear from these figures that  the Solberg-H{\o}iland criterion, summarized as conditions (i) and (ii) in the last paragraph, is satisfied by these models.

\begin{figure*}
\includegraphics[width=18pc]{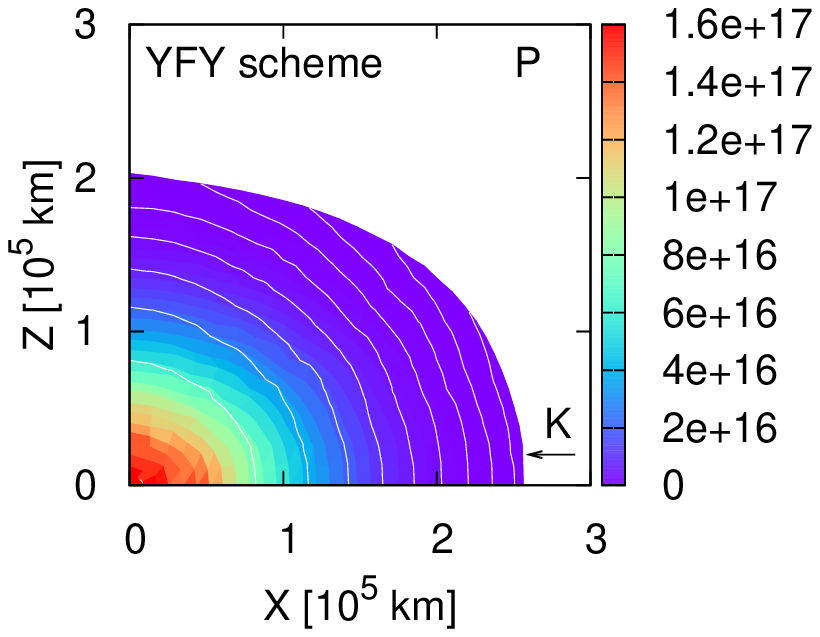}
\includegraphics[width=18pc]{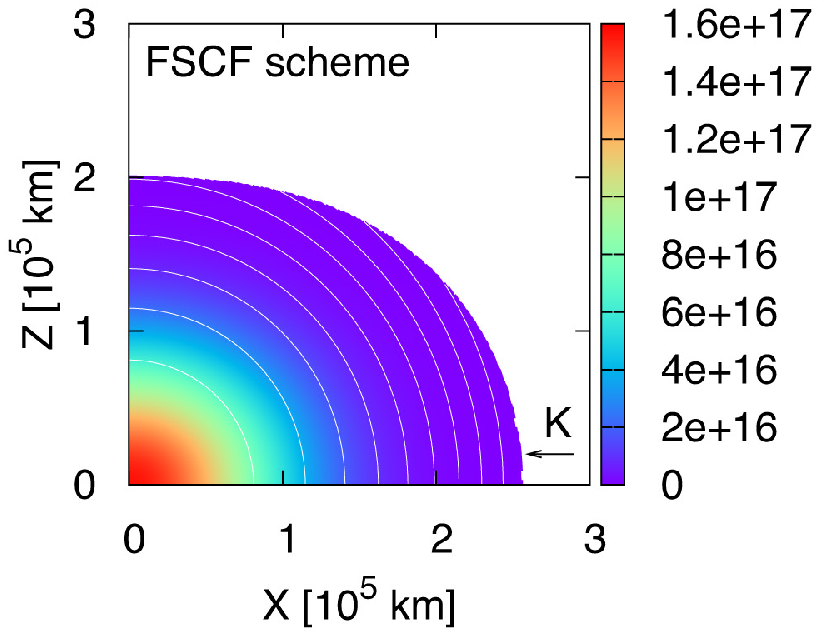}\\
\vspace{-10mm}
\includegraphics[width=18pc]{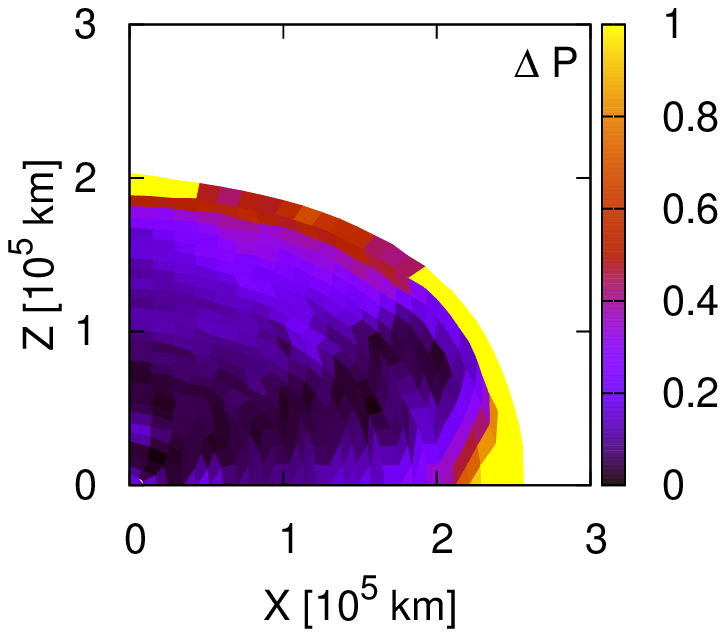}
\includegraphics[width=18pc]{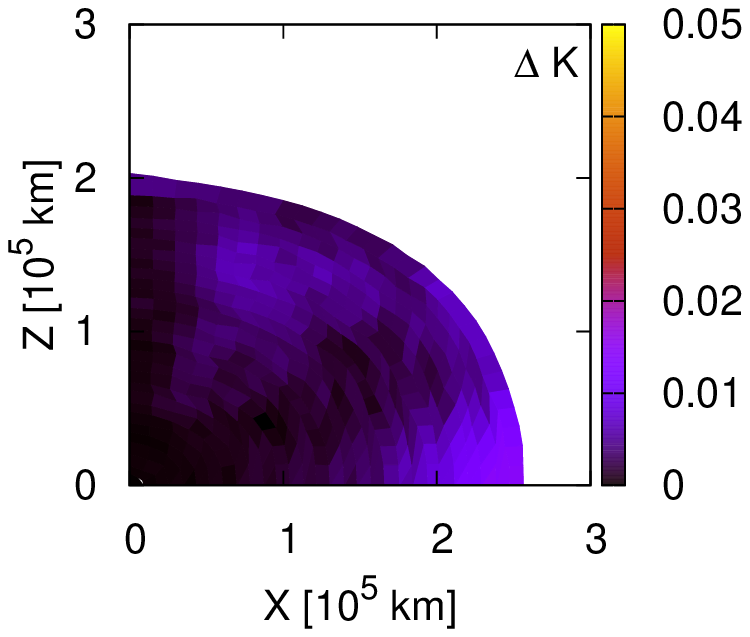}
\caption{\label{fig:pk_brcl} (colour on-line). 
(Upper panels) iso-$K$ surfaces on top of the pressure color-contours for the baroclinic configuration in Figs. ~\ref{fig:d_brcl}-\ref{fig:s_brcl}. The arrows indicate the direction, in which $K$ %decreases. 
increases. The left and right panels correspond to the results for the YFY and FSCF schemes, respectively. (Lower panels) the relative differences of pressure (left panel) and $K$ (right panel) the results for the two schemes.} 
\end{figure*}

%----- paragraph on FIG.15 Bjorkness-Rosseland rule -----
There is another well known rule we should care for: the Bjerknes--Rosseland rule claims that if $d\Omega/dz$ is positive at a given point, the isobaric surface is more oblate than the isopycnic one there~\citep{tassoul78}. This means in particular that {\it if $d\Omega/dz$ is positive everywhere, matter temperatures are lower at the poles than one on the equator on each isobaric surface}~\citep{tassoul78}. The rule is also confirmed for the results presented above. The isobaric surfaces and iso-$K$ (and, hence, iso-entropic) surfaces are shown in Fig.~\ref{fig:pk_brcl}. As is evident from the figure, matter on each isobaric surface has indeed a smaller value of $K$ at poles than on the equator. On the other hand, the gradient of specific angular moment $dj/dz$, and hence 
$d\Omega/dz$ is also slightly positive as can be seen in Fig.~\ref{fig:s_brcl}.

%----- FIG.10-----

%----- FIG.11-----

%----- FIG.12-----

%----- FIG.13-----

%----- FIG.14-----

%\clearpage
\bigskip

\begin{figure*}
\includegraphics[width=15pc]{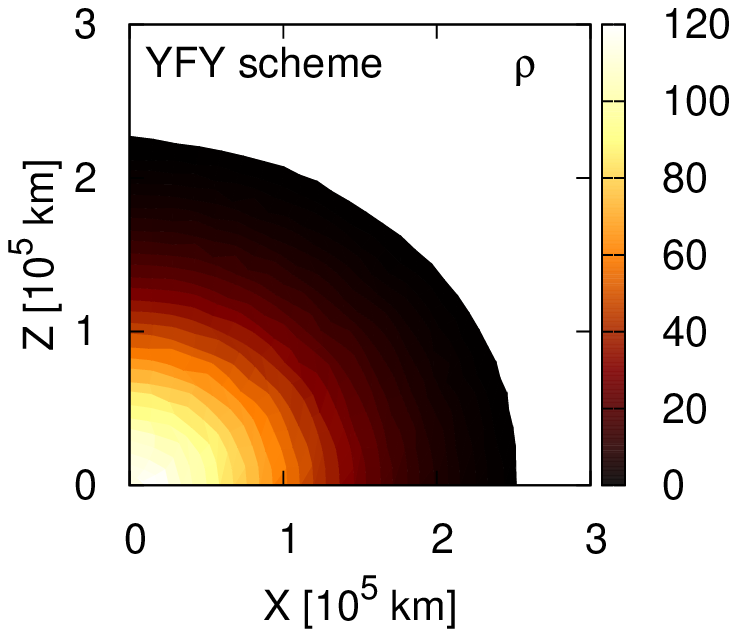}
\hspace{-13mm}
\includegraphics[width=15pc]{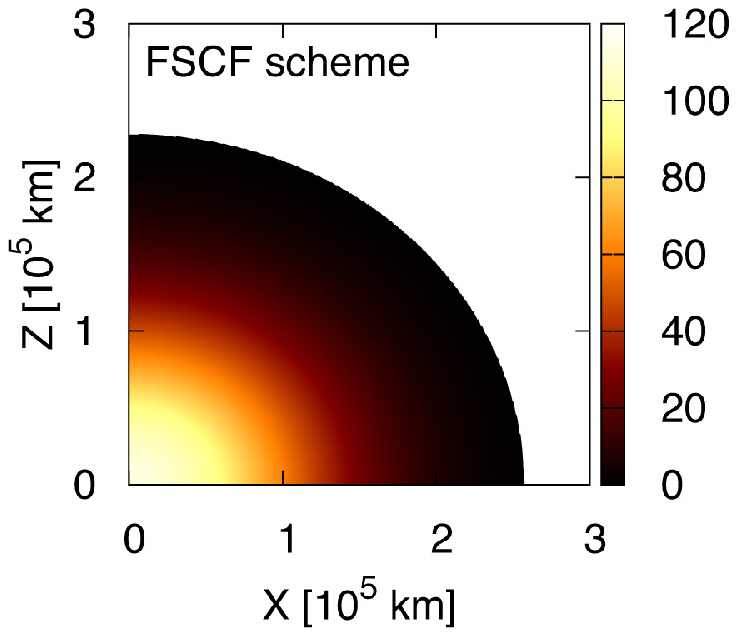}
\hspace{-13mm}
\includegraphics[width=15pc]{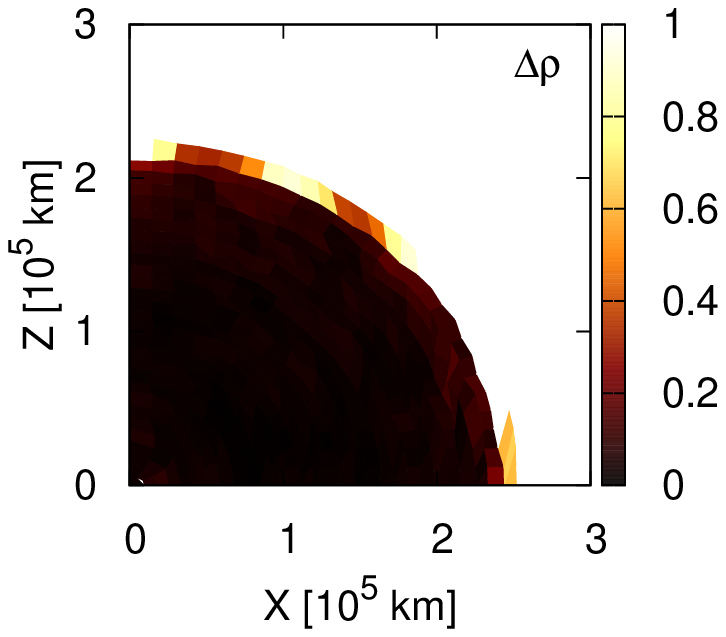}
\caption{\label{fig:d_shell} (colour on-line). 
Same as Fig.~\ref{fig:d_brcl} but for the oblate isentropic surfaces given in equation (\ref{eq:K_shell}). The maximum density and equatorial radius are set to $\rho_0 = 1.19\times 10^2$ g cm$^{-3}$ and $ R_e=2.57 \times 10^5$ km, respectively.} %In this renormalization, $K_0$ in equation (\ref{eq:K_brcl}) is given as $*** \times 10^{13}$ in cgs unit.} 
\end{figure*}

 %-------------- subsec 3-2-2 --------------
%%%%%% \subsubsection{Oblate isentropic surfaces models:  shellular-type rotation}  %%%%%%
%----- paragraph on FIG.16 -----
We move on to case (ii), i.e., with oblate isentropic surfaces as specified in equation (\ref{eq:K_shell}). What is important with this entropy distribution is that it leads to rotational equilibria with a shellular-type rotation demonstrated by~\citep{fujisawa15}. Figure~\ref{fig:d_shell} compares the isopycnic surfaces obtained with the YFY scheme~(left panel) and for the FSCF scheme~(middle panel). As quantified in the right panel, in which the relative difference is presented, the both results are in agreement within an error of $\sim$ 5 \% except near the surface as usual.
%The two normalization factors in FSCF scheme are set to $R_e = 2.57 \times 10^5$ km and $\rho_0 =  1.13 \times 10^2$ g cm$^{-3}$. 
%----- on FIG.17,18 -----
The behavior of the energy function and virial residual as well as the configuration itself is shown in  Figs. \ref{fig:his_brcl} and \ref{fig:edge_brcl}. It is quite similar to what we found for case (i) as well as for the barotropes. It is observed that the energy function is trapped erroneously to a local minimum around 130 sweeps as suggested by not-so-small values of the virial residual. The true minimum is reached only after 275 sweeps, where $V_C< 10^{-3}$ is satisfied. 

%----- paragraph on FIG.19,20 -----
We show in Fig.~\ref{fig:w_shell} the angular velocity distributions obtained with the two methods for the same model. It is obvious that they are not cylindrical but of {\it shellular-type}. The relative difference between the results obtained with the two schemes is $\sim$5 \% except near the rotation axis as shown in the right panel. As should be apparent from the left panel, the numerical interpolation on the triangulated grid that is employed by a graphic software in plotting the contour certainly contributes to the discrepancy. The relative differences for the specific angular momentum and entropy are given in the lower panels of Fig.~\ref{fig:s_shell}. The trends are quite similar to those for the angular velocity and density, respectively.

As mentioned repeatedly, the numerical construction of rotational equilibria with the shellular-type rotation law has important implications for stellar evolution theory, since most of the realistic calculations done so far have assumed shellular rotation ~\citep[e.g.][]{meynet97, talon97, heger00, mathis04, potter12a, potter12b, takahashi14}. The choice is based on the theory of the angular momentum transport induced by rotation advocated by ~\citet{zahn92}, who argues that horizontal turbulence should be much stronger than the vertical one, which will lead to a horizontal rearrangement of angular momentum, enforcing a uniform rotation on each concentric sphere and resulting in the ``shellular" rotation. It is stressed again that there have been not so many studies that have successfully obtained equilibria with shellular rotations in two-dimension because of the difficulty associated with baroclinicity. Although \cite{roxburgh06} and \cite{fujisawa15} are notable exceptions, their formulations are Eulerian, and will be unsuitable for stellar evolution, which could be a major break-through for the fully two-dimensional calculation of stellar evolution.

%----- paragraph on FIG.20,21 -----
The entropy distributions and the iso-angular momentum for the configuration presented in Fig.~\ref{fig:d_shell} are shown in Fig.~\ref{fig:s_shell}. Obviously they satisfy the two conditions of the Solberg and H{\o}iland criterion. In contrast to the previous case with the spherical isentropic surfaces, $d\Omega/dz$ is negative everywhere in the current case with the oblate isentropic surfaces as found in Fig.~\ref{fig:w_shell}. As expected and confirmed indeed in Fig.~\ref{fig:pk_shell}, the values of $K$ are larger at poles than on the equator this time. This model is hence also consistent with the Bjerkeness-Rosseland rule.  

%----- FIG.15-----

%----- FIG.16-----
 \begin{figure*}
\includegraphics[width=16pc]{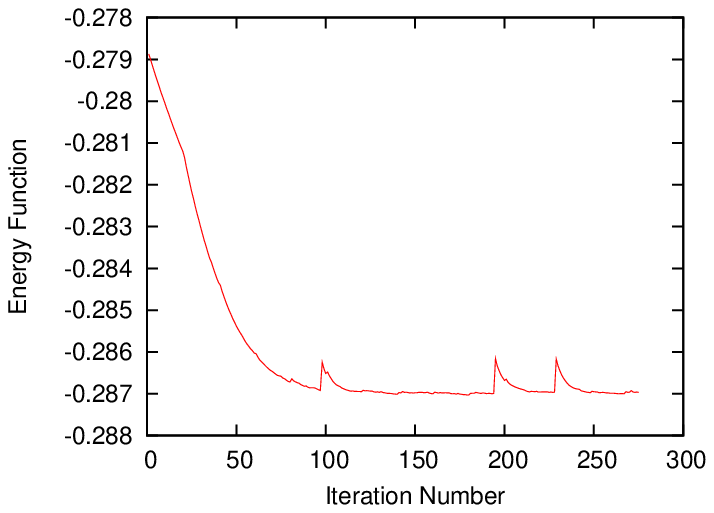}
\includegraphics[width=16pc]{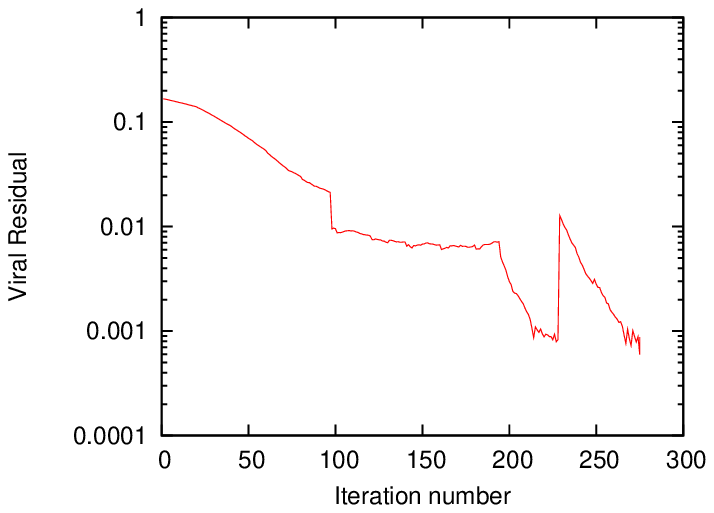}
\caption{\label{fig:his_shl} (colour on-line). 
Same as Fig.~\ref{fig:his_brcl} but for the configuration shown in Fig.~\ref{fig:d_shell}. The iteration is 
terminated at 275 sweeps.
} 
\end{figure*} 

%----- FIG.17-----
 \begin{figure*}
\includegraphics[width=16pc]{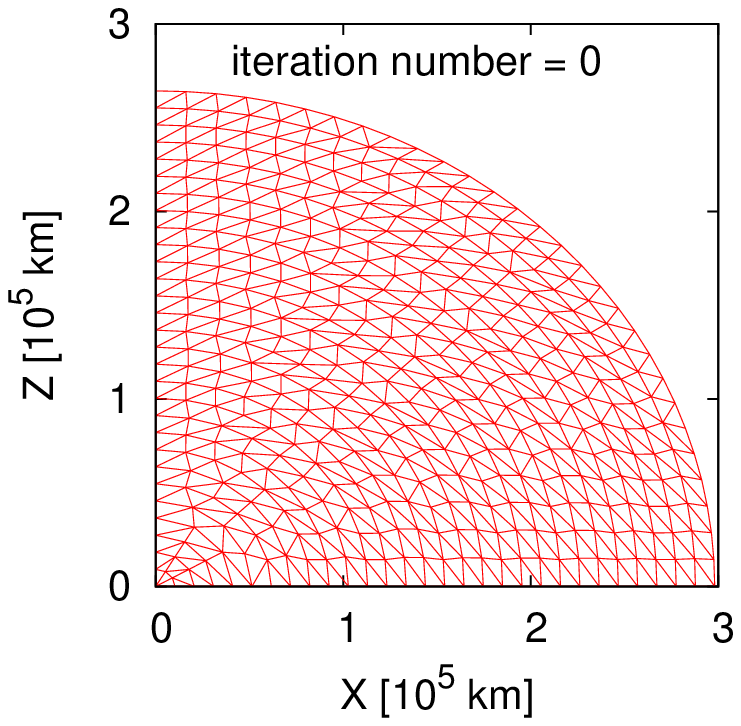}
\hspace{-15mm}
\includegraphics[width=16pc]{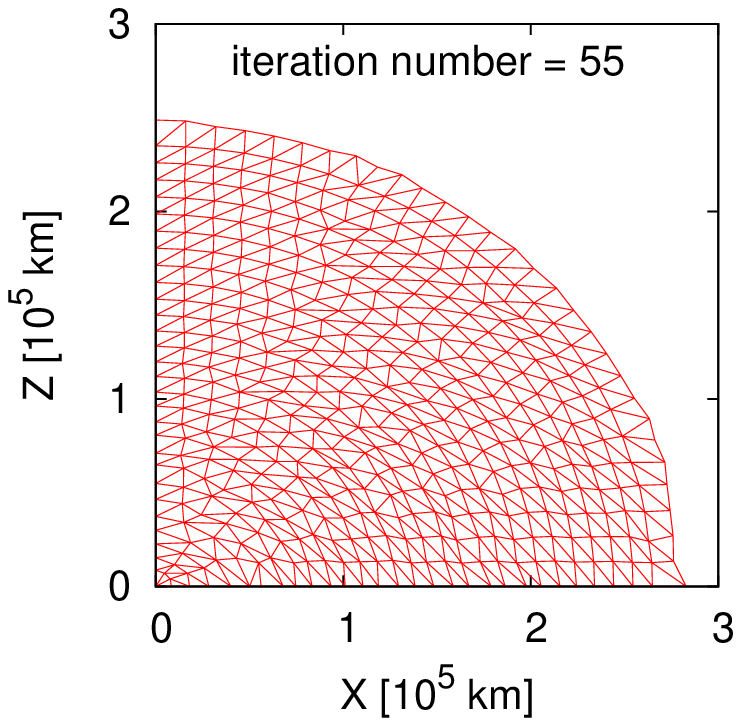}
\hspace{-15mm}
\includegraphics[width=16pc]{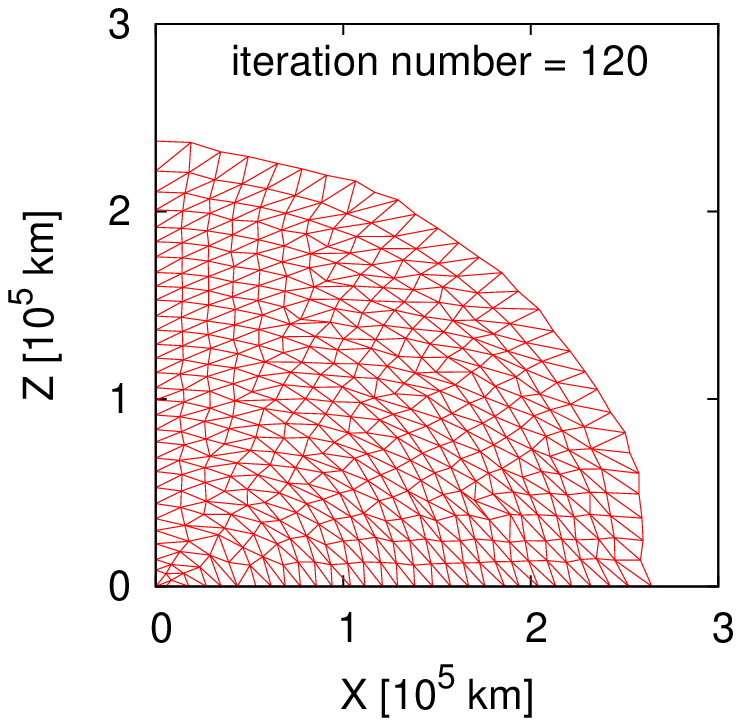}\\
\includegraphics[width=16pc]{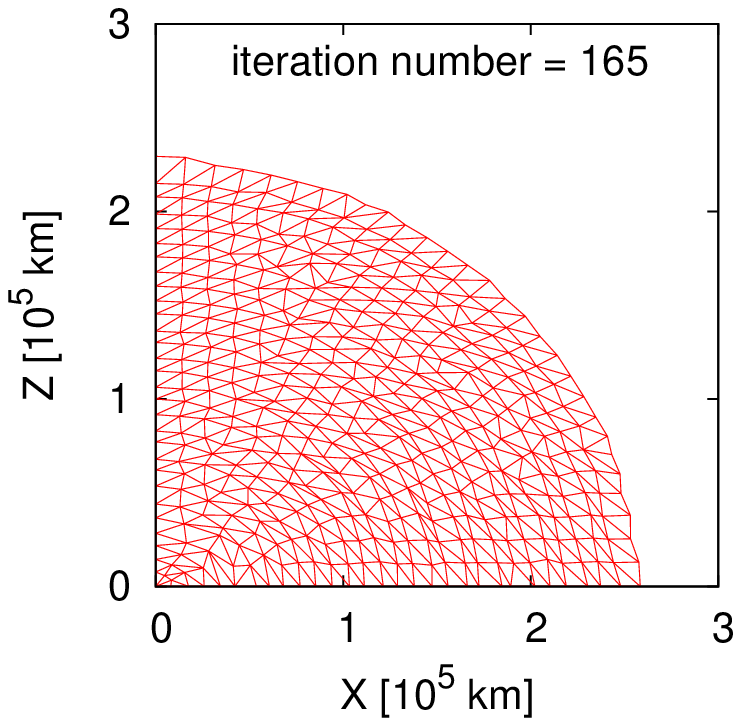}
\hspace{-15mm}
\includegraphics[width=16pc]{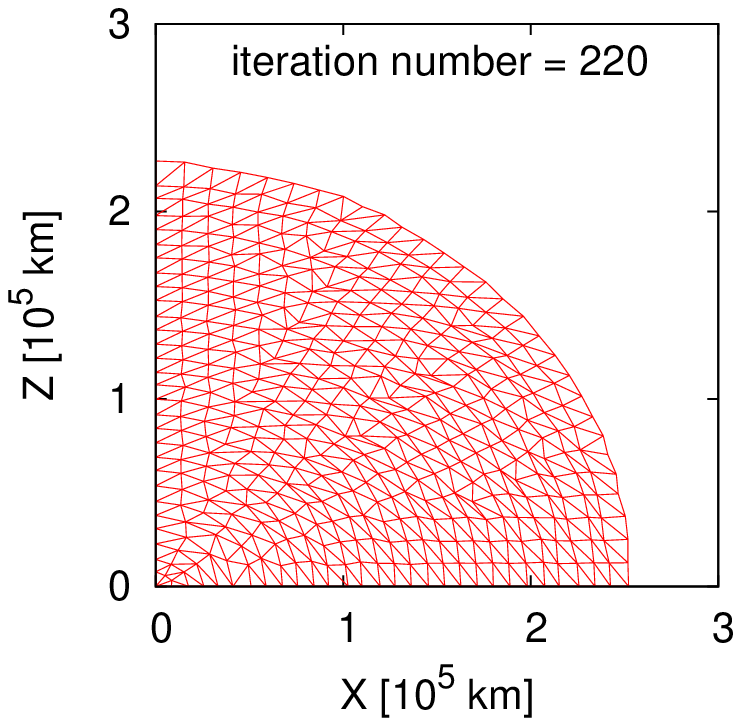}
\hspace{-15mm}
\includegraphics[width=16pc]{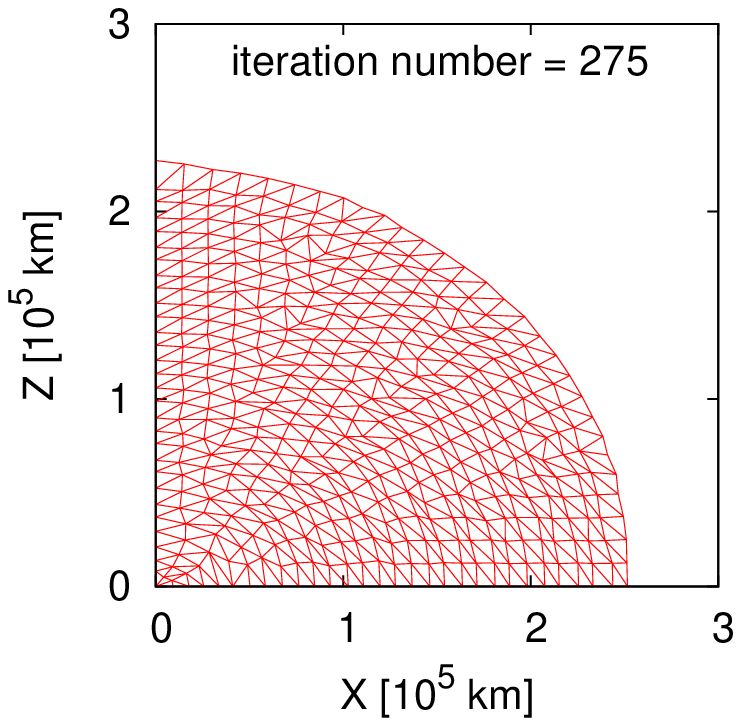}
\caption{\label{fig:edge_shl} (colour on-line). 
Same as Fig.~\ref{fig:edge_brcl} but for configuration displayed in Figs.~\ref{fig:d_shell} and \ref{fig:his_shl}. 
} 
\vspace{5mm}
\end{figure*} 

%----- FIG.18-----
\begin{figure*}
\includegraphics[width=14pc]{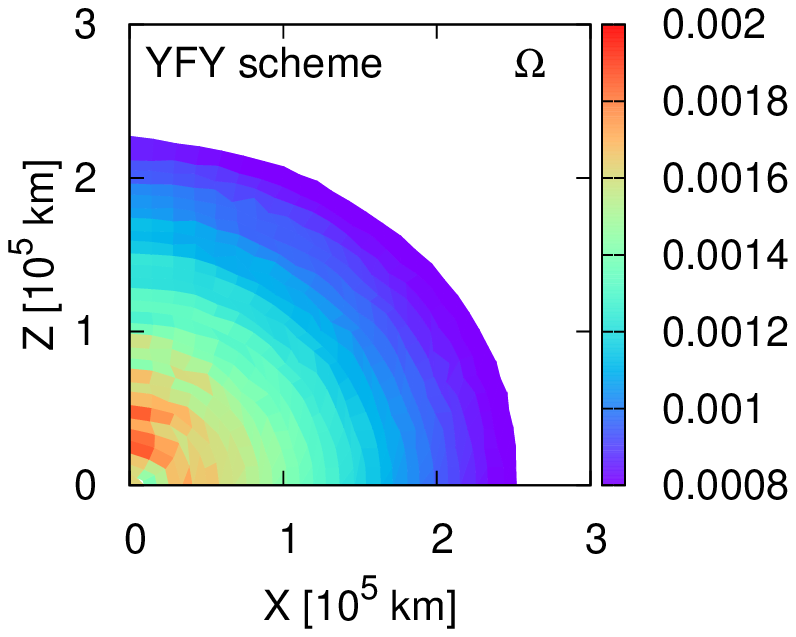}
\hspace{-8mm}
\includegraphics[width=14pc]{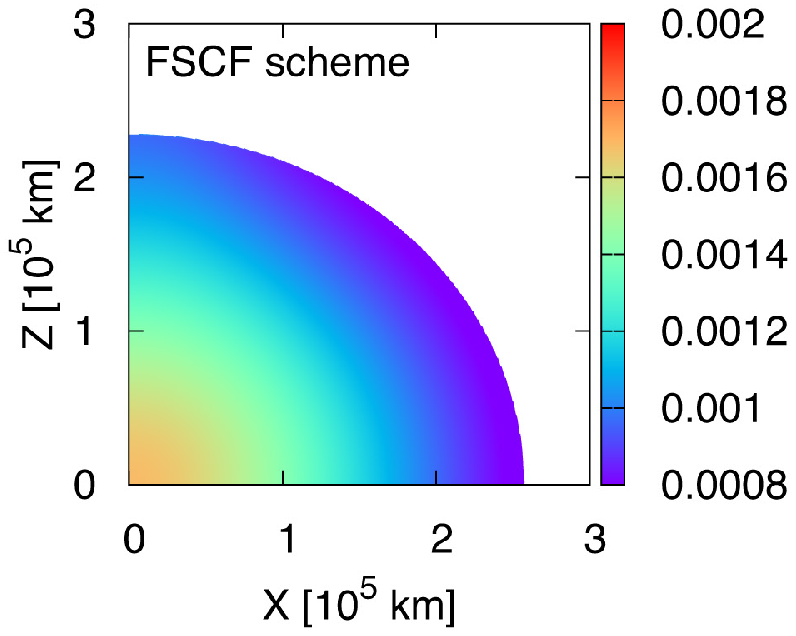}
\hspace{-8mm}
\includegraphics[width=14pc]{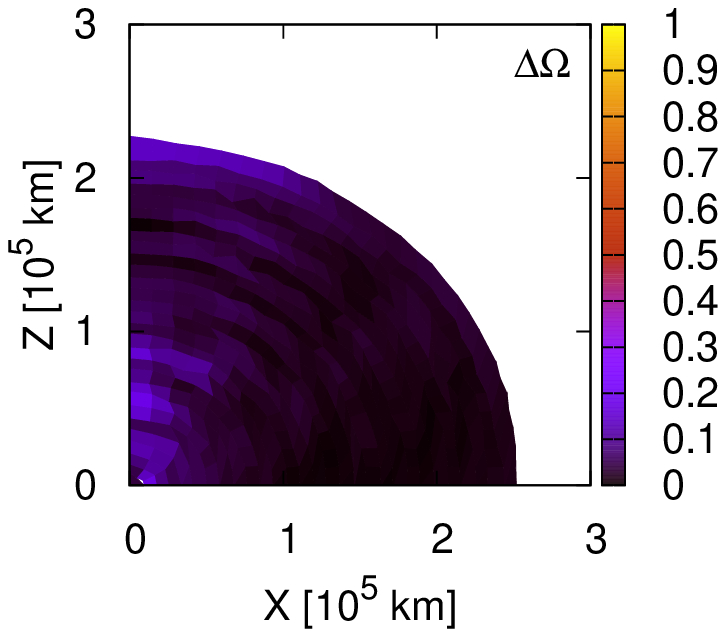}
\caption{\label{fig:w_shell} (colour on-line). 
The left and middle panels show the angular velocity distributions of shellular-type obtained for the configuration shown as in Figs.~\ref{fig:d_shell} - \ref{fig:edge_shl} with the YFY and FSCF schemes, respectively. The right panel exhibits the relative difference.}
\end{figure*}

%----- FIG.19-----
\begin{figure*}
\includegraphics[width=18pc]{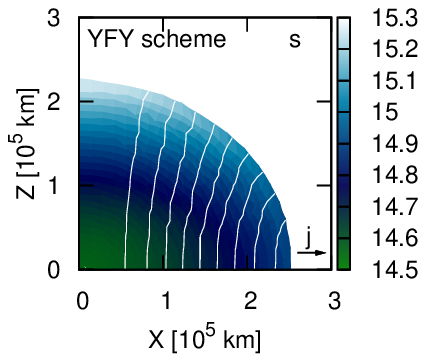}
\includegraphics[width=18pc]{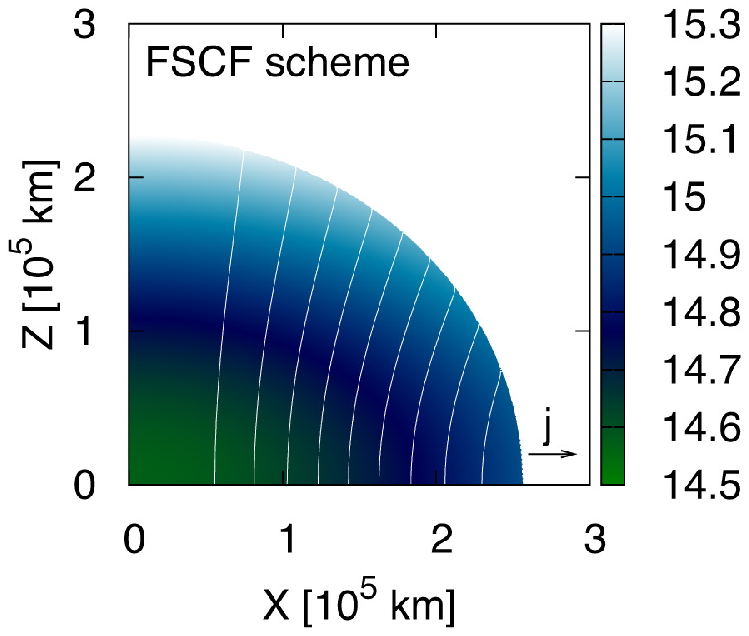}\\
\includegraphics[width=18pc]{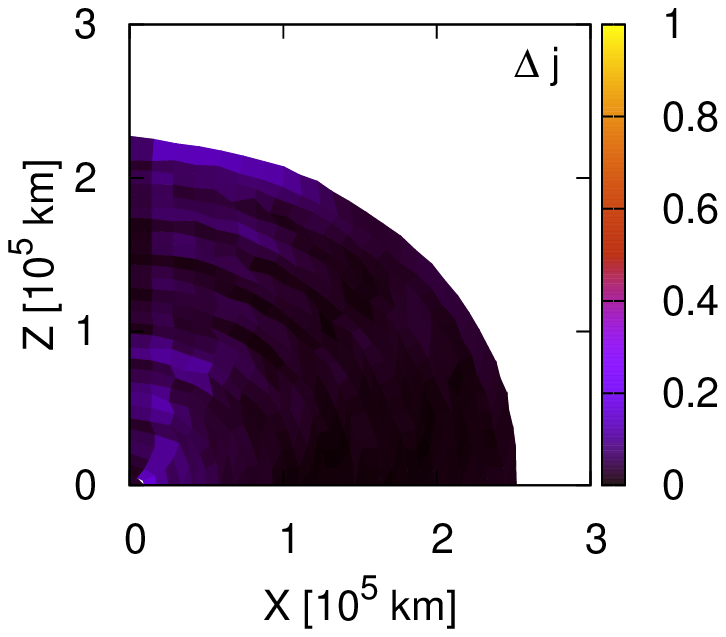}
\includegraphics[width=18pc]{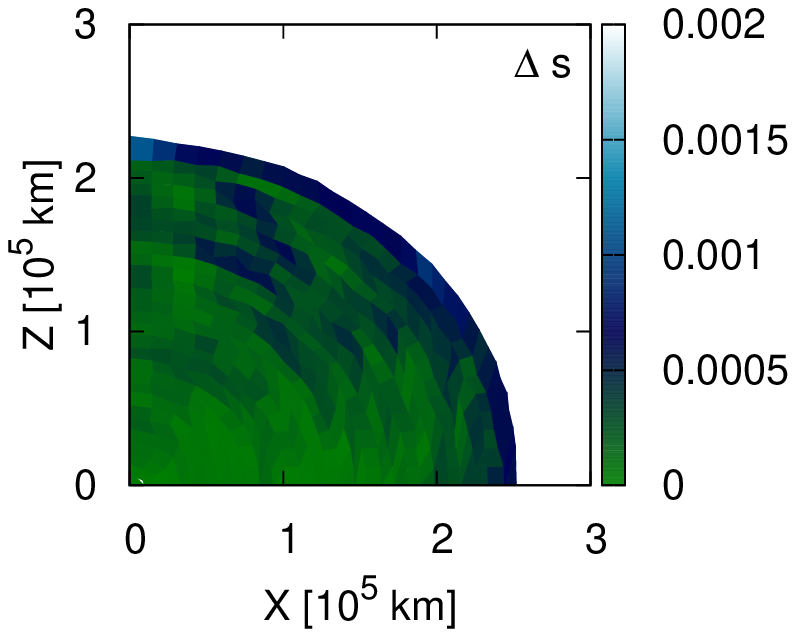}
\caption{\label{fig:s_shell} (colour on-line). Same as Fig.~\ref{fig:s_brcl} but for the configuration presented in Figs.~\ref{fig:d_shell}-\ref{fig:w_shell}. } 
\end{figure*}

%----- FIG.20-----
\begin{figure*}
\includegraphics[width=18pc]{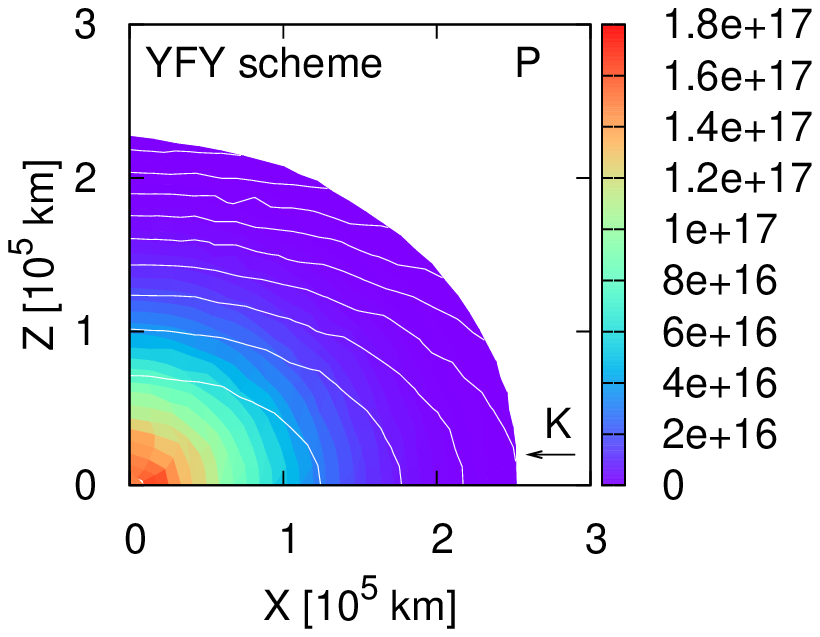}
\includegraphics[width=18pc]{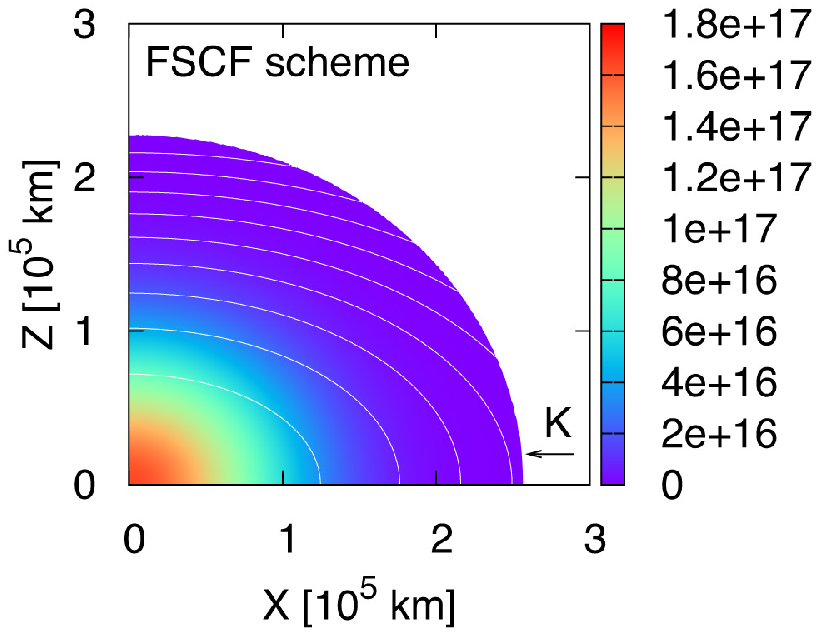}\\
\includegraphics[width=18pc]{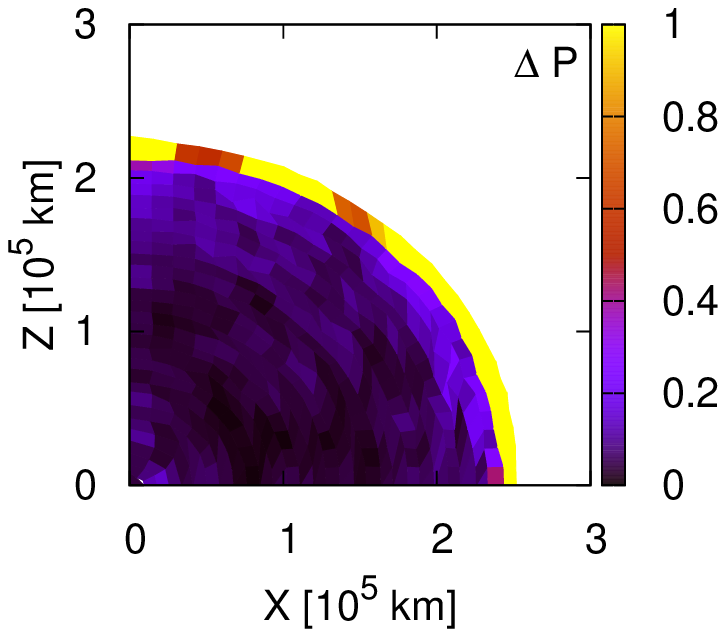}
\includegraphics[width=18pc]{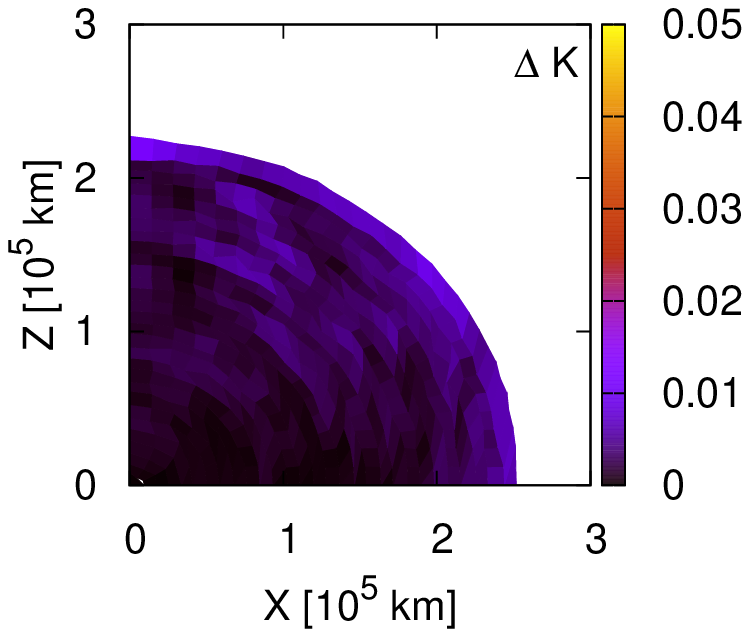}
\caption{\label{fig:pk_shell} (colour on-line). 
Same as Fig.~\ref{fig:pk_brcl} but for the the oblate isentropic surfaces given in equation (\ref{eq:K_shell}).} 
\end{figure*}

\newpage

%%%%%%%%%%%%%%%%%%%%%%%%%%%%%%%%%%%%%%%
\subsection{Reference configuration and smoothing}
\label{subsec:rearrange}
%%%%%%%%%%%%%%%%%%%%%%%%%%%%%%%%%%%%%%%

In the application of the YFY scheme, we have always constructed the reference configurations from the configurations obtained with the HSCF or FSCF scheme, so far, expanding them radially by a certain factor. This may raise a concern that the reference configuration must be prepared in this particular way initially for our scheme. In order to demonstrate that this is not the case in fact, we  try three more different constructions and see if the final configurations are the same or not.

To ensure that all models have the same angular momentum distribution in the final configuration, we still need to prepare the reference configurations by deforming this final configuration. Although it does not matter what configuration is employed for the latter, we adopt the model with the shellular-type presented in Figs. \ref{fig:d_shell}, \ref{fig:w_shell}, \ref{fig:s_shell} and \ref{fig:pk_shell}. This time we do not expand but shrink the final configuration for this model by 30 \% in three different ways: (A) radially, (B) horizontally, and (C) vertically. To the three reference configurations thus obtained, we apply the YFY scheme and see if the same equilibrium, which should be also identical to the configuration shown in Fig.\ref{fig:edge_shl}, is reached.

\begin{figure*}
\includegraphics[width=16pc]{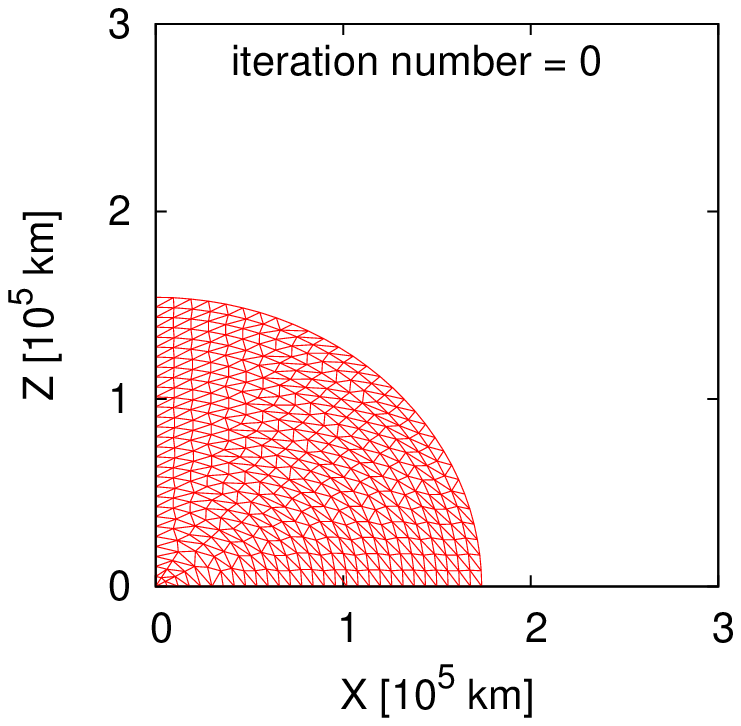}
\hspace{-15mm}
\includegraphics[width=16pc]{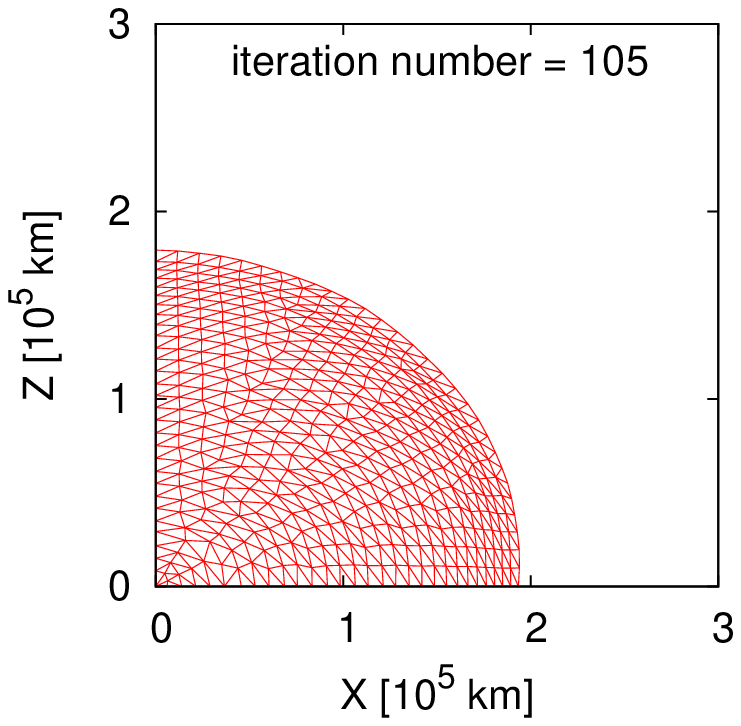}
\hspace{-15mm}
\includegraphics[width=16pc]{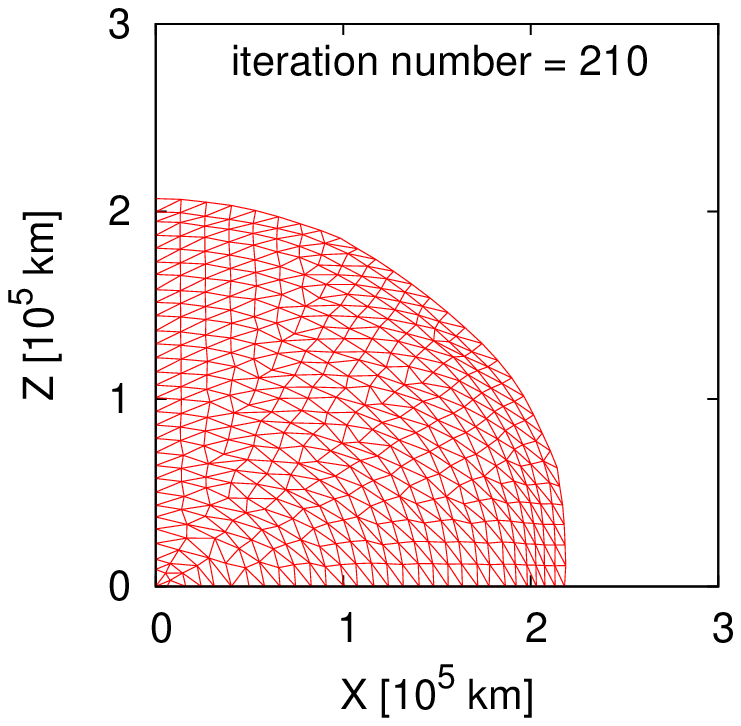}\\
\includegraphics[width=16pc]{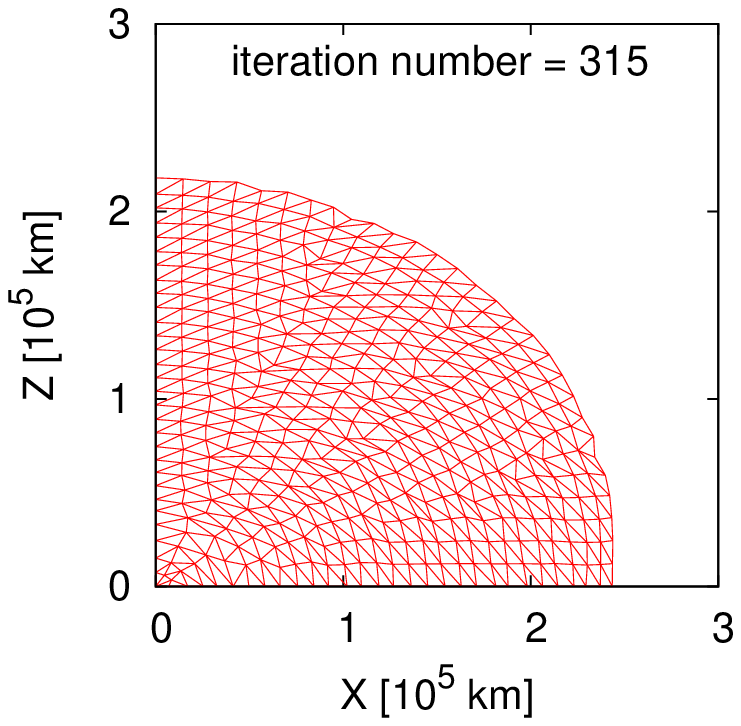}
\hspace{-15mm}
\includegraphics[width=16pc]{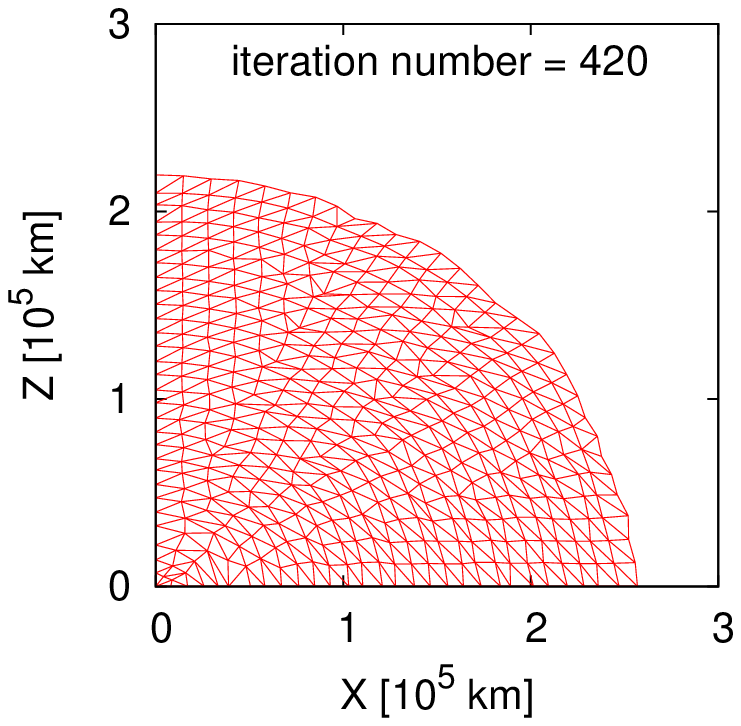}
\hspace{-15mm}
\includegraphics[width=16pc]{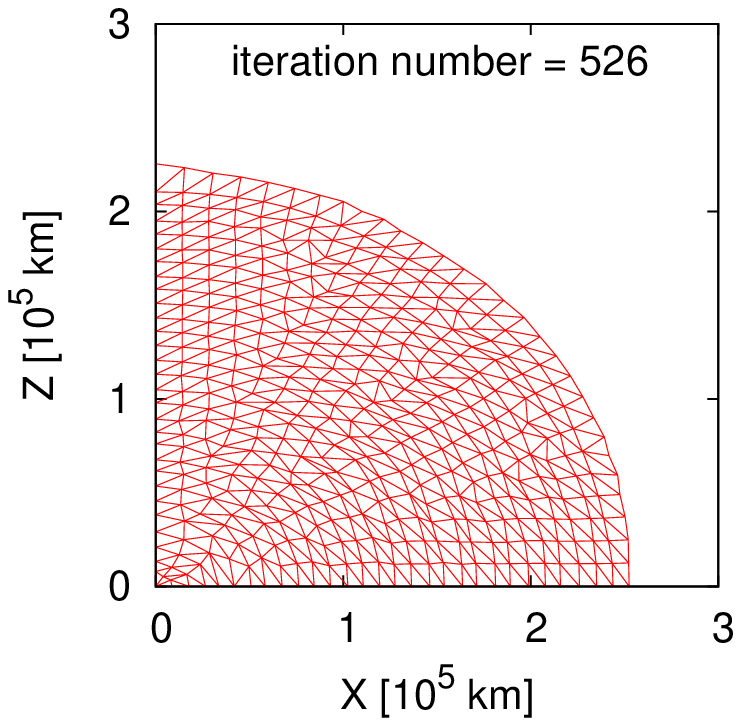}
\caption{\label{fig:r07edge} (colour on-line). 
Same as Fig.~\ref{fig:edge_shl} but for the reference configuration given by the deformation for case (A). } 
\vspace{5mm}
\end{figure*}

\begin{figure*}
\includegraphics[width=16pc]{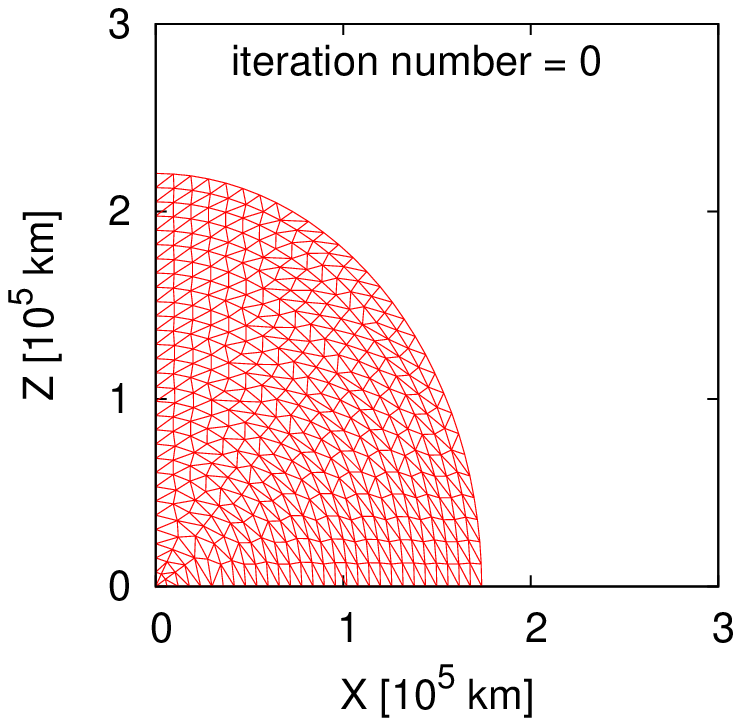}
\hspace{-15mm}
\includegraphics[width=16pc]{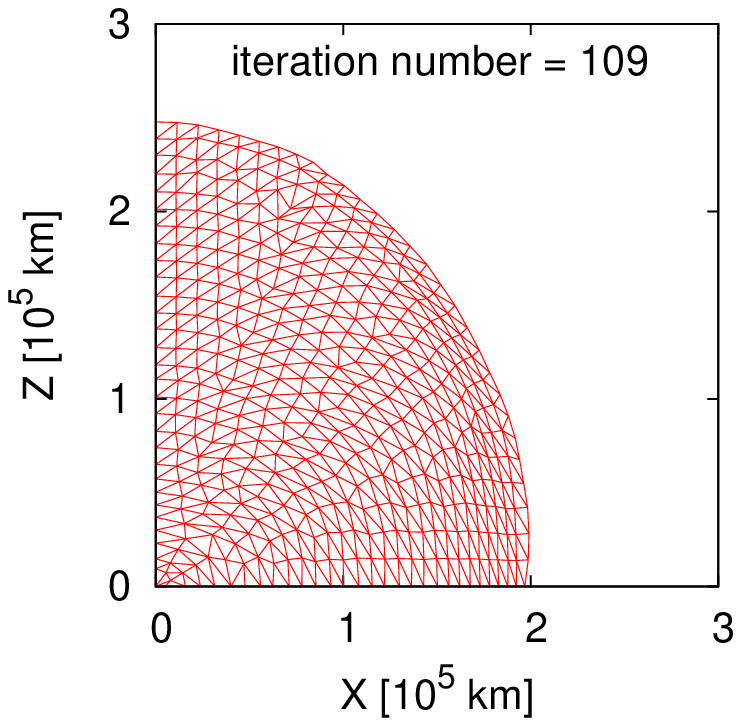}
\hspace{-15mm}
\includegraphics[width=16pc]{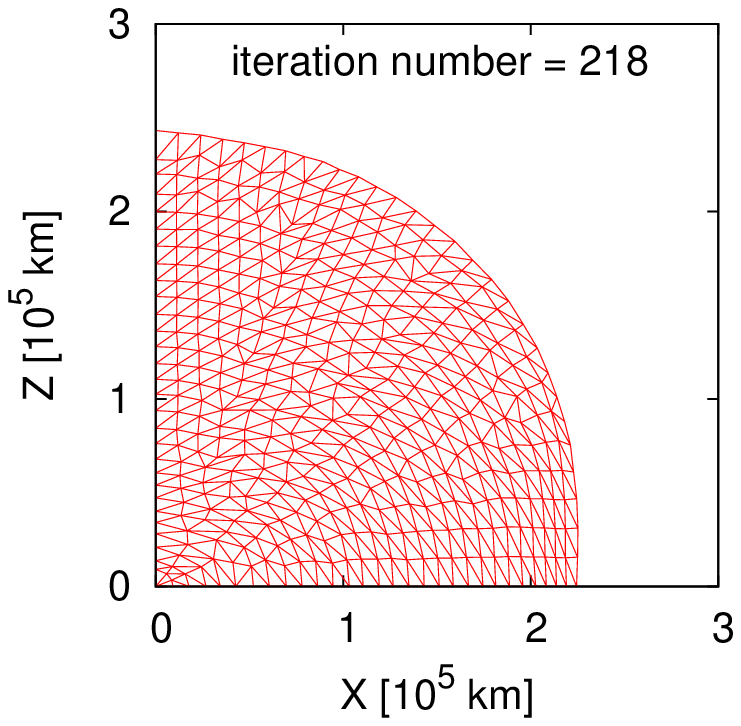}\\
\includegraphics[width=16pc]{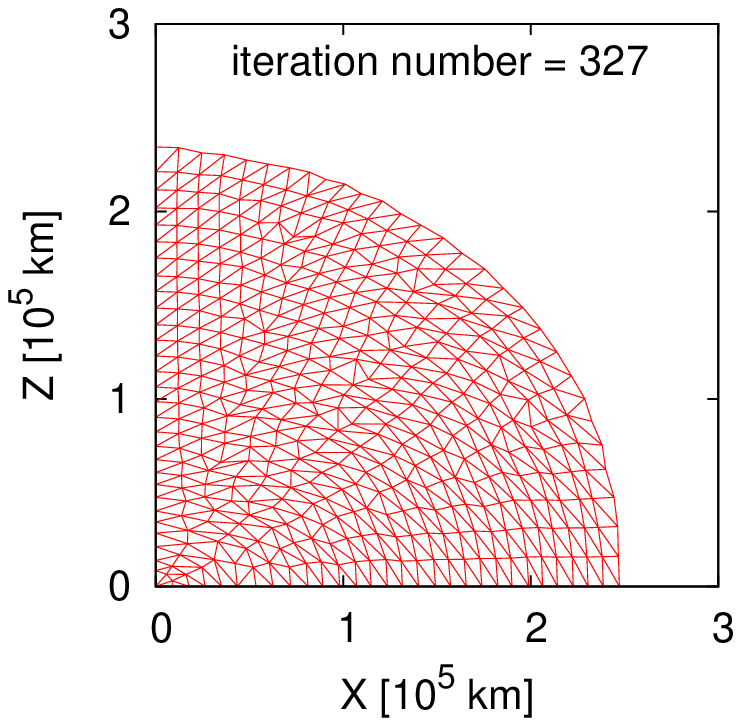}
\hspace{-15mm}
\includegraphics[width=16pc]{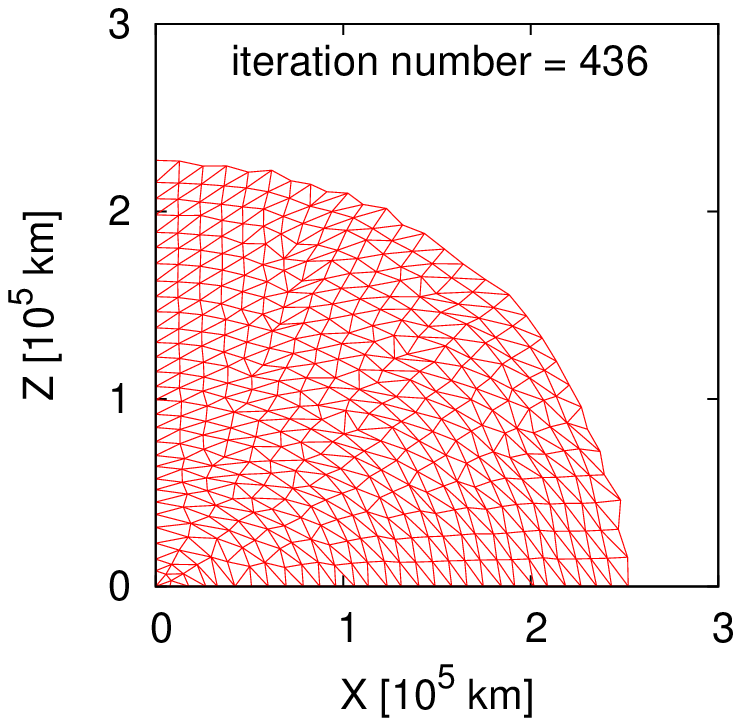}
\hspace{-15mm}
\includegraphics[width=16pc]{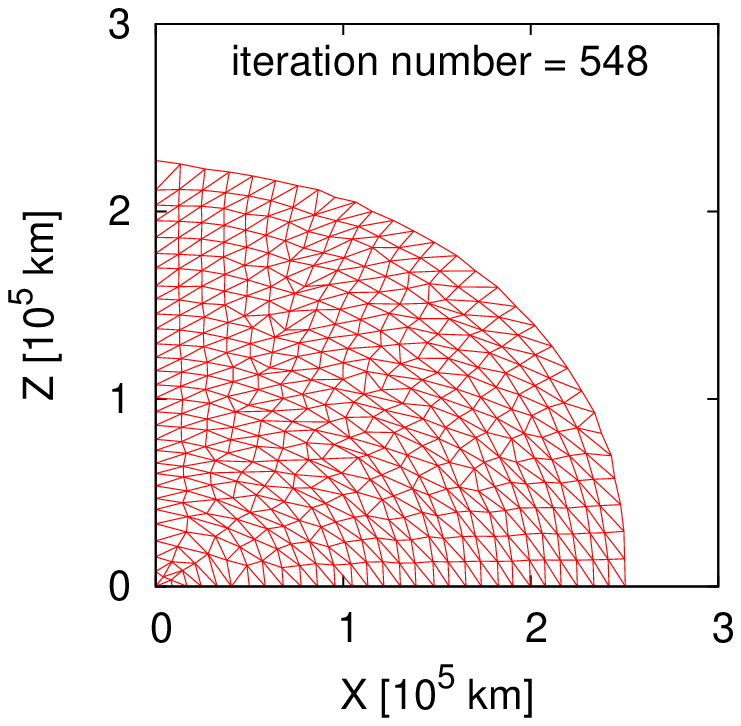}
\caption{\label{fig:x07edge} (colour on-line). 
Same as Fig.~\ref{fig:r07edge} but for case (B).} 
\vspace{5mm}
\end{figure*}

\begin{figure*}
\includegraphics[width=16pc]{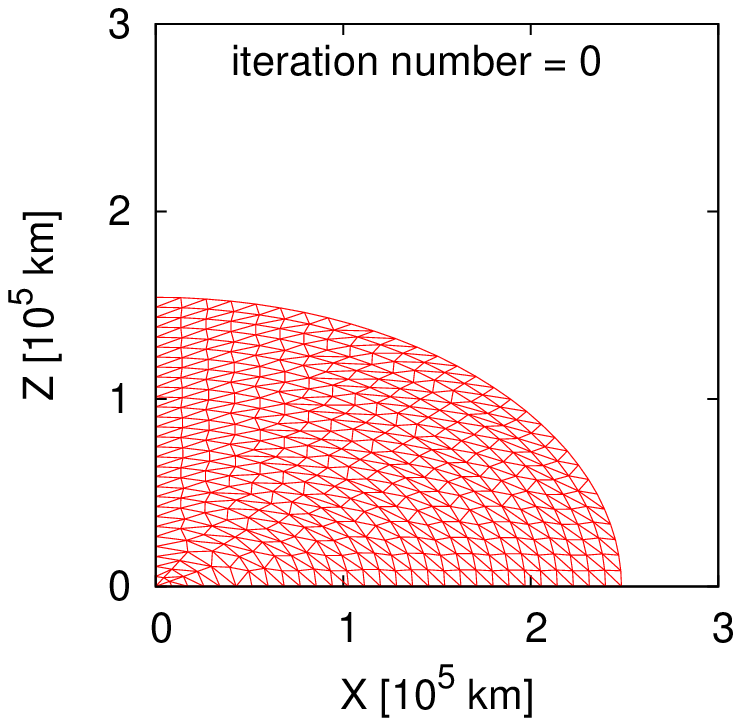}
\hspace{-15mm}
\includegraphics[width=16pc]{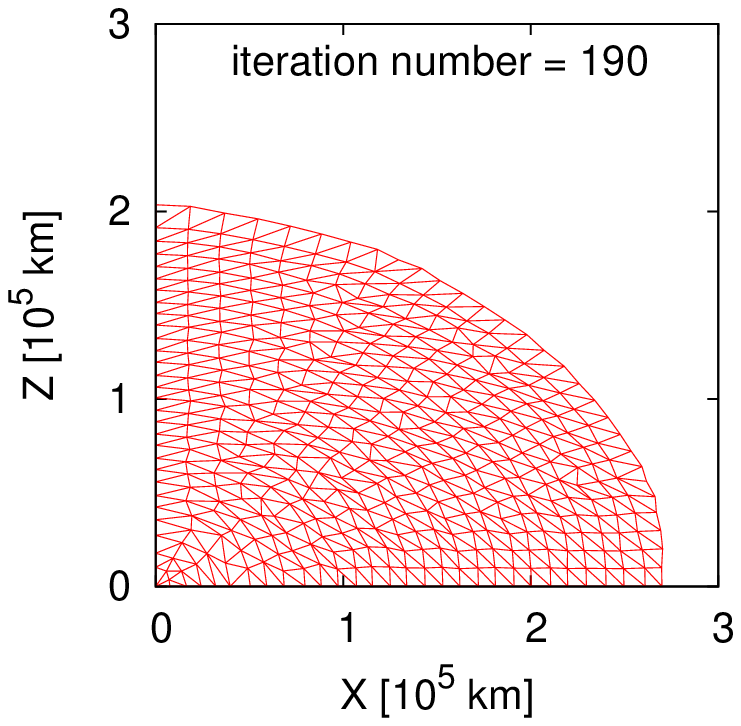}
\hspace{-15mm}
\includegraphics[width=16pc]{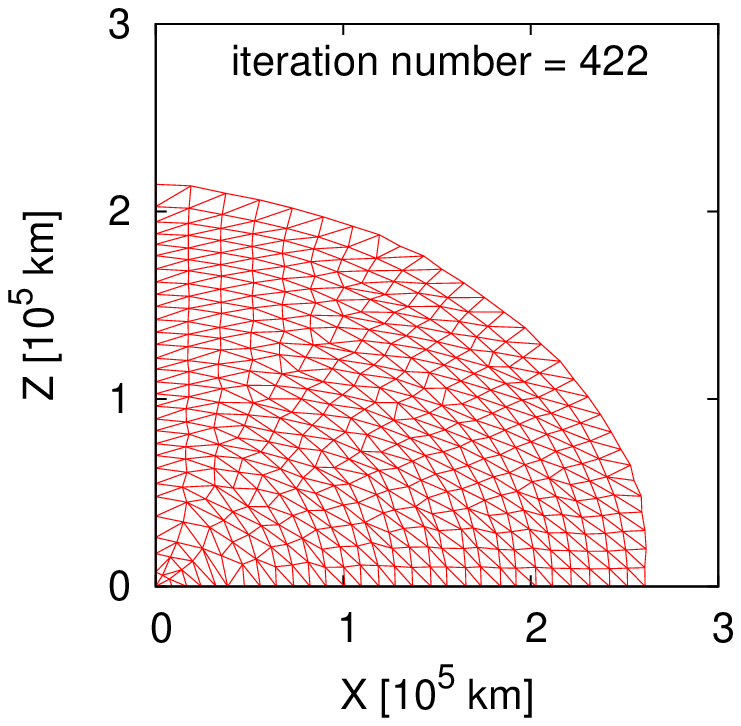}\\
\includegraphics[width=16pc]{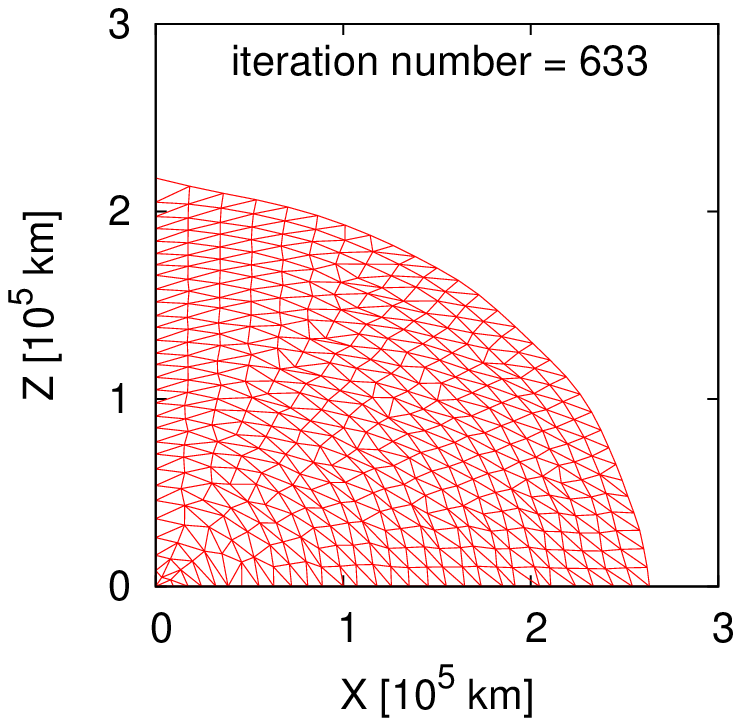}
\hspace{-15mm}
\includegraphics[width=16pc]{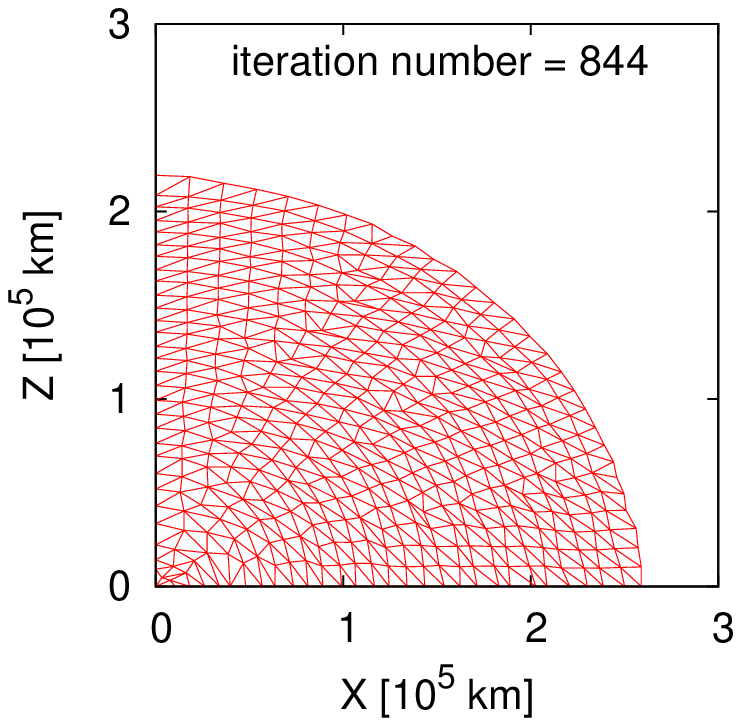}
\hspace{-15mm}
\includegraphics[width=16pc]{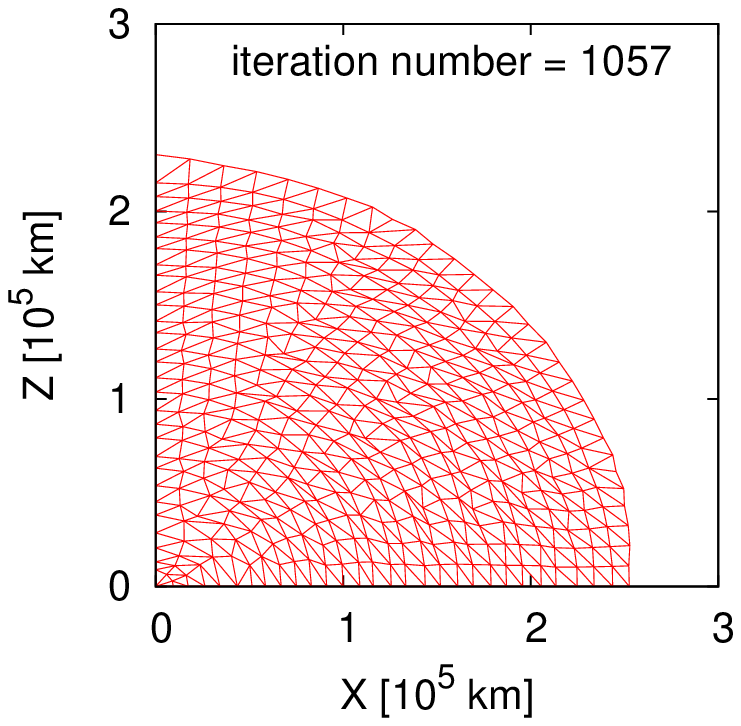}
\caption{\label{fig:z07edge} (colour on-line). 
Same as Fig.~\ref{fig:r07edge} but for case (C).}  
\end{figure*}

%----- paragraph on FIG.22-24 -----
We show the changes of the configurations with the Monte Carlo sweeps in Figs. \ref{fig:r07edge}-\ref{fig:z07edge}, which correspond to deformations (A)-({C}) respectively. The upper left panels in these figures exhibit the reference configurations, which are clearly different from the final configurations shown in the lower right panels. In particular, the reference configuration is prolate for case (B) whereas the rotational equilibrium is oblate owing to centrifugal forces. It is evident from the comparison of the lower right panels in the figures that the resultant configurations after the optimizations look almost identical with each other as well as with the original configuration given in Fig.\ref{fig:edge_shl}. We, hence, conclude that our scheme is robust to the variation of reference configuration. It should be also noted that in the application to the stellar evolution calculation, we will be able to use the stellar structure obtained at the previous time step as the reference configuration, which should be hence pretty close to the true equilibrium. 

It is one of the great advantages with the YFY scheme based on the Lagrange coordinates that we can record the history of the fluid element assigned to each node. This is true not only for the Monte Carlo sweep as has been demonstrated so far but also for the true evolution in time when the scheme is applied to the 2-dimensional calculation of stellar evolution. Since nuclear burnings occur locally in each fluid element, the chemical evolution is most easily treated on the Lagrange coordinates and hence by the YFY scheme. It should be also mentioned that since the specific entropy would be conserved for each fluid element in the absence of the generation and transport of heat, radiative transport of energy will be formulated most unambiguously on the Lagrange coordinates. This will be also the case for the transport of angular momentum. Convective transport of energy and angular momentum may be handled as well if it is justified to treat them approximately as diffusions. 

\begin{figure*}
\includegraphics[width=16pc]{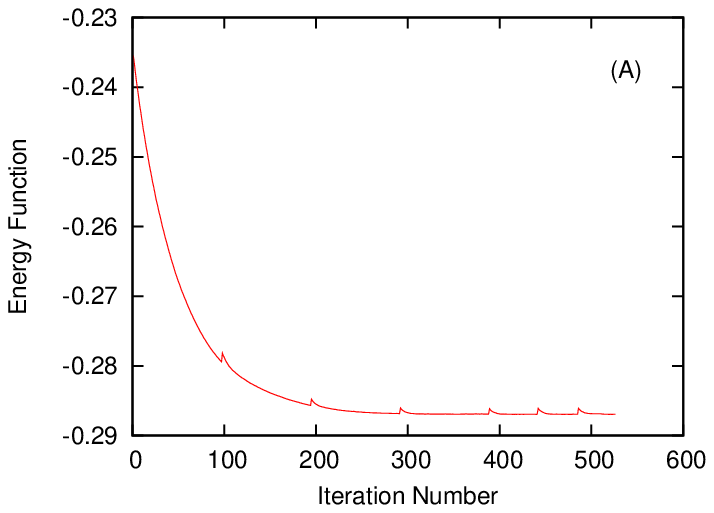}
\includegraphics[width=16pc]{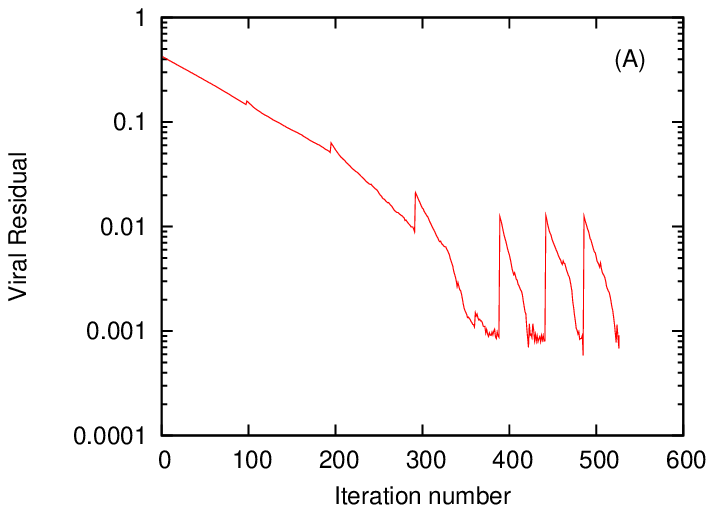}\\
\includegraphics[width=16pc]{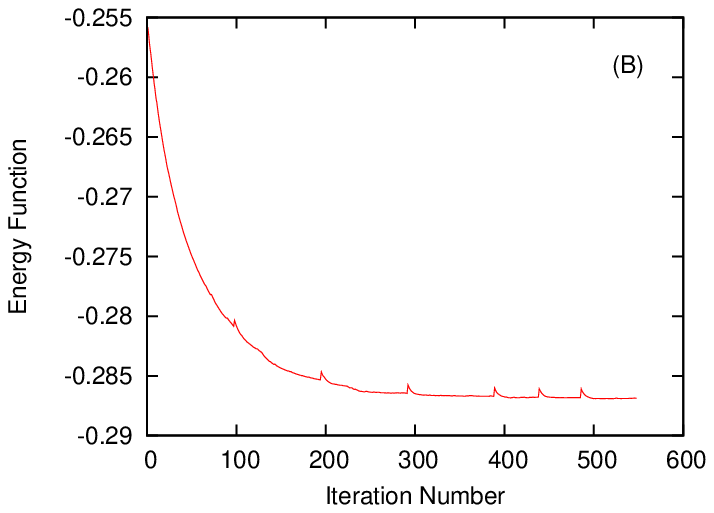}
\includegraphics[width=16pc]{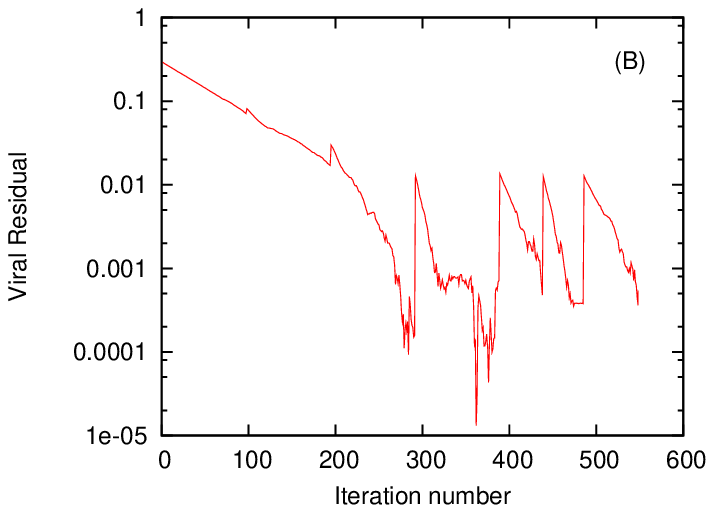}\\
\includegraphics[width=16pc]{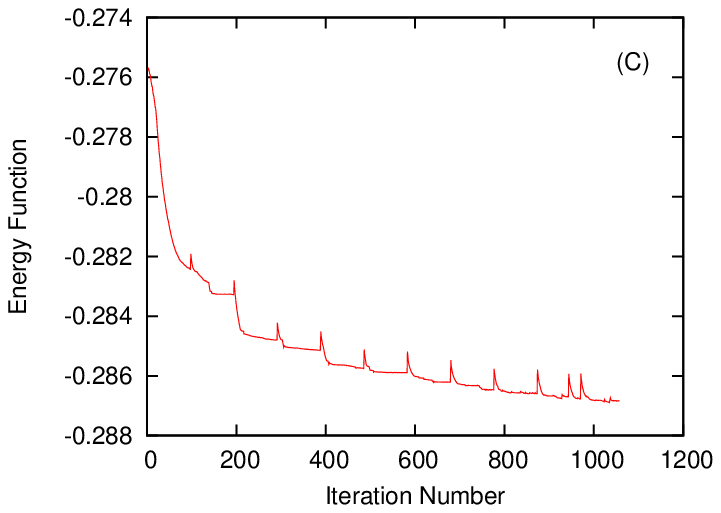}
\includegraphics[width=16pc]{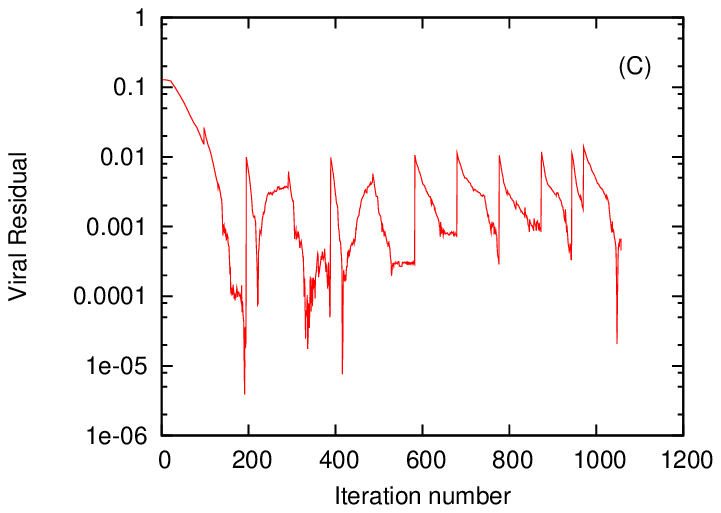}
\caption{\label{fig:vc} (colour on-line). 
Same as Fig.~\ref{fig:his_shl} but for the reference configurations given by the deformation for cases (A)-(C).} 
\end{figure*}

%----- paragraph on FIG.25 -----
The left three panels of Fig.~\ref{fig:vc} present the histories of the values of the energy function for the cases shown in Figs. \ref{fig:r07edge}-\ref{fig:z07edge}. 
The values of the energy function decrease in most of the time in the Monte Carlo sweep. Sporadical small glitches are due to the smoothings, which are necessary, as repeatedly mentioned, to ensure the convergence to the true minimum, avoiding the trap of false minima.

Shown in the right panels of Fig.~\ref{fig:vc} are histories of the virial residuals for the same models. In the previous sections, we used the condition $V_C < 10^{-3}$ as one of the diagnostics to judge if the equilibrium is reached as described in section 2. This condition works well also in the current cases as can be seen from the comparison of the right panels in Fig.~\ref{fig:vc} with Figs. \ref{fig:r07edge}-\ref{fig:z07edge}. The importance of the criterion is also understood from Fig.~\ref{fig:his_shl}, in which the virial residual is still large, $V_C > 10^{-3}$, after $\sim$120 iterations although the energy functional appears to have hit the minimum. The  configuration is still out of equilibrium indeed, which can be confirmed in the corresponding structure shown in Fig.~\ref{fig:edge_shl}.
%Otherwise ($V_C > 10^{-3}$), the structures do not reach to equilibrium state. Just after the first 150 iteration-steps in Fig.~\ref{fig:his_shl}, all virial residuals are $V_C > 10^{-3}$, and the corresponding structures in Figs. \ref{fig:r07edge}-\ref{fig:z07edge} are on the way of variation plainly.

It should be noted, however, that the condition on the virial residual alone is not sufficient for the judgement, since the virial residual is an integrated quantity and may become small accidentally. For example, the configuration presented in the upper-middle panel in Fig.~\ref{fig:z07edge} for the 190th iteration has not yet reached the equilibrium although the virial residual satisfies $V_C < 10^{-3}$ as shown in Fig.~\ref{fig:vc}. It is hence mandatory to take into account the convergence of energy function in addition to the condition on the virial residual. Considering the smoothing administered every $\sim$100 sweeps (see the last paragraph of this section), we impose the condition that the energy function at the $i-$th iteration, $E(i)$, should not lower than $E(i-100)$ $i-100$th to $i$th step. Although this simple criterion seems to work well at least for the models considered in this paper, it may need improvement.

\begin{figure*}
\includegraphics[width=20pc]{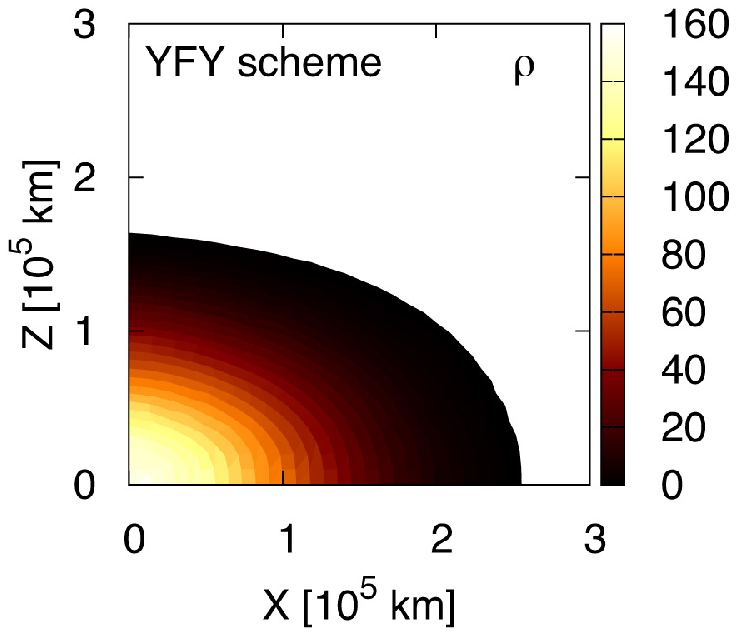}
\hspace{-8mm}
 \includegraphics[width=17.5pc]{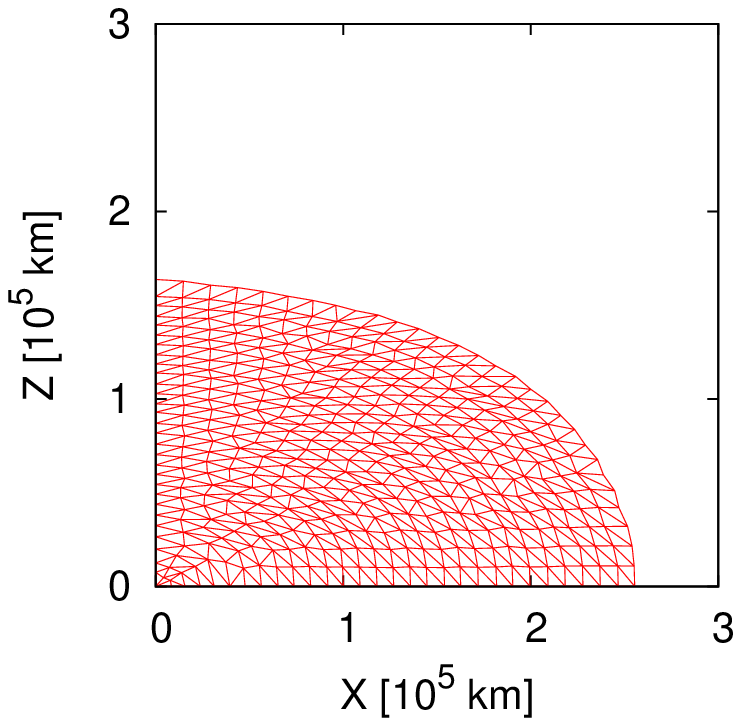}
\caption{ \label{fig:sd60} (colour on-line). 
The density distribution~(left panel), and the edges and nodes~(right panel) for shellular-type configuration for $R_p/R_e=0.60$ obtained with the YFY scheme. 
The result is essentially the same as the one derived with the FSCF scheme.
%, since the variational principle does not work properly at the mass-shedding limit. 
The unit of density is given in cgs. The central density, equatorial radius are $\rho_0 = 1.51 \times 10^2$ g cm$^{-3}$, $R_e = 2.57 \times 10^5$ km, respectively.} 
\end{figure*}

In this paper it is not our purpose to construct a sequence of rotational configurations with different angular momenta up to the mass shedding limit. As a matter of fact, the Monte Carlo method employed in this paper to find a configuration that gives the minimum to the energy functional is not suited very much for finding the limiting configuration for two reasons. Firstly, in the Monte Carlo method many (non-equilibrium) configurations are tried to reach the true equilibrium and in so doing some trial configurations give energies lower than the true minimum erroneously owing to numerical errors more often than not, which leads to a ``numerical" mass shedding and no convergence. Secondly, the equilibrium configuration may not be unique for a given distribution of specific angular momentum on the Lagrangian coordinates. It is well known in fact that the angular momentum is not a monotonic function of  the ratio of the polar to equatorial radii, $R_p/R_e$, near the mass shedding limit (see, for example, Table 3 in \citet{fujisawa15}, in which the angular momentum increases initially as the ratio decreases but it decreases after the peak that appears at $R_p/R_e \sim 0.55$). In this regime we are afraid that the Monte Carlo method may not work properly. As a matter of fact, we have confirmed for a model with $R_p/R_e=0.5$ that this is indeed the case. We emphasize, however, that our formulation can handle rapid rotators as demonstrated in Fig. \ref{fig:sd60}, in which we display a rotational configuration with the axis ration of $R_p/R_e=0.60$, which we find is essentially the same as the result obtained with the FSCF method. Incidentally, this model is in the regime, where the solution is unique in the above sense. 

\bigskip

\begin{figure*}
\includegraphics[width=14.5pc]{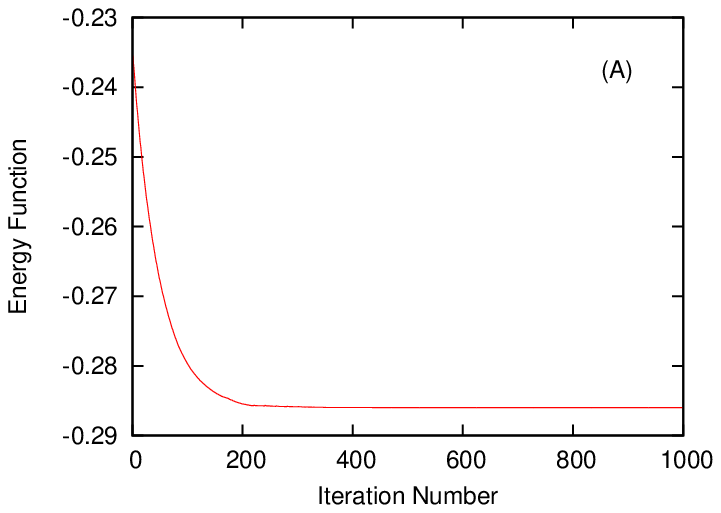}
\includegraphics[width=14.5pc]{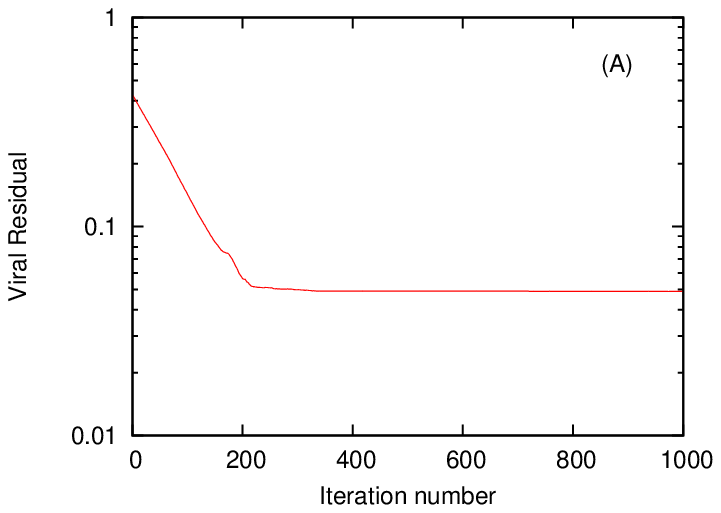}
\hspace{-10mm}
 \includegraphics[width=14.5pc]{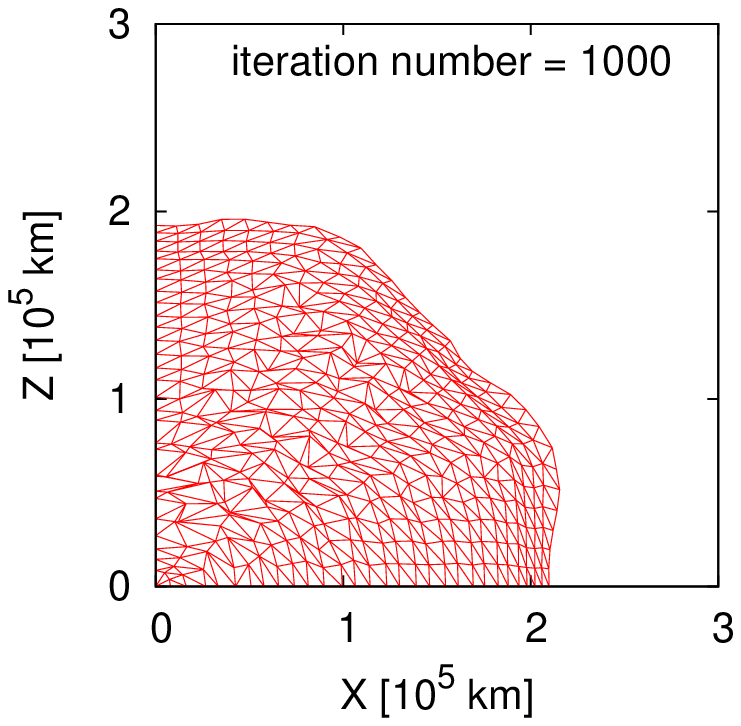}
\caption{\label{fig:fail} (colour on-line). 
The values of the energy function (left panel) and of the virial residual (middle panel) with initial model (A), but without the smoothing against acute-angled triangles.
The right panel corresponds to the resultant structure at 1000-th iteration number.} 
\end{figure*}

%----- paragraph on FIG.27 -----
At the end of section, we would like to discuss the importance of the smoothing. We employ the variational principle to obtain the hydrostatic equilibria. The idea is simple, but the implementation is not indeed. Without an appropriate smoothing, Monte Carlo sweeps tend to make too deformed triangles, which lead the generation of a false local energy minimum. In Fig.~\ref{fig:fail} we show the result of the calculation without the smoothing of acute-angled triangulations. There is no glitch in the panels of the energy function nor of the virial residual, since we do not take into account the smoothing. The energy function is settled after 1000 sweeps, although the virial residual does not satisfy $V_C < 10^{-3}$. This result suggests that the search of minimum energy is trapped in a local minimum. The resultant structure is far from the final results shown in Figs. \ref{fig:r07edge}-\ref{fig:z07edge}. This is the reason why the smoothing process is necessary. 

There will be many possible ways to alleviate highly acute triangles, which could be the origin of false local minima. Followings are the simple measures we employed in this study: 
\begin{description}
%%%%%%%%%%%%%インデントがいまいちらしい％％％％％％％％％％％％％％％
 \setlength{\leftskip}{4.0 mm}
\item \hspace{-8mm}(1) if there is a triangular cell that experiences more than 30 \% of change in area after a shift of a node during the Monte Carlo sweep,  we move this particular node of the triangle to the center of the polygon formed with its nearest neighboring nodes. 
\item \hspace{-8mm}(2) since the local shifts of some nodes in the first step sometimes produce zigzag distributions of the nodes that are initially located on one of the concentric surfaces (see the construction of the triangulated mesh in the reference configuration), we make them somewhat smoother by moving the grid on the top or bottom of the zigzag distribution toward the middle of  the two neighboring nodes; the first step is always followed by the second step.
\item \hspace{-8mm}(3) in rare cases, the order of the grid points on the axis or on the equator is changed by their large shifts in the Monte Carlo sweep; then we restore the original order by pulling back the grids in trouble toward the original positions appropriately; this measure is take only when it is needed.
\item \hspace{-8mm}(4) even if we do not find any large areal-change, we administer the smoothing prescribed in (1) to all nodes periodically; from our experience it should be done every $\sim$100 ($\sim$ the number of grid points / 5) sweeps; although this is mainly meant to be a preventive measure, it is also effective for the escape from a local minimum.
\end{description}
The above prescriptions are admittedly empirical and depend on the construction of the triangulated mesh as well as the direction of Monte Carlo sweep. Although there should be more sophisticated and systematic ways to treat the false local minimum, we defer a further improvement to the future work, since the measures given above appear to work well at least for the models considered in this paper.  

%\clearpage

%----- FIG.21-----

%----- FIG.22-----

%----- FIG.23-----

%----- FIG.24-----

%\clearpage

%----- FIG.25-----

%----- FIG.26-----

\newpage

%----------------------
\section{Summary and Discussions}
%----------------------

We have developed a new formulation to obtain rotational equilibria numerically. The scheme can handle not only barotropic but also baroclinic configurations, which are critically important for the application to  realistic stars. Such an achievement is itself a major break through to the status quo, in which previous works are not many and, more importantly, they are all based on the Eulerian description~\citep{uryu94, uryu95, espinosa07, espinosa13, rieutord16}. Our method, on the other hand, employs the Lagrangian description just as in one-dimensional counterparts for non-rotating stars and hence stellar evolution calculations, since we can trace the potentially complex movements of each fluid element. Our formulation is based on the variational principle, in which rotational equilibria are obtained as the configurations that optimize the energy functional for given distributions of mass, specific entropy, and angular momentum on the Lagrangian coordinates. In this paper all physical quantities are discretized on triangulated grids. 
 
In order to validate our scheme, we compare the configurations obtained by our scheme with those by other Eulerian schemes: the HSCF scheme developed by Hachisu~\citep{hachisu86} for the barotropic configuration, and the FSCF scheme conceived by Fujisawa~\citep{fujisawa15} more recently for the baroclinic one. 
We have confirmed that all the configurations including the ones with shellular-type rotations obtained in this paper are linearly stable against axisymmetric perturbations and comply with the Bjorkness-Rosseland rule.

In this comparison, we have reproduced the equilibria a shellular-type rotation, which were obtained only recently by Fujisawa although the shellular rotation is commonly assumed in the one-dimensional stellar evolution calculation with rotation being taken into account only approximately. It is found that the result obtained with the YFY scheme agree with those derived by other schemes with an error of 5 $\sim$ \%,  which we think is reasonable, considering the relatively small number of nodes (489) we used in this paper. 
%The numerical accuracy is lowered in the regions near the surface and rotation axis, since they contribute little to the energy functional. 
%---- Yamada-san ----
It is then a legitimate question to ask how the accuracy is improved with the number of nodes. In order to address this issue, we have conducted additional computations, changing the node number from $\sim$ 200 to $\sim$1000, the latter of which is the maximum node number we can afford at present, since the numerical cost is proportional to $N^2$ in our currently unparallelized code, where $N$ is the node number. 

\begin{figure*}
 \hspace{-5mm}
 \includegraphics[width=14.pc]{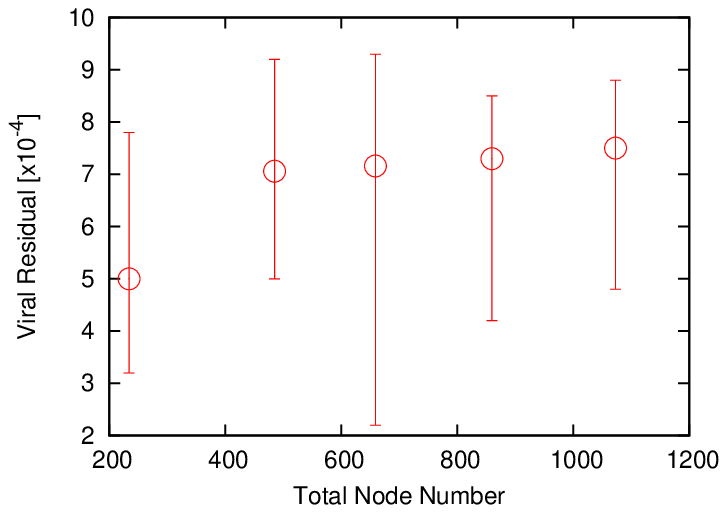}
\includegraphics[width=14.pc]{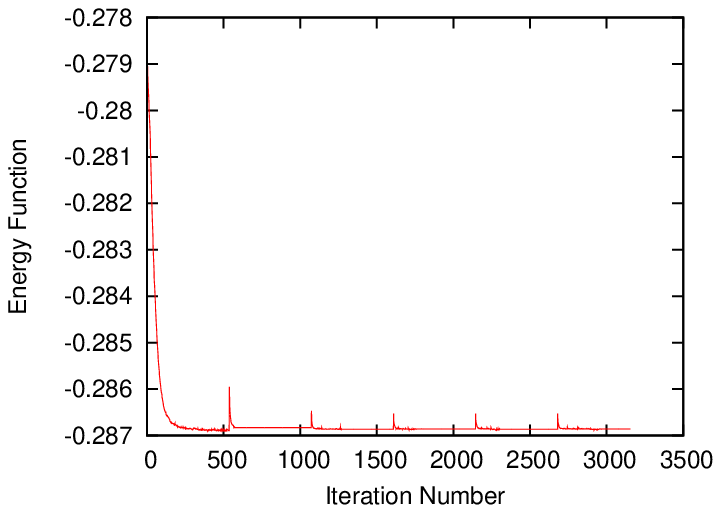}
\includegraphics[width=14.pc]{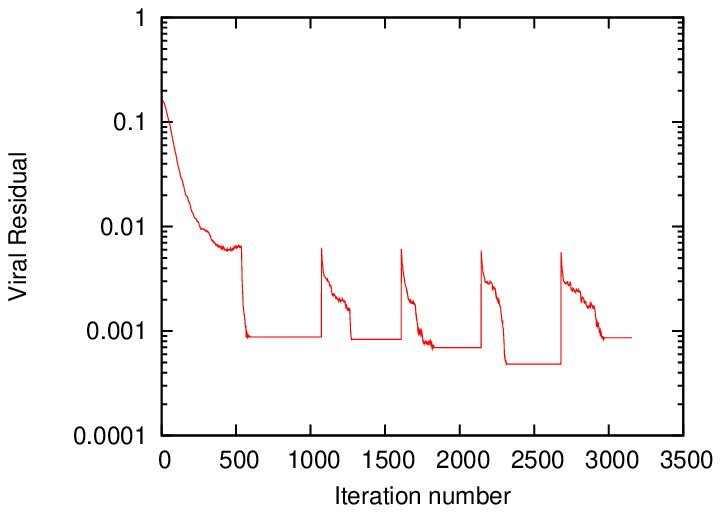}
\caption{\label{fig:ebl} (colour on-line). 
The range of converged virial residual for each resolution (left panel). The middle and the right panels show an example of the energy functional and the virial residual with 1073 nodes. 
%The converged virial residual is determined at ${\partial E}/{\partial N_i}=0$ and ${\partial V_C}/{\partial N_i}=0$. In this meaning, there are 12 types of converged values of the virial residuals in the middle and the right panels. 
} 
\end{figure*}

We find that the higher resolutions with $\sim$800 and $\sim$1000 nodes do not improve the accuracy at a recognizable level as shown in the leftmost panel of Fig.\ref{fig:ebl}. As demonstrated for the highest resolution model in the middle and rightmost panels of this figure, as the final configuration is approached, the values of energy function and virial residual repeat a sudden jump followed by a settlement to some values which satisfy our convergence criteria. This is due to the smoothing that we administer regularly to avoid the trapping by false minima (see the last part of the last section). Because of the probabilistic nature of the Monte Carlo method, the final value of the virial residual after each repetition fluctuates around the mean value. In the left panel of Fig.\ref{fig:ebl} we present these mean values and variances as dots and bars, respectively, for different node numbers.

%----- FIG.27-----

The reason why the precision is not improved is the following: in our formulation we search for the rotational configuration that gives the lowest value to the energy functional for given mass, specific entropy and angular momentum; it is not surprising then that  some nodes contribute more than others; in particular, the nodes close either to the surface or to the rotational axis are the least contributors, since they have either small densities or volumes; this means in turn that their positions are very difficult to obtain accurately; this may be understood if one scrutinizes the right panels of Figs. 2, 3, 6, 9, 10, 15, 18 as well as the lower panels of Figs. 14, 19, 20; on the hand, these nodes give contributions of the order of $10^{-4}$ to the virial residual. 

In principle, if the number of nodes is sufficiently large particularly in the regions of our concern, then the virial equality should be improved. The current resolutions we can afford are way too small, though, as can be understood from the figures showing the node configurations. This may be remedied if we implement the multi-layer treatment. 

Multi-layer treatment may be useful to improve the accuracy\footnote{Such a treatment is already employed in their scheme by Espinosa Lara and Rieutord \citep{espinosa13} and, combined with the spectral method, achieves very high accuracy. See also Kiuchi et al. \citep{kiuchi10}},  which will be also required in applying the new scheme to rotating stars in advanced evolutionary stages. 
%It is stressed that the our formulation can be extended to treat such multi-layer structures without any difficulties. {\bf The previous methods need to be imposed some boundary conditions  between layers appropriately. We, then, need to set the boundary conditions between the layers depended on the stage of evolution. On the other hand, our method does not need.  The only need to do is just to prepare the initial profiles of mass, angular momentum, entropy, and elementary fractions, and the multi-layers appear as the optimal result for the initial guesses: it does not have the concept of layers namely. }
As explained earlier, the numerical error comes mainly from the regions, whose contributions to the energy functional are minor. If the multi-layer treatment is employed, these minor regions can be treated separately from the major regions. Then the minor contributions are no longer minor. The important thing is that the variational principle can be applied to each region individually, with other regions serving as fixed gravitational potentials. In order to obtain the global equilibrium, we need to find an equilibrium configuration in each region consecutively and iteratively until all these configurations do not change any longer simultaneously. In addition to the minor regions mentioned above, the central region may also need to be treated with a special care, since the nodes are rather sparse  there as can be seen, e.g., in the right panel of Fig. 1.
 
We have demonstrated that our scheme is robust, obtaining the same final configuration in equilibrium irrespective of the reference configurations assumed initially. We have also observed, on the other hand, that the Monte Carlo sweeps to get to the minimum of the energy functional are prone to be trapped by false local minima, which are generated by deformations of the numerical grid and that the smoothing should be properly administered to escape them. In order to distinguish the true minimum from a false one, the virial residual is found to be a good diagnostic. Although we have not made any attempt in this paper to reconstruct a triangulated mesh when it becomes too deformed, such re-gridding will be necessary when applying the YFY scheme to the evolution of rotating stars.
%---- 意味不明。 ---- 
%The concept of our scheme is quite simple: {\it ``Find a energy minimum for optimal stellar-structures"}. We, however, find two more arts to find the solutions empirically. One is the virial residual, which is a necessary condition for the hydrostatic equilibrium. The convergence of energy is not the sufficient for the force balance because of numerical errors, e.g. the local minimum problems come from the acute triangle mesh. The another is the smoothing procedure to reduce the acute-angled triangulations grids artificially, otherwise the system will be trapped in the local energy minimum. We fix the adjacency matrix in the smoothing process in this paper, but there may be some alternatives for that. We then need to take into account the opacities to obtain a realistic atmosphere models as employed in 1D simulations.  Otherwise we can not discuss the mass loss in stellar evolution calculations.

The next step toward the application to the stellar evolution calculation will be to incorporate radiation transport in rotational equilibria. Unlike the previous works~\citep{uryu94, uryu95, roxburgh06, espinosa07, espinosa13, rieutord16}, we will solve time-dependent diffusion equation on the same triangulated grid, which will not be a difficult task. The meridional circulation is another important ingredient~\citep{zahn92} when considering the application of our scheme to actual rotating stars, since it is supposed to play a major role in the re-distribution of angular momentum and elements (e.g. \citealt{mathis09}). Although such motions cannot be handled directly with our scheme, they may be incorporated either as advections or as diffusions as just as in current 1D calculation. 

Convection is even more difficult to treat. The redistributions of angular momentum and elements as well as entropy in this case may be approximated as diffusions or advections like the meridional circulation but the real difficulty with the convection is the fact that the convectively unstable configurations do not correspond to the minimum of the energy functional but to a saddle point. A couple of ideas are currently being tested and will be reported elsewhere. 

Since our scheme is applicable only to axisymmetric configurations, in which the specific angular momentum is conserved in each fluid element, extension of our formalism to 3D configurations such as triaxial equilibria~\citep{tassoul78} is much beyond the scope of this paper.

Possible applications of our scheme are not limited to the ordinary stars, though. They will be extended to e.g., compact stars, proto-stars and proto-planets to mention a few. Incorporation of magnetic fields and/or general relativity should be considered in due course. 

\section*{Acknowledgements}
We are grateful to K. Suzuki, T. Yamasaki, M. Okamoto, and T. Maruyama for fruitful discussions. This work was partially supported by JSPS KAKENHI Grant Numbers 25105510, 24244036 and Grant-in-Aid for Scientific Research on Innovative Areas, No.24103006.

\appendix

%---- How to interpolate ------
\section{Transformation matrix}\label{appendix:trns}
In this appendix we give the transformation matrix $T_{lm}$ that is used in the expression of the discretized energy functional, equation (\ref{eq:eq2}). It is actually introduced in association with the interpolation of physical quantities in each triangular cell.

\begin{figure}
\begin{center}
\includegraphics[width=20pc]{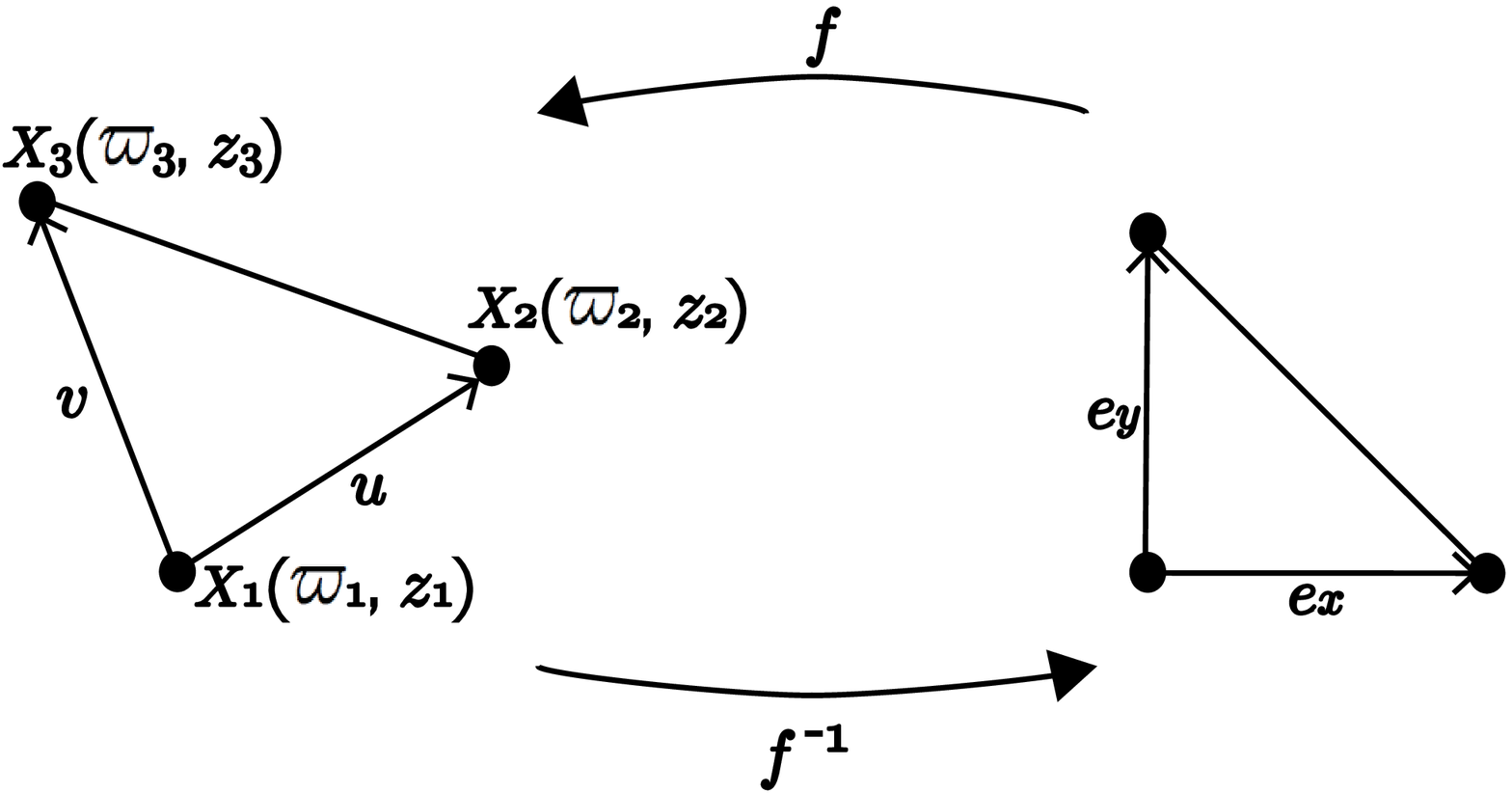}
\end{center}	
\caption{\label{fig:trans_mtrx} The diffeomorphic mapping that gives the Jacobian. } 
\end{figure}

Suppose that a quantity $X$ is given by on all nodes and we interpolate them in each triangular cell as
\begin{equation}
X = \alpha_1^X  z + \alpha_2^X \varpi + \alpha_3^X,
\end{equation}
in which $\varpi$ and $z$ are the cylindrical coordinates of an arbitrary point in the cell and $\alpha$'s are numerical coefficients to be determined. Evaluating the above equation at three apexes (labeled with $l$) of the $n$-th triangular cell, one obtains the following equations:
\begin{eqnarray}
&&{X_l(n)} = {T_{lm}(n)}{\alpha_m^X(n)}, 
\end{eqnarray}
where $l,~m=1, 2, 3$. The matrix ${\bm T(n)}$ in this expression is referred to {\it the transformation matrix} in this paper, in which defined as 
\begin{equation}
{\bm T(n)} =
\left(
    \begin{array}{ccc}
      z_1(n) & \varpi_1(n) & 1\\
      z_2(n) & \varpi_2(n) & 1\\
      z_3(n) & \varpi_3(n) & 1
    \end{array}
\right).
\end{equation}
This matrix is useful in {\it the finite element method} employed here. For example, the volume of the triangular cell $V_\Delta$ is evaluated as,  
\begin{equation}
V_\Delta =\int_{\rm cell} 2 \pi \varpi ~d \varpi~ dz = 2 \pi \cdot \frac{|{J}|}{2} \cdot \frac{\varpi_1+\varpi_2+\varpi_3}{3}.
\end{equation}
Here $|{J}|$ is the Jacobian for the diffeomorphic mapping $f$ of an isosceles right triangle constructed 
 by the unit fundamental vectors ${\bf e_x},~{\bf e_y}$ of the two-dimensional Cartesian coordinates to
 the triangle with three apexes $ X_1 X_2 X_3$ as shown in Fig.~\ref{fig:trans_mtrx}.
The Jacobian is then equal to the determinant $|{\bm T}(n)|$. This is obtained from the following relation between two forms: Here, the areal element under the mapping is shown as  
\begin{eqnarray}
{du} \wedge {dv}= J(f) dx \wedge dy, 
\end{eqnarray}
where ${\bm u}$ and ${\bm v}$ are the vectors $\overrightarrow{X_1X_2}$ and $\overrightarrow{X_1X_3}$, respectively. 
${|{\bm T(n)}|}/{2}$ in equation (A4) hence is the area of the triangular-cell whereas $(\varpi_1+\varpi_2+\varpi_3)/3$ is the $\varpi$-coordinate of the geometrical centre of the triangule.

%----- FIG.A1-----

%%%%%%%%%%%% REFERENCE %%%%%%%%%%%%


\begin{thebibliography}{21}
\expandafter\ifx\csname natexlab\endcsname\relax\def\natexlab#1{#1}\fi
\expandafter\ifx\csname bibnamefont\endcsname\relax
  \def\bibnamefont#1{#1}\fi
\expandafter\ifx\csname bibfnamefont\endcsname\relax
  \def\bibfnamefont#1{#1}\fi
\expandafter\ifx\csname citenamefont\endcsname\relax
  \def\citenamefont#1{#1}\fi
\expandafter\ifx\csname url\endcsname\relax
  \def\url#1{\texttt{#1}}\fi
\expandafter\ifx\csname urlprefix\endcsname\relax\def\urlprefix{URL }\fi
\providecommand{\bibinfo}[2]{#2}
\providecommand{\eprint}[2][]{\url{#2}}

\bibitem[Burbidge et al.(1957)]{b2fh57} 
Burbidge K. M., Burbidge G. R., Fowler W. A., Hoyle F.
\ 1957, Review of Modern Physics, 29, 4 

\bibitem[Cameron (1957)]{cameron57} 
Cameron, A. G. W., \ 1957, Chalk River Report, CRL-41


\bibitem[Chandrasekhar(1969)]{chandrasekhar} Chandrasekhar, S.\ 1969, 
The Silliman Foundation Lectures, New Haven: : {\it Ellipsoidal figures of equilibrium}, Yale University Press  


\bibitem[Endal \& Sofia(1976)]{endal76} Endal, A.~S., \& Sofia, S.\ 1976, \apj, 210, 184 
%\bibitem[Endal 
%\& Sofia(1978)]{endal78} Endal, A.~S., \& Sofia, S.\ 1978, \apj, 220, 279 

\bibitem[{{Espinosa Lara} \& {Rieutord}(2007)}]{espinosa07}
{Espinosa Lara} F., {Rieutord} M., 2007, \aap, 470, 1013

\bibitem[Espinosa Lara \& Rieutord(2013)]{espinosa13}
---, 2013, \aap, 552, A35

\bibitem[{{Eriguchi} \& {Mueller}(1985)}]{eriguchi85}
{Eriguchi} Y., {Mueller} E., 1985, \aap, 146, 260

\bibitem[Friedman \& Schutz(1978)]{friedman} Friedman, J.~L., \& Schutz, B.~F.\ 1978, \apj, 221, 937 

\bibitem[Fukuda(1982)]{fukuda82} Fukuda, I.\ 1982, \pasp, 94, 271

\bibitem[Fujisawa(2015)]{fujisawa15} Fujisawa, K.\ 2015, MNRAS 454, 3060

%\bibitem[Hachisu \& Eriguchi (1984)]{hachisu84} Hachisu, I., Eriguchi, Y., \ 1984, AP\&SS, 99, 71 

\bibitem[Hachisu(1986)]{hachisu86} Hachisu, I.\ 1986, ApJS, 61, 479 

\bibitem[Hayashi(1961ab)]{hayashi61}
Hayashi, C., \ 1961, PASJ, 13, 442: Hayashi, C., \ 1961, PASJ, 13, 450. 

\bibitem[Heger et al.(2000)]{heger00} Heger, A., Langer, N., \& Woosley, S.~E.\ 2000, \apj, 528, 368 

\bibitem[Henyey et al.(1964)]{henyey64} Henyey, L.~G., Forbes, 
J.~E., \& Gould, N.~L.\ 1964, \apj, 139, 306  

\bibitem[Kiuchi et al.(2010)]{kiuchi10} Kiuchi, K., Nagakura, H., \& Shoichi, Y.\ 2010, \apj, 717, 666  

\bibitem[Maeder 
\& Meynet(2000)]{maeder00} Maeder, A., \& Meynet, G.\ 2000, Astron. Astrophys., Suppl. Ser., 38, 143 

\bibitem[Maeder et 
al.(2013)]{maeder13} Maeder, A., Meynet, G., Lagarde, N., \& Charbonnel, C.\ 2013, \aap, 553, A1 
%\bibitem[Mathis; Rieutord \& Espinosa Lara(2013)]{LNP} Mathis, S. (Chap.2), Rieutord, M. \& Espinosa Lara, F. (Chap. 3) \ 2013, in Lecture Notes in Physics, Berlin Springer Verlag, 865, {\it Seismology for Studies of Stellar Rotation and Convection}, ed. Goupil,M., Belkacem, K., Neiner, C., Lignieres, F., \& Green, J. J. 49 %, astro-ph/1208.4926

\bibitem[Mathis (2009)]{mathis09} Mathis, S.\ 2009, \aap, 506, 811 

\bibitem[Mathis (2013)]{LNP} Mathis, S. 2013, 
in Goupil M., Belkacem K., Neiner C., Lignieres F., Green
J. J., eds, Lecture Notes in Physics, Vol. 865, Seismology for Studies of
Stellar Rotation and Convection. Springer-Verlag, Berlin, p. 49

\bibitem[Mathis et al.(2004)]{mathis04} Mathis S., Palacios A., Zahn J.-P., 2004, \aap, 425, 243

\bibitem[Mathis et al.(2013)]{mathis13} Mathis, S., Decressin, T., Eggenberger, P., \& Charbonnel, C.\ 2013, \aap, 558, A11 

\bibitem[Meynet \& Meynet (1997)]{meynet97} 
Meynet G. \& Maeder A. \ 1997, \aap, 321, 465 

\bibitem[Ostriker \& Mark(1968)]{ostriker} Ostriker, J.~P., \& Mark, J.~W.-K.\ 1968, \apj, 151, 1075 


\bibitem[Potter et al.(2012a)]{potter12a} Potter A. T., Tout C. A., Brott I., 2012a, MNRAS, 423, 1221 

\bibitem[Potter et al.(2012b)]{potter12b} Potter A. T., Tout C. A., Eldridge J. J., 2012b, MNRAS, 419, 748

\bibitem[Rieutord et al.(2016)]{rieutord16} Rieutord, M., Espinosa Lara, F., Putigny, B. 2016, J. Computational Phys., 318, 277

\bibitem[Roxburgh(2006)]{roxburgh06} Roxburgh, I.~W., 2006, \aap, 454, 883

\bibitem[Takahashi et al.(2014)]{takahashi14}
Takahashi K., Umeda H., Yoshida T., 2014, ApJ, 794, 40

\bibitem[Talon et al.(1997)]{talon97} Talon, S., Zahn J.-P., Maeder A., Meynet G., 1997, \aap, 332, 209

\bibitem[Tassoul(1978)]{tassoul78} Tassoul, J.-L.\ 1978, 
Princeton Series in Astrophysics, Princeton: {\it Theory of rotating stars}, University Press, and references therein 

\bibitem[Uryu \& Eriguchi(1994)]{uryu94} Uryu, K., \& Eriguchi, Y.\ 1994, MNRAS, 269, 24 

\bibitem[{{Uryu} \& {Eriguchi}(1995)}]{uryu95}
---, 1995, \mnras, 277, 1411

\bibitem[Woosley et al.(2002)]{woosley02} Woosley, S.~E., Heger, A., \& Weaver, T.~A. \ 2002, Reviews of Modern Physics, 74, 1015, and references therein 

\bibitem[Yasutake et al. (2015)]{yasutake15} 
Yasutake, N., Fujisawa K., Yamada S. \ 2015, MNRAS letter, 446, L56

\bibitem[Zahn (1992)]{zahn92} 
Zahn, J.-P., \ 1992, \aap, 265, 115

\end{thebibliography}
\end{document}